\documentclass[fleqn]{2023SCGE}
\usepackage{orcidlink}
\usepackage{multirow}
\usepackage{adjustbox}
\usepackage{placeins}
\usepackage[percent]{overpic} 
\usepackage[T1]{fontenc}

\newcommand{\blue}
{}%{\textcolor{blue}}
\begin{document}

\ensubject{subject}

%%%%%%%%%%%%%%%%%%%%%%%%%%%%%%%%%%%%%%%%%%%%%%%%%%%%%%%
%%% Authors do not modify the information below
%%% ????????????????
%%% ??????????, ????????????{}, ???????????????????
%Letter to the Editor??Article%??????
\ArticleType{Invited Review}%??Invited Review
%\SpecialTopic{SPECIAL TOPIC: }%???????
\Year{2025}
%\Month{August}
%\Vol{63}
%\No{8}
%\DOI{https://doi.org/10.1007/s11433-019-1516-y}
%\ArtNo{280411}
%\ReceiveDate{November 11, 2019}
%\AcceptDate{January 21, 2020}
%\OnlineDate{April 13, 2020}
%%%%%%%%%%%%%%%%%%%%%%%%%%%%%%%%%%%%%%%%%%%%%%%%%%%%%%%

%%% title: ????
%%%   \title{title}{title for citation}
\title{Gravitational wave standard sirens: A brief review of cosmological parameter estimation}{Gravitational wave standard sirens: A brief review of cosmological parameter estimation}

%%% Corresponding author: ???????
%%%   \author[number]{Full name}{{email@xxx.com}}
%%% General author: ???????
%%%   \author[number]{Full name}{}
\author[1,2,3]{Shang-Jie Jin \orcidlink{0000-0003-3697-3501}}{} 
\author[1]{Ji-Yu Song \orcidlink{0009-0003-8111-0470}}{}
\author[1]{Tian-Yang Sun\orcidlink{0009-0002-5109-6420}}{}
\author[1]{Si-Ren Xiao\orcidlink{0009-0008-6742-0145}}{}
\author[4,5]{\\He Wang \orcidlink{0000-0002-1353-391X}}{}
\author[6]{Ling-Feng Wang \orcidlink{0000-0001-6221-2867}}{}
%\author[1]{Geng-Chen Wang}{}
%\author[1]{\\Tian-Nuo Li}{}
%\author[1]{Yue Shao}{}
%\author[1]{Tao Han}{}
\author[1]{Jing-Fei Zhang \orcidlink{0000-0002-3512-2804}}{}
\author[1,7,8]{Xin Zhang \orcidlink{0000-0002-6029-1933}}{{zhangxin@mail.neu.edu.cn}}
%\Authorfootnote

%%% Author information for page head. ?¨¹?§Ö????????
%%% ??????????????, ??????????author???
\AuthorMark{Shang-Jie Jin}%\authorcr????????

%%% Authors for citation. ????????§Ö????????
%%% ??????????????, ??????????author???
\AuthorCitation{Shang-Jie Jin et al.}

%%% Address. ???
%%%   \address[number]{Address, City {\rm Postcode}, Country}
\address[1]{Liaoning Key Laboratory of Cosmology and Astrophysics,
College of Sciences, Northeastern University, Shenyang 110819, China}
\address[2]{Department of Physics, University of Western Australia, Perth WA 6009, Australia}
\address[3]{Research Center for the Early Universe (RESCEU), Graduate School of Science, The University of Tokyo, Tokyo 113-0033, Japan}
\address[4]{International Centre for Theoretical Physics Asia-Pacific, University of Chinese Academy of Sciences, Beijing 100190, China}
\address[5]{Taiji Laboratory for Gravitational Wave Universe (Beijing/Hangzhou),
University of Chinese Academy of Sciences, Beijing 100190, China}
\address[6]{School of Physics and Optoelectronic Engineering, Hainan University, Haikou 570228, China}
\address[7]{MOE Key Laboratory of Data Analytics and Optimization for Smart Industry, Northeastern University, Shenyang 110819, China}
\address[8]{National Frontiers Science Center for Industrial Intelligence and Systems Optimization, Northeastern University, Shenyang 110819, China}

%\contributions{}%????????

%%% Abstract. ??
\abstract{Gravitational wave (GW) observations are expected to serve as a powerful and independent probe of the expansion history of the universe. By providing direct and calibration-free measurements of luminosity distances through waveform analysis, GWs provide a fundamentally different and potentially more robust approach to measuring cosmic-scale distances compared to traditional electromagnetic (EM) observations, which is known as the standard siren method. In this review, we present an overview of recent developments in GW standard siren cosmology, 
\blue{including up-to-date $H_0$ constraints: the re-analysis bright siren GW170817 \(H_{0} = 78.4^{+25.7}_{-12.0}\,\mathrm{km\,s^{-1}\,Mpc^{-1}}\) (employing the same methodology as the O4a dark and spectral siren studies), the most recent O4a dark-siren analysis \(H_{0} = 81.6^{+21.5}_{-15.9}\,\mathrm{km\,s^{-1}\,Mpc^{-1}}\), and their combination \(H_{0} = 76.6^{+13.0}_{-9.5}\,\mathrm{km\,s^{-1}\,Mpc^{-1}}\)}, and prospects for constraining cosmological parameters using future GW detections \blue{($H_0$ is expected to be constrained to the sub-percent level in a 10-year observation of the third-generation GW detectors).} We first introduce standard sirens based on how redshift information is obtained and outline the Bayesian framework used in cosmological parameter estimation. We then review the measurements on the Hubble constant from the LIGO-Virgo-KAGRA network and present the potential role of future standard siren observations in cosmological parameter estimations. A central focus of this review is the unique ability of GW observations to break cosmological parameter degeneracies inherent in the EM observations. Since the cosmological parameter degeneracy directions of GW and EM observations are quite different (roughly orthogonal in some cases), their combination can significantly improve constraints on cosmological parameters. This complementarity is expected to become one of the most critical advantages for GW standard siren cosmology. \blue{We also briefly highlight the impact of systematic uncertainties, such as detector calibration, weak lensing, peculiar velocities, and host-galaxy catalog completeness, and corresponding potential mitigation strategies, which currently limit the constraint precision of cosmological parameters.} Looking forward, we highlight the importance of combining GW standard sirens with other emerging late-universe cosmological probes such as fast radio bursts, 21 cm intensity mapping, and strong gravitational lensing to forge a precise cosmological probe for exploring the late universe. \blue{Finally, we introduce the challenges and the role of machine learning in searching for more signals, ensuring reliable parameter inferences, and accelerating the inference process for cosmological parameters.}
}

%%% Keywords. 


\maketitle

%\tableofcontents%?????

%%%%%%%%%%%%%%%%%%%%%%%%%%%%%%%%%%%%%%%%%%%%%%%%%%%%%%%
%%% The main text. ???????
%???????????????????\cref{fig1}
%\twocolumn\onecolumn
%%%%%%%%%%%%%%%%%%%%%%%%%%%%%%%%%%%%%%%%%%%%%%%%%%%%%%%
\begin{multicols}{2}
\section{Introduction}\label{sec1}

Throughout history, humanity has been driven by a profound  curiosity to unravel the mysteries of the universe. This quest reached a pivotal point in the 1920s when Edwin Hubble made a groundbreaking discovery that the universe is expanding~\cite{Hubble:1929ig}. This revelation fundamentally shifted our understanding, challenging the then-dominant belief in a static universe~\cite{Einstein:1917ce}. 
This discovery was not only a great achievement of the 20th century but also a catalyst for the development of Big Bang cosmology.
For many years, scientific consensus held that the expansion of the universe would slow down due to gravitational forces. In light of this, accurately determining two essential cosmological parameters, the deceleration parameter ($q_0$) and the Hubble constant ($H_0$), became a central focus of research in the field. %\cite{sandage1970cosmology}.

In 1998, two independent supernova research teams made a groundbreaking discovery. By observing Type Ia supernovae, they precisely measured the deceleration parameter and unexpectedly found it to be negative~\cite{SupernovaSearchTeam:1998fmf,SupernovaCosmologyProject:1998vns}. This unexpected result revealed that the universe's expansion is accelerating rather than decelerating, challenging the prevailing understanding of cosmology. This discovery was later honored with the 2011 Nobel Prize in Physics for the discovery of the accelerating expansion of the universe through observations of distant supernovae. 
%As a result, accurately determining $H_0$ has become a central focus of modern cosmological research.

$H_0$, first proposed in the 1920s, is fundamental in understanding the universe's scale, including its age, size, the evolution of cosmic structures, and the nucleosynthesis of light isotopes. The value of $H_0$ can be inferred from both early-universe and late-universe observations. In the early 21st century, both types of observations yielded consistent estimates of $H_0$, but these early measurements carried substantial uncertainties. However, as precision improved, a significant tension has emerged. This tension has an approximately 10\% discrepancy exceeding a $5\sigma$ level of significance, posed a major challenge to cosmology. The Planck cosmic microwave background (CMB) observation, based on the standard $\Lambda$CDM model at a redshift of $z\sim 1100$, infers a lower value of $H_0$ at $H_0=67.4\pm 0.5~\rm kms^{-1}Mpc^{-1}$~\cite{Planck:2018vyg}. In contrast, local universe observations, such as the SH0ES team, report a higher value of $H_0=73.04\pm 1.04~\rm kms^{-1}~Mpc^{-1}$ based on the Cepheid-supernovae Ia sample~\cite{Riess:2021jrx}.
The Hubble tension is now widely viewed as a severe crisis in cosmology~\cite{Verde:2019ivm,Riess:2019qba} (see e.g., Refs.~\cite{DiValentino:2021izs,Abdalla:2022yfr,Kamionkowski:2022pkx,Perivolaropoulos:2021jda,Li:2013dha,H0LiCOW:2019pvv,Freedman:2021ahq,Dainotti:2021pqg,Vagnozzi:2023nrq,Poulin:2023lkg,Hu:2023jqc,Poulin:2018cxd,Zhang:2014dxk,Zhang:2014nta,Zhao:2017urm,Feng:2017nss,Li:2010xjz,Zhang:2014ifa,Guo:2018ans,Gao:2021xnk,Cai:2021wgv,Yang:2018euj,DiValentino:2020zio,DiValentino:2019ffd,DiValentino:2019jae,Liu:2019awo,Zhang:2019cww,Ding:2019mmw,Li:2020tds,Wang:2021kxc,Vagnozzi:2021tjv,Vagnozzi:2021gjh,Vagnozzi:2019ezj,Guo:2019dui,Feng:2019jqa,Lin:2020jcb,Gao:2022ahg,Zhao:2022yiv,Song:2025ddm,Han:2024sxm,Zhang:2024rra,Feng:2024mfx,Fu:2024cjj,Pan:2024xoj,Xiao:2024nmi,Feng:2024lzh,Dong:2024bvw,Han:2025fii,CosmoVerse:2025txj,Pedrotti:2024kpn,Jiang:2024xnu,Jiang:2024viw,Ling:2025lmw,Bian:2025ifp} and references therein for recent discussions).
While the Hubble tension may suggest the need for new physics beyond $\Lambda$CDM, no extended model has yet gained consensus as a definitive resolution, see e.g., Refs.~\cite{Zhang:2012uu,Li:2012spm,Zhang:2014nta,Feng:2016djj,Zhang:2015rha,Ma:2007av,Fu:2011ab,Zhang:2014ifa,Feng:2009hr,Feng:2017usu,Li:2025muv,Wu:2025vfs,Zhou:2025nkb,Du:2025xes,Li:2025htp,Zhang:2025dwu} for discussions about extended models. In this context, the search for promising and powerful cosmological probes for independently measuring $H_0$ has become increasingly crucial. Among these, the gravitational wave (GW) standard siren method stands out as a particularly promising avenue to address the Hubble crisis; see e.g., Ref.~\cite{Chen:2024gdn} and references therein for related discussions. 

Furthermore, the enigmatic nature of dark energy remains one of the most profound mysteries in cosmology. A potential breakthrough in understanding dark energy could come from precise measurement of the equation of state (EoS) of dark energy, represented as $w={p/\rho}$, where $p$ and $\rho$ are the pressure and density, respectively. Nevertheless, current measurements of 
$w$ are not yet sufficient to reveal the true characteristics of dark energy~\cite{DES:2025bxy,DESI:2025zgx}. In particular, the recent analysis from DESI suggests that the EoS of dark energy evolves over time, with constraints in the $w_0-w_a$ plane deviating from the $\Lambda$CDM model by up to $4.2\sigma$~\cite{DESI:2025zgx}, widely discussed in e.g., Refs.~\cite{Li:2024qso,Du:2024pai,Li:2024qus,Li:2025owk,Feng:2025mlo,Giare:2024gpk,Jiang:2024viw,Jiang:2024xnu,Dinda:2024ktd,Reboucas:2024smm,2024arXiv241013627P,2024arXiv240814787P,Wang:2024tjd,Park:2025azv,Li:2025eqh,Li:2025dwz,Li:2025ula}.
In this quest, GW standard sirens are also anticipated to play a pivotal role, potentially bringing us closer to unraveling this mystery. \blue{Motivated by these unresolved tensions, GW standard sirens matter because they can provide calibration-independent luminosity distances without relying on the distance ladder, which makes them sensitive to $H_0$. Therefore, GW standard sirens can be treated as an independent cross-check on $H_0$. Moreover, the different parameter degeneracy direction between GW and other EM observations allow their combination to break the cosmological parameter degeneracies and help measure the EoS of dark energy.}

In this review, we categorize standard sirens based on how their redshift information is obtained and outline the Bayesian framework used for cosmological parameter inferences. We review recent progress in using GW standard sirens to constrain $H_0$ and discuss the promising prospects of next-generation ground-based detectors, space-based detectors, and the nano-hertz PTA observatories for constraining the cosmological parameters, particularly for $H_0$ and EoS of dark energy $w(z)$. A central focus of this review is the unique capability of GW standard sirens to break cosmological parameter degeneracies that inherent in the traditional EM observations. Furthermore, we highlight the importance of the joint cosmological analysis utilizing GW standard sirens and other promising cosmological probes to forge a precise cosmological probe to explore the expansion history of the universe. 

This paper is organized as follows. In Sec.~\ref{sec1}, we introduce standard sirens based on how redshift information is obtained and outline the Bayesian framework for the cosmological parameter estimations.
Sec.~\ref{sec3} summarize the current constraints on $H_0$ from the LIGO-Virgo-KAGRA (LVK) network. In Sec.~\ref{sec4}, we focus on the prospects for constraining cosmological parameters using future GW standard siren observations. In Sec.~\ref{sec5}, we highlight the ability of GW to break the cosmological parameter degeneracies inherent in the current EM observations. We also explore the synergy between GW and other cosmological probes in cosmological parameter estimations. In Sec.~\ref{sec6}, we outline the key challenges related to standard siren measurements and highlight the emerging role of machine learning techniques in addressing these issues. Conclusion is given in Sec.~\ref{sec7}.

%%%%%%%%%%%%%%%%%%%%%%%%%
\section{Methodology in GW standard siren cosmology}\label{sec2}
GWs are ripples in space-time, a groundbreaking concept first predicted by Albert Einstein in 1916 based on his theory of general relativity~\cite{Einstein:1916vd}. The detection of GW spanned nearly a century. In a landmark event in 2015, the LIGO-Virgo collaboration announced the first-ever detection of a GW signal, named GW150914, originating from a binary black hole (BBH) merger~\cite{LIGOScientific:2016aoc}. This discovery marked the advent of GW astronomy era. Just two years later, the detection of the binary neutron star (BNS) merger event GW170817~\cite{LIGOScientific:2017vwq}, along with its electromagnetic (EM) counterparts across various band ushered in the new era of muti-messenger and time-domain astronomy~\cite{LIGOScientific:2017ync}. In the following section, we shall have a brief overview of GW standard sirens, introduce the methods for inferring redshift of GW, and the Bayesian framework for each method for cosmological parameter estimations.

\subsection{Overview of GW standard sirens}\label{sec2.1}

GW has clear predictions for the waveform of a compact binary coalescence from general relativity, making it a natural messenger to probe the cosmic expansion history. The GW frequency $f$ and its frequency derivative $\dot f$ are related to the chirp mass $\mathcal{M}_{\rm c}$:
\begin{equation}
\mathcal{M}_{\rm c} \propto  f^{-11/5} \dot{f}^{3/5}, \quad 
\mathcal{M}_{\rm c} = \frac{(m_1 m_2)^{3/5}}{(m_1 + m_2)^{1/5}},
\end{equation}
where $m_1$ and $m_2$ are the component mass of the binary system. The chirp mass characterizes the intrinsic strength of the GW signal. The GW strain $h_{\rm gw}$ considering the cosmic expansion is given by
\begin{equation}
h_{\rm gw} \propto \frac{\mathcal{M}_{\rm c}^{5/3}}{d_{\rm L}(z)}, \quad 
d_{\rm L}(z) = c(1+z) \int_0^z \frac{\mathrm{d}z'}{H(z')},\label{eq:hgwdl}
\end{equation}
where $d_{\rm L}$ is the luminosity distance and $H(z)$ is the Hubble parameter at redshift $z$ which is given by
\begin{align}
H(z) &= H_0 \left[ \Omega_{\rm m}(1+z)^3 + (1-\Omega_{\rm m}) \exp\left( 3 \int_0^z \frac{1 + w(z')}{1+z'} \, {\rm d}z' \right) \right]^{1/2}, \label{eq:hz_wz}
\end{align}
where $w(z)$ is the EoS parameter of dark energy. For $w(z)=-1$, Eq.~(\ref{eq:hz_wz}) is reduced to the standard $\Lambda$CDM model, while for $w(z)=\mathrm{constant}$, Eq.~(\ref{eq:hz_wz}) corresponds to the $w$CDM model, and for $w(z)=w_0+w_az/(1+z)$, Eq.~(\ref{eq:hz_wz}) describes the $w_0w_a$CDM model. 
Moreover, GW detectors are laser interferometers that respond directly to waveform strain, not intensity, enabling the absolute measurement of luminosity distance, known as ``standard sirens'', as evident in Eq.~\ref{eq:hgwdl}.

The term ``standard sirens'' is analogous to ``standard candles'' and ``standard rulers'' used in cosmology. The concept, initially proposed by Schutz in 1986~\cite{Schutz:1986gp} and further elaborated by Holz and Hughes in 2005~\cite{Holz:2005df}. Unlike traditional EM-based distance measurements that typically measure relative distances and are subject to unavoidable systematic errors, standard sirens offer a significant advantage by providing direct and calibration-independent measurements of cosmic-scale distances. However, due to the degeneracies between GW parameters, the standard siren approach also faces challenges. In the next subsection, we shall introduce the methodologies that can potentially break the degeneracies.

\subsection{Methods for breaking GW parameter degeneracies}\label{sec2.2}
A strong parameter degeneracy exists between the luminosity distance and the inclination angle of the binary, making it challenging to independently and precisely measure both parameters. This issue not only limits the understanding of the source's physical properties but also poses obstacles to the precise measurement of cosmological parameters. To address this, researchers have explored various methods, such as introducing eccentricity, memory effects, higher-order modes, and precession effects, to provide additional information and break the degeneracies between GW parameters.

Eccentricity introduces additional harmonic components into the GW signal, leading to complex nontrivial coupling between parameters such as distance and angular parameters, thereby effectively breaking their degeneracies~\cite{Yang:2022tig,Yang:2022iwn,Yang:2022fgp,Yang:2023zxk}. For example, in mid-frequency (0.1–10 Hz) band, binary systems with eccentricity can achieve significant improvements in the precise measurements of luminosity distance and source localization, particularly in cases of small inclination angles, where the improvement can reach several orders of magnitude~\cite{Yang:2022tig}. Furthermore, the presence of eccentricity enables more precise identification of host galaxies, laying the foundation for an increase in the number of standard sirens and significantly improving the measurement precision of cosmological parameters such as $H_0$ and the EoS of dark energy.

The memory effect provides additional observational features through permanent changes in space-time following a GW event, which can effectively break the degeneracy between distance and inclination angle~\cite{Gasparotto:2023fcg,Xu:2024ybt}. Furthermore, studies have shown that incorporating the memory effect can reduce systematic errors in parameter estimation, whereas neglecting this effect may result in estimates deviating significantly from the true values, thereby impacting the precise measurement of key parameters~\cite{Sun:2022pvh,Goncharov:2023woe,Sun:2024nut,Xu:2024ybt}.

Higher-order modes introduce additional harmonic features into waveform modeling, enhancing the signal's distinguishability and playing a crucial role in the GW parameter estimations, such as mass, spin, and inclination angle~\cite{LIGOScientific:2020stg,Mills:2020thr,Gong:2023ecg,Liu:2024jkj}. While higher-order modes help break degeneracies and improve luminosity distance estimation, their relative contribution diminishes for more distant sources due to reduced signal-to-noise ratios (SNRs)~\cite{Liu:2024jkj}. Compared to using only lower-order modes, higher-order modes enhance the accuracy of template libraries and the sensitivity of signal detection, optimizing parameter reconstruction and providing an effective solution to degeneracy issues that cannot be avoided in traditional models.

Precession, by altering the temporal evolution of the waveform in binary black hole systems, is critical to improve the accuracy of distance estimation~\cite{Yun:2023ygz,Lin:2024pkr}. Studies have shown that precession can improve the measurement precision of distant GW sources' parameter estimations by up to an order of magnitude~\cite{Yun:2023ygz}. More importantly, precession greatly reduces the candidate range of host galaxies, significantly enhancing the accuracy of $H_0$ measurements, which is essential for exploring the history of cosmic expansion.

In conclusion, eccentricity, memory effect, higher-order modes, and precession provide powerful optimization tools for GW signal analysis across multiple aspects, including waveform modeling, template matching, and parameter estimation. These methods not only effectively resolve the degeneracy between key parameters such as distance and inclination angle but also significantly improve the measurement precision of parameter estimation. In cosmological applications, they lay a solid foundation for precise measurements of $H_0$, investigations into the EoS of dark energy, and explorations of the history of cosmic expansion. Meanwhile, these advancements also provide rich theoretical support and technical tools for the future development of GW astronomy.

\subsection{Redshift inference and Bayesian framework}\label{sec2.3}
% For cosmological applications, measurements of luminosity distance alone are insufficient. As shown in Eq.~(\ref{dl}), the application of standard sirens in cosmological parameter estimations requires both distance and redshift measurements, with the latter playing a particular important role. In the following section, we briefly introduce methods for determining redshifts using different sirens and their Bayesian inferences of cosmoloical parameters.
\subsubsection{Bright sirens: redshift from EM counterparts}\label{sec2.3.1}
The most promising and direct methods to measure the redshift of the GW source is to identify the EM counterparts of GW events, which is commonly referred to as the ``bright siren'' method. The bright siren approach is fundamentally grounded in the multi-messenger characteristics of certain GW events, i.e., BNS mergers and some neutron star–black hole (NSBH) mergers, which are anticipated to generate unique EM counterparts, including short gamma-ray bursts (SGRBs), kilonovae, and broadband afterglows~\cite{Metzger:2011bv,Nissanke:2012dj}. By identifying these EM signals, one can achieve much higher accuracy localization than that solely detected by GW detectors. Currently, the LVK network can typically localize the GW events in tens to thousands of square degrees~\cite{LIGOScientific:2018mvr,LIGOScientific:2020ibl,KAGRA:2021vkt,Singer:2016eax}, which corresponds to tens of thousands of candidate galaxies within the localized sky volume. This large uncertainty often prevents direct redshift measurements. However, due to the detection of an EM counterpart, the exact host galaxy can possibly be pinpointed which is in the arcsecond precision, enabling the spectroscopic follow-up observations to measure the galaxy's redshift. 

The identification of EM counterparts includes a step-by-step observational process. First, the optical and near-infrared telescopes rapidly scan the sky localization region which is provided by the GW observatories. The aim is to identify the transient sources which are consistent with expected EM signals from the binaries. Then the multi-wavelength telescopes in, e.g., X-ray and radio band can confirm the identification and increase the confidence in the host galaxy association.

Once the host galaxy is identified, its redshift can be measured using well-established spectroscopic techniques. Since the galaxy's spectral lines shift according to the cosmic expansion, the observation of such spectral lines, such as hydrogen Balmer lines, oxygen forbidden lines, and other nebular emissions, can provide a direct and robust measurement of the GW source's redshift~\cite{Im:2017scv}. Combined with the measurement of the luminosity distance from the GW waveform, the cosmological parameters can be estimated directly using the distance-redshift relation, including $H_0$, and other cosmological parameters like the matter density or dark energy equation of state. In the following, we shall introduce the Bayesian inference of cosmological parameters using bright sirens.

For bright sirens, the information from EM observations should be incorporated alongside GW data. Within this framework, the parameters associated with individual GW events are denoted by $\boldsymbol{\theta}$, while the population-level parameters are denoted by $\boldsymbol{\Lambda}$, which includes the cosmological parameters. Since GW and EM detections are statistically independent, the overall likelihood function is given by~\cite{Mastrogiovanni:2021wsd,KAGRA:2021duu}
\begin{equation}
    \mathcal{L}(\{x\} \mid \boldsymbol{\Lambda}) \propto \prod_{i=1}^{N_{\text {obs }}} \frac{\int \mathcal{L}_{\rm EM}(x_i|z,\Omega)\mathcal{L}_{\rm GW}\left(x_{i} \mid \boldsymbol{\theta}, \boldsymbol{\Lambda}\right) \frac{{\rm d}N}{{\rm d}\theta_{\rm d}{\rm d}t_{\rm d}}(\boldsymbol{\Lambda}) \mathrm{~d} \boldsymbol{\theta}}{\int p_{\rm{det}}^{\rm EM}(z,\Omega)p_{\rm{det}}^{\rm GW}(\boldsymbol{\theta}, \boldsymbol{\Lambda}) \frac{{\rm d}N}{{\rm d}\theta_{\rm d}{\rm d}t_{\rm d}}(\boldsymbol{\Lambda}) \mathrm{~d} \boldsymbol{\theta}},
\end{equation}
where $N_{\rm obs}$ is the number of observed GW events in the GW data set $\{x\}$, $\mathcal{L}_{\rm EM}(x_i|z,\Omega)$ denotes the likelihood of the EM observations, characterizing uncertainties in redshift and sky localization, and $\mathcal{L}_{\rm GW}\left(x_{i} \mid \boldsymbol{\theta}, \boldsymbol{\Lambda}\right)$ is the GW likelihood of a single GW event, quantifying the uncertainties in measuring $\boldsymbol{\theta}$ through the waveform analysis. ${\rm d}N/({\rm d}\theta_{\rm d}{\rm d}t_{\rm d})$ is the event rate in the detector frame, which can be converted to the GW event in the source frame ${\rm d}N/({\rm d}\theta_{\rm s}{\rm d}t_{\rm s})$ via
\begin{equation}
    \frac{{\rm d}N}{{\rm d}\theta_{\rm d}{\rm d}t_{\rm d}}=\frac{{\rm d}N}{{\rm d}\theta_{\rm s}{\rm d}t_{\rm s}}\frac{1}{1+z}\frac{1}{\frac{\partial d_{\rm L}}{\partial z}(1+z)^2},
\end{equation}
where $1/(1+z)$ comes from the difference between source-frame and detector-frame time, and $\partial d_{\rm L}/\partial z(1+z)^2$ is the Jacobian for the change of variables between the detector and the source frames. 

$p_{\rm det}^{\rm GW}(\boldsymbol{\theta}, \boldsymbol{\Lambda})$ characterizes the detection probability of a GW event with parameters $\boldsymbol{\theta}$ and under population-level parameters $\boldsymbol{\Lambda}$~\cite{Gray:2019ksv}, and $p_{\rm det}^{\rm EM}(z,\Omega)$ is the detection probability of EM signals at redshift $z$ and sky position $\Omega$. 
% In current LVK observations, the detection range of GW events is significantly smaller than that of EM observations, making it reasonable to approximate $p_{\rm{det}}^{\rm EM} \approx 1$. However, this assumption may no longer hold as GW detectors increase their sensitivity in the future. % 

\begin{figure*}[!htbp]
\includegraphics[width=0.8\textwidth]{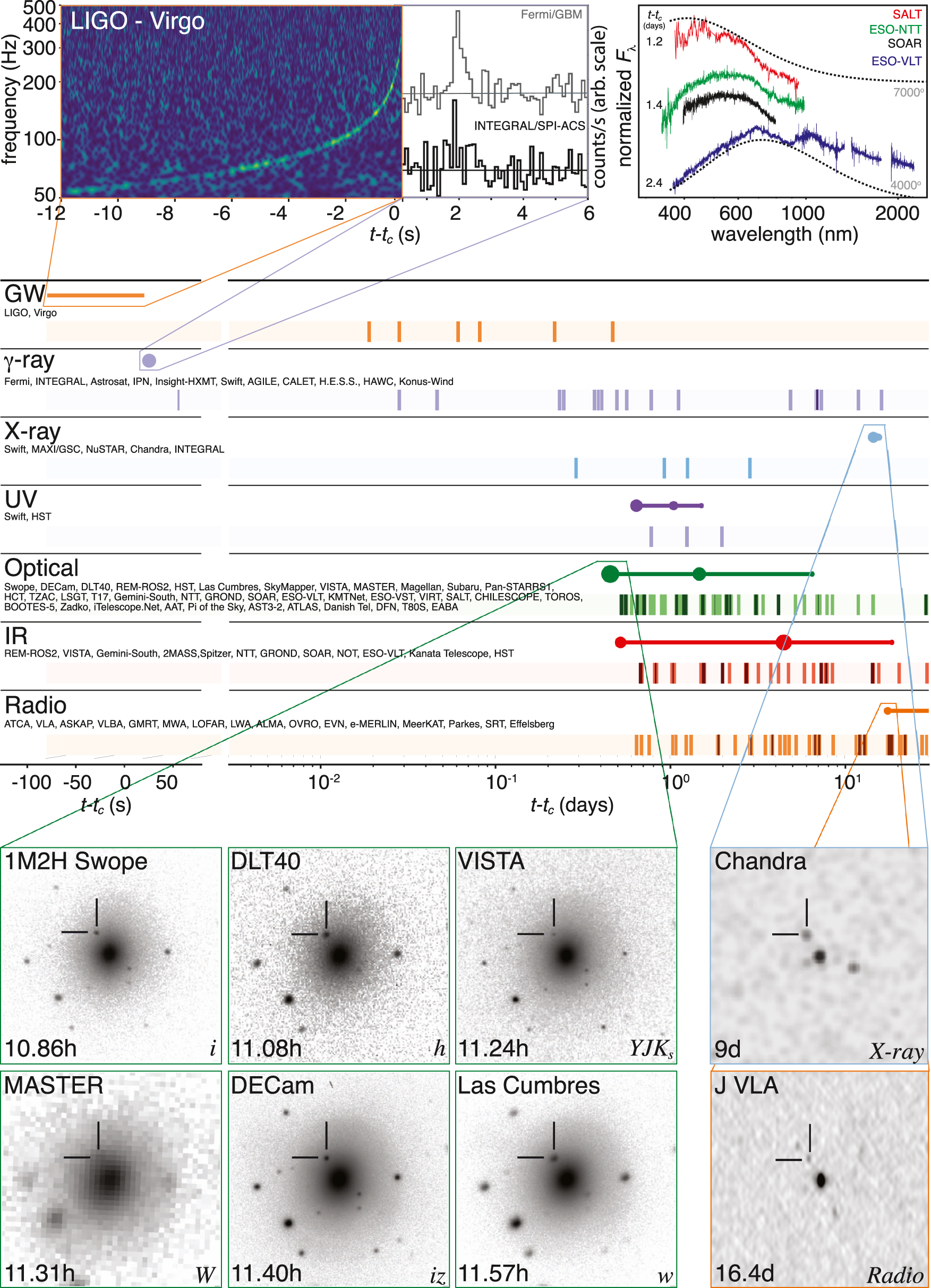}
\centering
\caption{Timeline of the discovery of GW170817, GRB~170817A, and SSS17a/AT~2017gfo, along with follow-up observations, shown by messenger and wavelength relative to the GW event time. The figure is from Ref.~\cite{LIGOScientific:2017ync} and reproduced by permission of the AAS.}
\label{fig:multimessenger}
\end{figure*}

\begin{figure*}[!htbp]
\centering
\includegraphics[width=1.7\columnwidth]{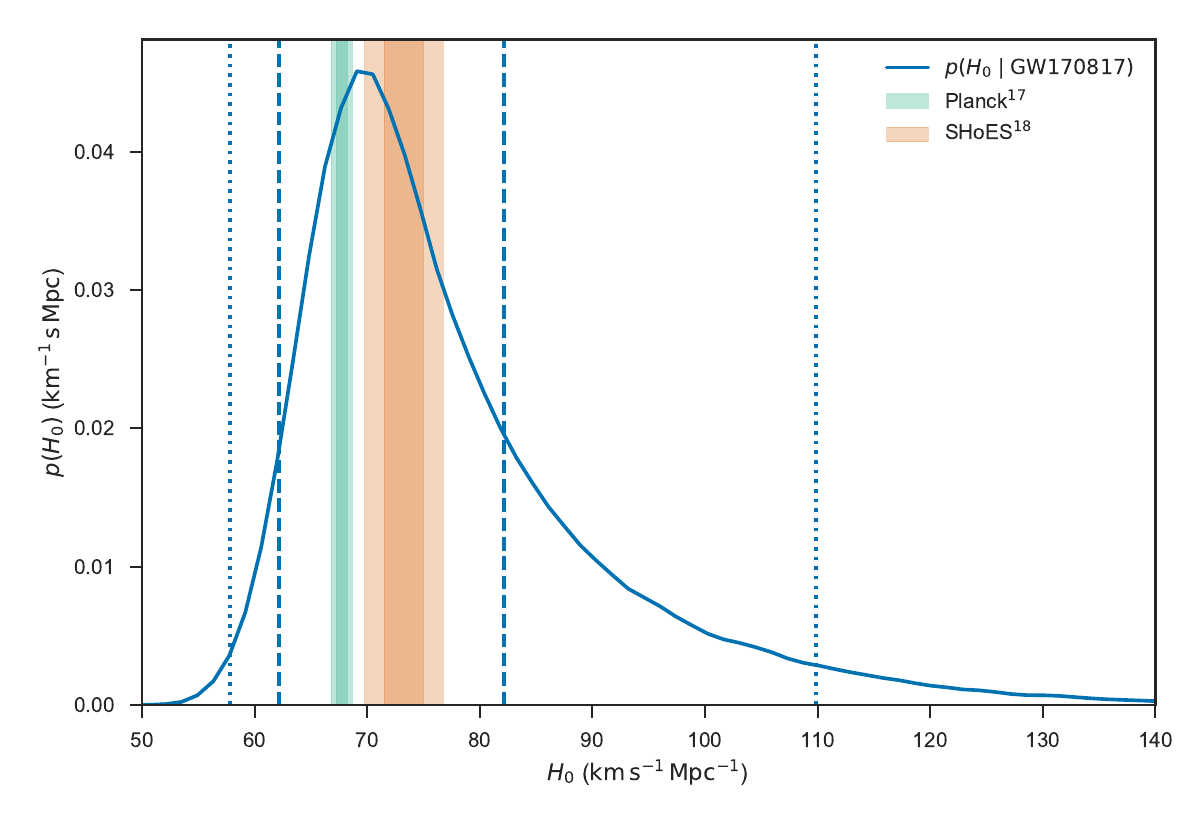}
\caption{Posterior distributions for $H_0$ using the bright siren method from GW170817. The 68.3\% ($1\sigma$) and 95.4\% ($2\sigma$) minimal credible intervals from GW170817 are indicated by dashed and dotted lines. Constraints on $H_0$ at $1\sigma$ and $2\sigma$ from Planck and distance ladder measurement are shown in green and orange. Reproduced from Ref.~\cite{LIGOScientific:2017adf} with permission.}
\label{fig:gw170817h0}
\end{figure*}

In current LVK observations, the detection volume of EM observatories (especially optical telescopes) is significantly larger than that of GW detectors. This justifies the common approximation $p_{\rm det}^{\rm EM}(z,\Omega) \approx 1$ under the assumption that if a GW event is detected and has an EM counterpart, the EM signal will also be observed. However, as GW detectors increase their sensitivity in future observing runs and probe higher redshifts, this assumption may no longer hold, and EM selection effects must be carefully accounted for. 
This likelihood framework provides a statistically rigorous approach for jointly analyzing multi-messenger data from bright sirens, enabling precise and direct measurements of cosmological parameters.

Despite its advantages, the bright siren method faces several observational and astrophysical challenges. First, the EM emission associated with GW events can be highly anisotropic, short-lived, and faint, depending on the source properties and orientation. For example, strongly-beamed SGRBs result in only a small fraction being detected at inclination angles $\iota \leq 20^{\circ}$~\cite{Rezzolla:2011da,Chen:2020zoq,Kocsis:2005vv}. Although kilonovae are more isotropic, the rapid fading and intrinsic faintness make them challenging to detect. In addition, due to the highly anisotropic nature of EM counterparts, properly accounting for the selection effects of EM observations in the likelihood function requires sufficient prior knowledge of the EM emission structure; otherwise, significant systematic errors may arise in the cosmological constraints~\cite{Chen:2020dyt}.
Furthermore, some GW sources, such as BBH mergers, are commonly not expected to produce detectable EM counterparts, limiting the application of the bright siren method.

\subsubsection{Dark sirens: redshift from galaxy catalog association}\label{sec2.3.2}
Although the bright siren method is promising, it highly relies on the sensitivity of the observatories, as well as the properties and precise localization of the GW source. The current observed GW events by the LVK network, except for GW170817 have not detected their associated EM counterparts. Actually, such absence is common for BBH mergers since they are not expected to emit EM signals. Moreover, the EM counterparts of some BNS mergers also cannot be detected due to the limited sensitivity of the current observatories. Both of the above factors make GW170817 the only GW event for which its EM counterparts have been detected. In order to use various GW standard sirens without EM counterparts, the ``dark siren'' method has been developed.
% Therefore, the ``dark siren'' method has been developed for these ``dark'' GW events by a statistical framework to infer redshift.

The fundamental premise of the dark siren method is to associate GW events with potential host galaxies identified within the three-dimensional localization volume determined solely from GW observation. Through GW parameter estimation, the sky location and the luminosity distance of the GW source could be obtained. By cross-matching the GW localization volume with the galaxy catalog, the potential host galaxies could be obtained, typically the number on the order of $\mathcal{O}(10^5)$. Actually, a single GW event can give the constraints on cosmological parameters by statistically averaging the redshifts of all potential host galaxies to account for the uncertainty regarding the true host galaxy, with improved precision achieved through the analysis of multiple GW events~\cite{Chen:2017rfc,DelPozzo:2011vcw,Gair:2022zsa,Borghi:2023opd}.

Compared to the likelihood of bright sirens, the likelihood of the dark sirens does not have terms corresponding to EM observations, and is given by
\begin{equation}
    \mathcal{L}(\{x\} \mid \boldsymbol{\Lambda}) \propto \prod_{i=1}^{N_{\text {obs }}} \frac{\int \mathcal{L}_{\rm GW}\left(x_{i} \mid \boldsymbol{\theta}, \boldsymbol{\Lambda}\right) \frac{{\rm d}N}{{\rm d}\theta_{\rm d}{\rm d}t_{\rm d}}(\boldsymbol{\Lambda}) \mathrm{~d} \boldsymbol{\theta}}{\int p_{\rm{det}}^{\rm GW}(\boldsymbol{\theta}, \boldsymbol{\Lambda}) \frac{{\rm d}N}{{\rm d}\theta_{\rm d}{\rm d}t_{\rm d}}(\boldsymbol{\Lambda}) \mathrm{~d} \boldsymbol{\theta}}.
\end{equation}

The GW event rate in the source frame is written as~\cite{Mastrogiovanni:2023zbw,Mastrogiovanni:2023emh}
\begin{equation}
\begin{aligned}
    &\frac{{\rm d}N}{{\rm d}z{\rm d}m_{\rm 1,s}{\rm d}m_{\rm 2,s}{\rm d}\chi{\rm d}t_{\rm s}{\rm d}\Omega}\propto\Psi(z|\boldsymbol{\Lambda})p_{\rm pop}(m_{\rm 1,s},m_{\rm 2,s}|\boldsymbol{\Lambda})p_{\rm pop}(\chi|\boldsymbol{\Lambda})\times\\
    &\bigg[\frac{{\rm d}V_{\rm c}}{{\rm d}z}\phi_{*}(H_0)\Gamma_{\rm inc}(\alpha+\epsilon+1,x_{\rm max},x_{\rm thr})+\\&
    \frac{1}{\Delta\Omega}\sum^{N_{\rm in(\Omega)}}_{j}10^{0.4\epsilon(M_{*}(H_0)-M(H_0,m_j,z))}p(z|\hat{z_{j}},\sigma_{z,j})\bigg],
\end{aligned}
\end{equation}
where $\Psi(z|\boldsymbol{\Lambda})$, $p_{\rm pop}(m_{\rm 1,s},m_{\rm 2,s}|\boldsymbol{\Lambda})$, and $p_{\rm pop}(\chi|\boldsymbol{\Lambda})$ represents population models describing the redshift, mass, and spin distributions of GW sources in the source frame, respectively. The two terms within the bracket correspond to the contribution from the incomplete portion of the catalog and the complete portion, respectively. In the first term, $\phi_{*}(H_0)=\phi_{*}(H_0/100=1)\times(H_0/100)^3$ denotes the rescaled normalization of the Schechter galaxy luminosity function. $\Gamma_{\rm inc}$ is the incomplete gamma function, in which $\alpha$ is the faint-end slope of the Schechter function, and $\epsilon$ characterizes the correlation between galaxy luminosity and the probability of hosting a GW source. $x_{\rm max/thr}=10^{0.4(M_{*}(H_0)-M_{\rm max/thr}(H_0))}$, with $M_{*}(H_0)$ and $M_{\rm max}(H_0)$ denoting the rescaled knee and maximum absolute magnitudes of the Schechter function, respectively, given by
\begin{equation}
    M_{*/{\rm max}}(H_0)=M_{*/{\rm max}}(H_0/100=1)+5{\rm log}_{10}^{(H_0/100)}.
\end{equation}
$M_{\rm thr}(H_0)=M(H_0,m_{\rm thr}(\Omega), z)$ is converted from the apparent magnitude threshold, $m_{\rm thr}(\Omega)$, of the galaxy catalog at a given sky location $\Omega$. In the second term, we sum over all potential host galaxies within the GW sky localization region. $M(H_0,m_j,z)$ is converted from the apparent magnitude of the $j$-th galaxy. The term $p(z|\hat{z_{j}},\sigma_{z,j})$ is given by
\begin{equation}
    p(z|\hat{z_{j}},\sigma_{z,j})\propto\mathcal{N}(\hat{z_{j}},\sigma_{z,j}),
\end{equation}
where $\hat{z_j}$ and $\sigma_{z,j}$ is the mean value and the 1-$\sigma$ measurement error of the $j$-th galaxy's redshift.

% The dark siren method has the advantage of allowing the GW events without EM counterparts to be used in cosmological parameter estimations.
% However, the ability is strongly affected by the localization error of the GW source, the completeness, size, and the depth covered by the galaxy catalog. If the catalog misses some possible host galaxies, it can lead to biased results. Also, uncertainties in galaxy redshifts, especially when using photometric data, can reduce the precision of the cosmological parameter estimations.

\blue{The galaxy host identification method is currently the most precise approach for obtaining redshifts of GW standard sirens in the absence of EM counterparts. This method incorporates redshift information from galaxy catalogs. Nevertheless, it is subject to various sources of systematic error. The first concerns the completeness of galaxy catalogs. Due to the observational limitations of survey telescopes, catalogs cannot include all galaxies without omission. Current practice typically involves assuming a luminosity function to estimate the completeness of galaxies within the GW localization region and then compensating for the missing portion. Such compensation often assumes that galaxies are uniformly distributed within the comoving volume, an oversimplification, since the actual distribution of galaxies is influenced by the large-scale structure of the universe and traces the underlying matter density field \cite{Dalang:2024gfk,Leyde:2024tov,Gupta:2024sss,Datrier:2025zdf}. In addition, for modeling redshift measurement errors, it is common to approximate galaxy redshift posteriors with Gaussian distributions. However, in real observations, the redshift posteriors are often non-Gaussian, which may also introduce systematic errors \cite{Turski:2023lxq}. Furthermore, the dark siren method relies on assumptions about the population model of GW sources. Since the population distribution of GW sources is essentially a phenomenological model inferred from both present and past observational data, it may introduce biased estimates of cosmological parameters if the modeling is inaccurate \cite{Dhani:2024jja,Agarwal:2024hld}.}

\begin{figure*}[!htbp]
\centering
\includegraphics[width=\columnwidth]{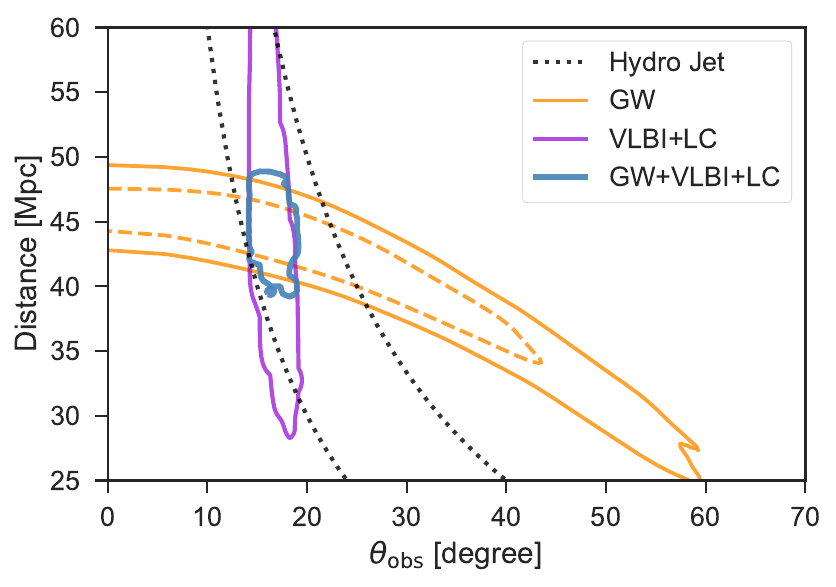}
\includegraphics[width=\columnwidth]{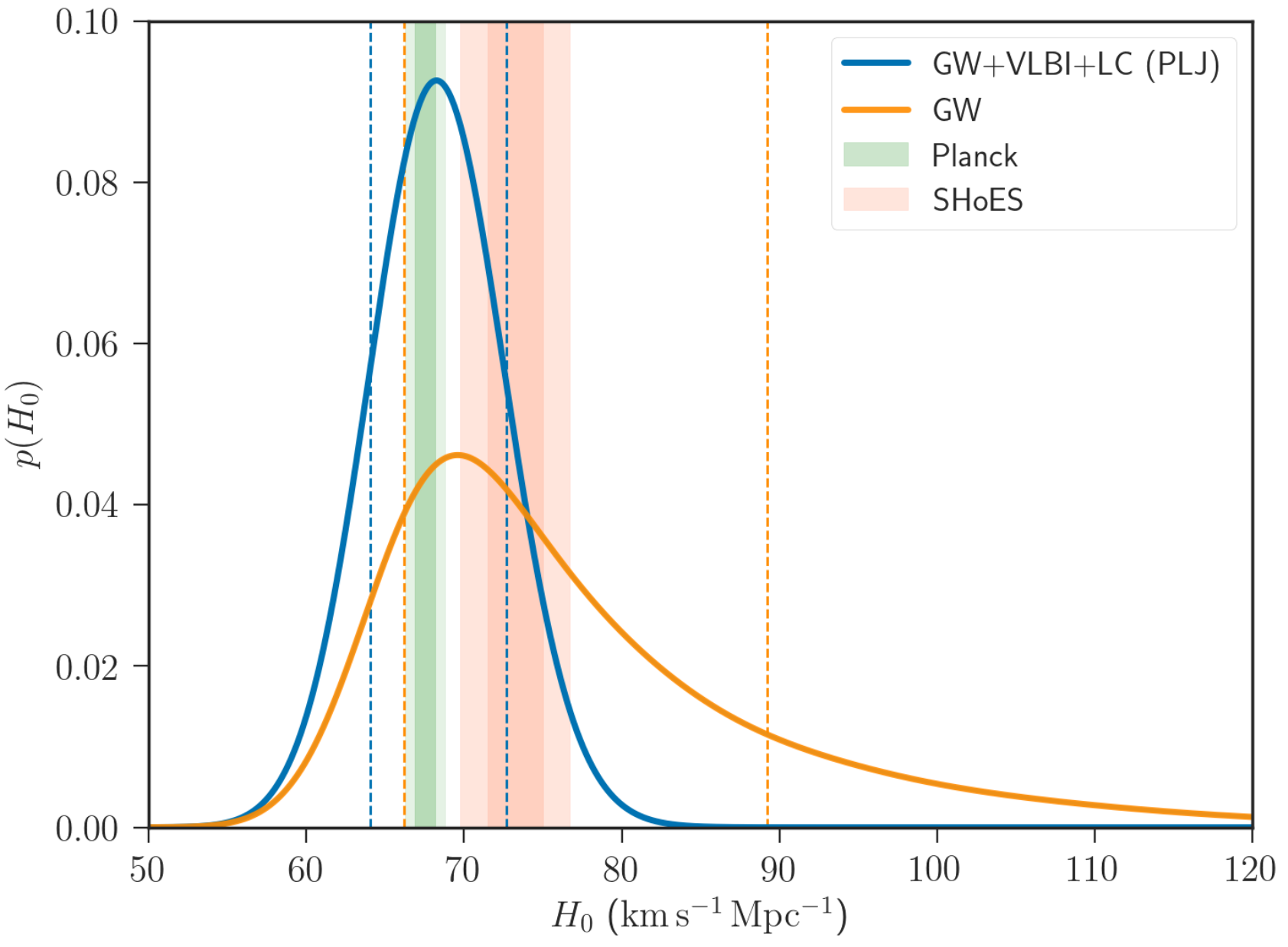}
\caption{Update constraints on the inclination, distance, and $H_0$ from GW170817. Left panel: two-dimensional marginalized posterior contours for the inclination and distance using the Hydro jet (95\% confidence level), GW170817 only (68\% and 95\% confidence level), afterglow light curve (LC) and centroid motion through VLBI (95\% confidence level), and their combination with GW170817 assuming power-law jet (PLJ) model (95\% confidence level). Right panel: posterior distributions for $H_0$ using GW+VLBI+LC and GW. The $1\sigma$ and $2\sigma$ constraints from Planck CMB and distance ladder are also shown in green and orange vertical bands. Reproduced from Ref.~\cite{Hotokezaka:2018dfi} with permission.}
\label{fig:vlbi}
\end{figure*}

\subsubsection{Cross-correlation sirens: redshift from spatial correlation}\label{sec2.3.3}
In addition to using the spatial localization of GW detection to search for potential host galaxies in the galaxy catalog, another method of combining GW detection and galaxy surveys is to use the spatial cross-correlation between the GW sources and the galaxies to infer cosmological parameters~\cite{Oguri:2016dgk,Bera:2020jhx,Mukherjee:2020hyn,Ghosh:2023ksl}. Since GW sources and galaxies trace the matter density, they are spatially correlated through the underlying matter field. By analyzing the three-dimensional spatial correlation between GW sources and redshift-known galaxies, we can infer the host redshift shells for GW sources. 

\subsubsection{Spectral sirens: redshift from GW population models}\label{sec2.3.4}
When EM counterparts are unavailable or when GW source localization is poor or not covered by galaxy catalogs, GW waveforms can still provide valuable information about the redshift of the source through the so-called ``spectral siren'' method. Since the GW waveforms encode intrinsic properties of the binary system, including component masses, which are affected by cosmic redshift due to the expansion of the universe, the observed mass in the detector frame are redshifted:
\begin{equation}
    m_{\rm d}=m_{\rm s}(1+z),
\end{equation}
where $m_{\rm d}$ is the redshifted mass inferred from the waveform in the detector frame, $m_{\rm s}$ is the intrinsic mass in the mass frame, and $z$ is the cosmological redshift. Meanwhile, the distance of the source is also redshifted. If we could find the mass scale from the data of the mass spectrum or the theoretical prediction, the reference scale can be used to reconstruct the redshift at the distance. This approach can be analogized to the absorption or emission lines in galaxy spectra, known as the spectral siren approach.

The spectral siren method has the same Bayesian framework as the dark siren method, but using only the population models to describe the GW event rate in the source frame.
\begin{equation}
    \frac{{\rm d}N}{{\rm d}\theta_{\rm s}{\rm d}t_{\rm s}}\propto{\Psi(z|\boldsymbol{\Lambda})}p_{\rm pop}(m_{\rm 1,s},m_{\rm 2,s}|\boldsymbol{\Lambda})p_{\rm pop}(\chi|\boldsymbol{\Lambda})\frac{{\rm d}V_{\rm c}}{{\rm d}z}.
\end{equation}
More details about population models of GW events can be found in Refs.~\cite{LIGOScientific:2020kqk,KAGRA:2021duu}.

The application of the spectral siren approach is affected by whether the mass distribution has clear features or not~\cite{Ezquiaga:2022zkx}. Since the mass distributions of BNSs are narrow and likely to have a cutoff\footnote{The maximum mass of a NS cannot exceed the mass that the star becomes a BH and the minimum mass of a NS cannot be less than the mass at which the star becomes a white dwarf.}, BNSs are the first candidates for the spectral siren approach~\cite{Chernoff:1993th,Taylor:2011fs}. The currently most detected events are BBH mergers, making some features of BBH's mass distribution being revealed. For example, the number of BBH at high mass is very little~\cite{Fishbach:2017zga,KAGRA:2021duu} and there is an excess over the power-law at $\sim 35~M_{\odot}$~\cite{Farah:2023vsc}. Resolving the origin of these features are quite important for the spectral siren cosmology with the help of the current detectors and next-generation detectors. Note here that the feature shape, location, or some theoretical astrophysical prior can also be assumed, but it cannot represent the real data, potentially making the wrong estimations of cosmological parameters~\cite{Agarwal:2024hld}.

\begin{figure*}[!htbp]
\centering
\includegraphics[width=1.5\columnwidth]{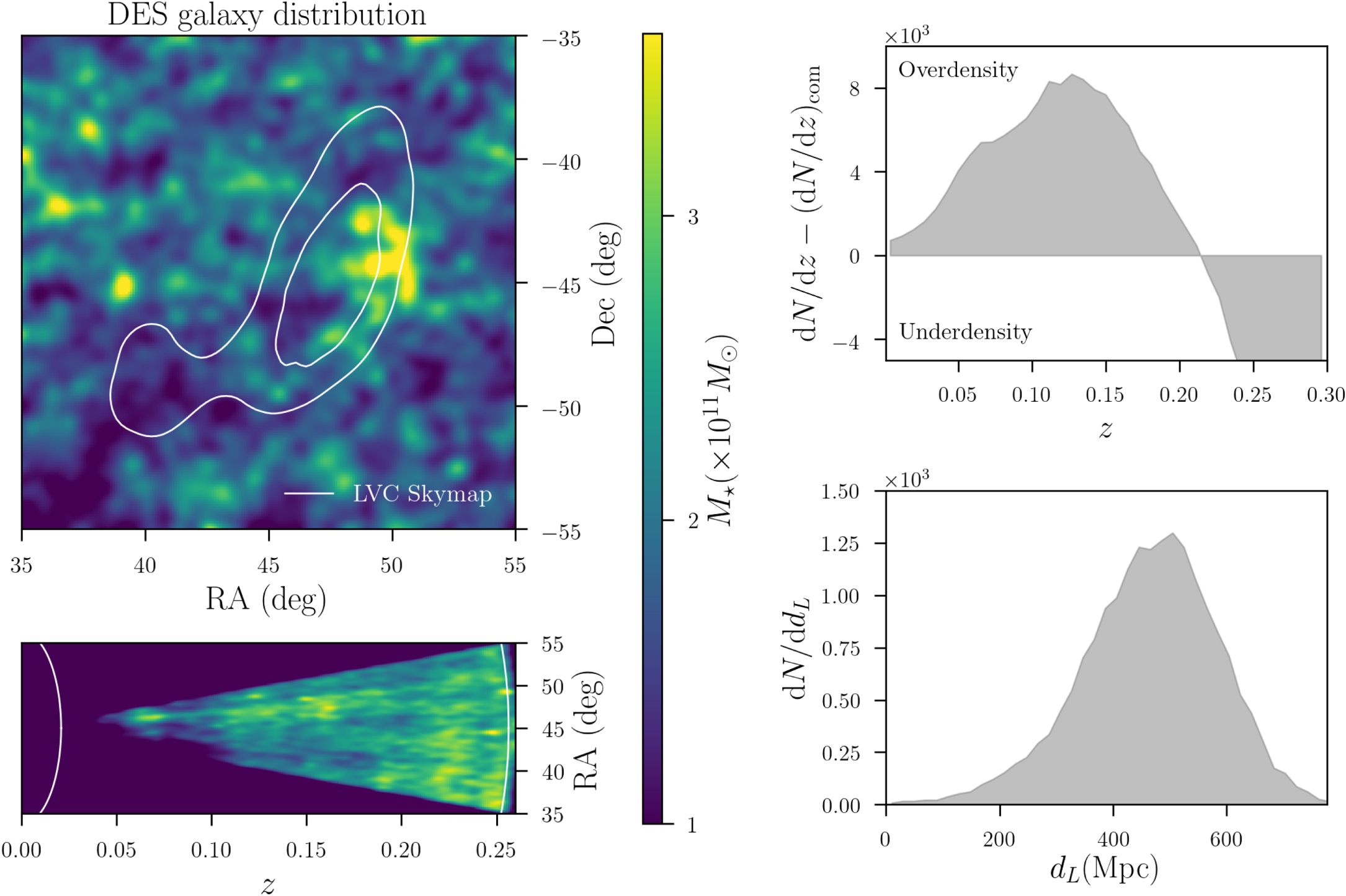}
\caption{GW170814 localization, DES galaxy distribution, and the distributions of the DES galaxy redshifts and luminosity distance in the localization region. Left panel: the GW170814 localization region at 50\% and 90\% confidence level and the used stellar mass distribution of DES galaxies. The galaxy's stellar mass distribution in RA and redshift, projected over DEC. Right panel: the distributions of the DES galaxy redshifts within the region of interest and the luminosity distance in HEALPIX pixels from the distance likelihood. The figure is from Ref.~\cite{DES:2019ccw} and reproduced by permission of the AAS.}
\label{fig:gw170814local}
\end{figure*}

\subsubsection{Love sirens: redshift from tidal effects in BNS}\label{sec2.3.5}
For NS mergers, except for the bright siren and spectral siren approach, there is another approach to directly measure redshift from GW waveform. When NSs are part of the compact binary coalescences, either BNS or NSBH mergers, they could have tidal deformations during the late stages of inspiral. The deformations could affect the phase evolution at the fifth and sixth post-Newtonian (PN) order. However, the coefficients of NSs are comparable in magnitude with the 3 PN and 3.5 PN phase terms~\cite{Messenger:2011gi}. The tidal deformation contribution is related to the tidal deformation parameters $\Lambda_1$ and $\Lambda_2$, which is related to the source-frame masses and EoS of NSs. If given by a fiducial EoS of NSs, the source-frame mass of NS can be estimated when $\Lambda_1$ and $\Lambda_2$ can be measured. Consequently, the redshift can be inferred solely through GW waveform analysis~\cite{Messenger:2011gi}, known as the ``love siren'' approach. Note that we refer to this kind of sirens as dark sirens in the following.

Limited by the sensitivity, the current GW observatories are unable to give precise measurement of tidal deformation parameters, but can only give an upper limit~\cite{LIGOScientific:2017vwq}. Even for the next-generation GW detectors, only $\tilde{\Lambda}$, which is related to $\Lambda_1$, $\Lambda_2$, and source-frame mass, are expected to be precisely measured, not for $\Lambda_1$ and $\Lambda_2$~\cite{Smith:2021bqc}. 

\section{Current constraints on $H_0$ from standard sirens}\label{sec3}
In this section, we shall report the current measurements of $H_0$ using different standard siren methods, including the bright siren, dark sirens, and spectral sirens.

\subsection{Constraints from the bright siren GW170817}\label{sec3.1}
On August 17, 2017, the advanced LIGO and advanced Virgo observed the first ever BNS merger event, named as GW170817~\cite{LIGOScientific:2017vwq}. Less than 2 seconds after the merger, the GRB event GRB170817A was detected by the Fermi GRB monitor within the sky region consistent with the localization from the LIGO-Virgo detection~\cite{Goldstein:2017mmi,LIGOScientific:2017zic}. In order to obtain a well-constrained and three-dimensional localization of GW170817, a series of EM observatories across the EM spectrum are launched to have an extensive follow-up observation. Less than 11 hours after the merger, a bright optical transient AT 2017gfo was discovered in NGC4993 (at $\sim 40 \rm Mpc$) by the 1M2H team~\cite{Coulter:2017wya} and also confirmed by other follow-up observations~\cite{LIGOScientific:2017ync,DES:2017kbs,Valenti:2017ngx,Smartt:2017fuw}. In Fig.~\ref{fig:multimessenger}, we show the timeline of the discovery of GW170817 and its EM emissions by the follow-up observations relative to the GW event time. GW170817 ushered in a new era of multi-messenger and time-domain astronomy. 

\FloatBarrier
\begin{figure*}[!t]
\centering
\includegraphics[width=1.5\columnwidth]{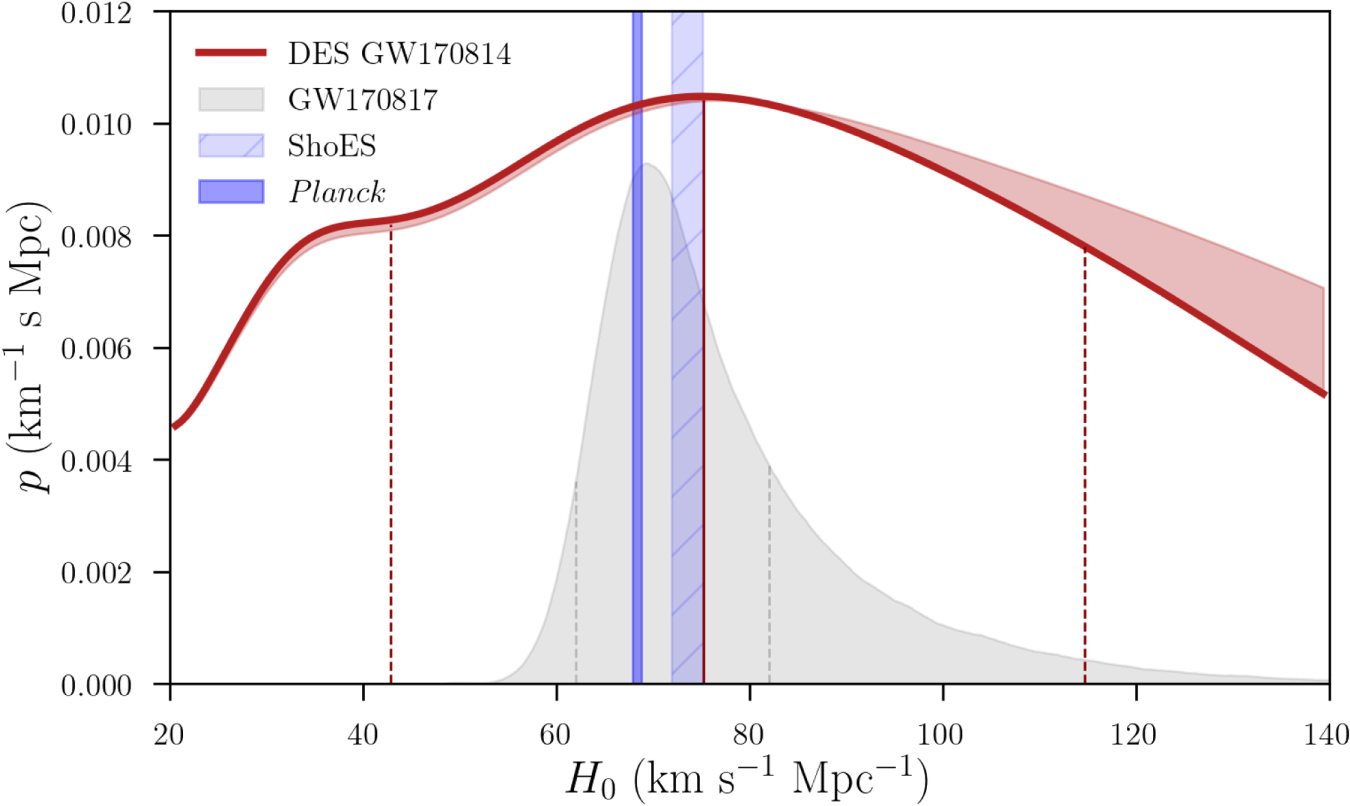}
\caption{The posterior distributions for $H_0$ using the DES GW170814 dark siren, GW170817 bright siren, SH0ES distance ladder, and Planck CMB observation. The figure is from Ref.~\cite{DES:2019ccw} and reproduced by permission of the AAS.}
\label{fig:gw170814h0}
\end{figure*}

Combined with the luminosity distance measurement from its GW waveform, the first measurement of $H_0$ using the standard siren method was obtained: $H_0=70^{+12}_{-8}~\rm km~s^{-1}~Mpc^{-1}$ with a 14\% precision~\cite{LIGOScientific:2017adf}. In Fig.~\ref{fig:gw170817h0}, we show the marginalized posterior distribution for $H_0$ using the bright siren GW170817. We can see that the measurement of $H_0$ has a relatively large error bar, which encompasses the results obtained from both the Planck CMB observation and the SH0ES distance ladder measurements. This broad error range is partly caused by a single data point and relatively large error of luminosity distance $d_{\rm L}=43.8^{+2.9}_{-6.9}~\mathrm{Mpc}$ in 68.3\% confidence level~\cite{LIGOScientific:2017adf} and also the strong parameter degeneracy between luminosity distance and the inclination angle in the measurement~\cite{LIGOScientific:2016vlm}. 

Fortunately, the radio jet of GW170817 is observed, and thus the degeneracy is broken~\cite{Hotokezaka:2018dfi}. Through the analysis of the afterglow light curve (LC) and centroid motion through very-long baseline interferometry of GW170817, the measurement of $H_0$ is also improved to $H_0=68.3^{+4.4}_{-4.3}~\rm km~s^{-1}~Mpc^{-1}$ with a 6.7\% precision by assuming PLJ model~\cite{Hotokezaka:2018dfi}, see e.g., Refs.~\cite{Vitale:2018wlg,Mastrogiovanni:2020ppa,Bulla:2022ppy,Xie:2022brn,Gong:2023ecg,Wang:2022apn,Wong:2023kkh} for the discussion about breaking the luminosity distance and inclination angle degeneracy. In the left panel of Fig.~\ref{fig:vlbi}, we show the posterior distributions for the inclination angle and distance using the Hydro jet, GW170817 only, afterglow LC and centroid motion through VLBI, and their combination with GW. In the right panel of Fig.~\ref{fig:vlbi}, we show the posterior distributions for $H_0$ using GW+BLBI+LC, GW170817 only, and the 1 and 2-$\sigma$ regions from Planck CMB (TT,TE,EE+lowP+lensing) and SH0ES distance ladder. We can clearly see that the parameter degeneracy between distance and inclination angle is effectively broken, resulting in a significant improvement in measurement accuracies of inclination angle and distance. Furthermore, the measurement precision of $H_0$ also has great improvement. 

Although the current precision is far from resolving the Hubble tension, some forecasts show that the precision can be improved to $\sim 2\%$ if 50 GW170817-like bright sirens can be detected by the advanced LVK network~\cite{Chen:2017rfc,Nissanke:2013fka,Feeney:2018mkj}. If about 15 additional GW170817-like events can be well-localized and accompanied by radio imaging and LC data with comparable SNRs and viewing angles, the precision on $H_0$ could be improved to 1.8\%~\cite{Hotokezaka:2018dfi}.
%If 15 more localized GW170817-like events can be detected their radio images and LC data with comparable SNRs and similar orientations, they can improve the precision of $H_0$ to 1.8\%~\cite{Hotokezaka:2018dfi}. 

\begin{figure*}[!htbp]
\centering
\includegraphics[width=\linewidth]{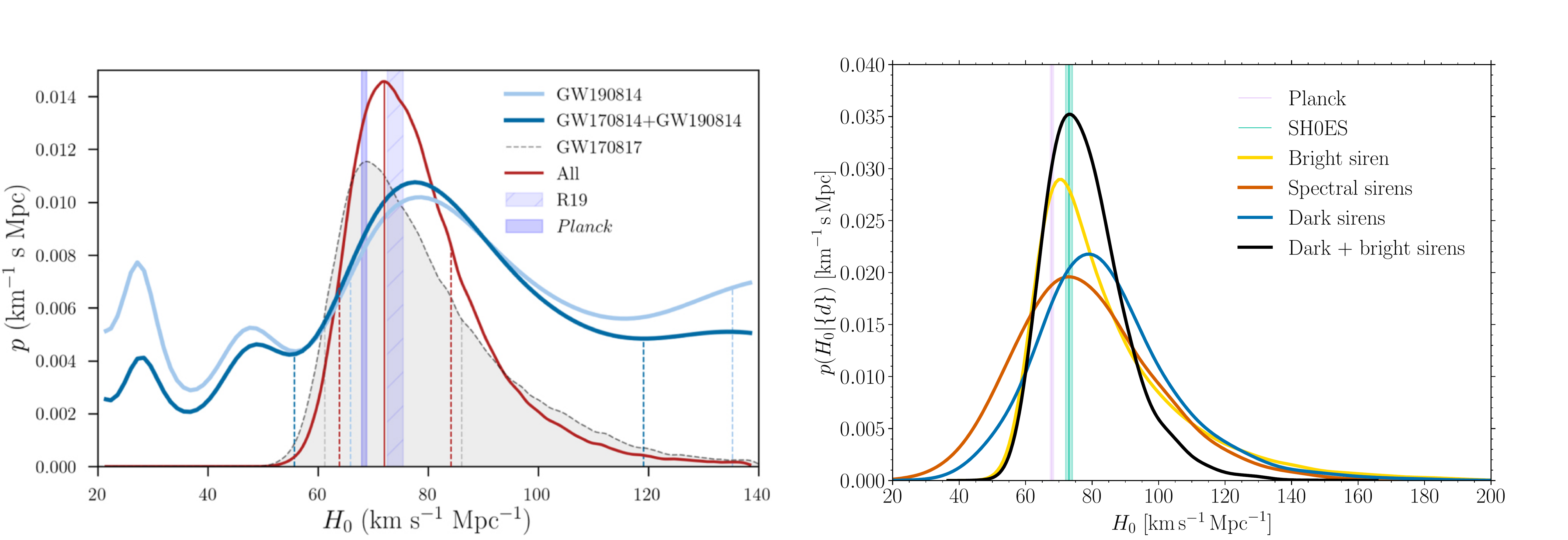}
\caption{Posterior distributions for $H_0$ using dark sirens. Left panel: posterior distributions for $H_0$ using GW190814 dark siren, GW170814 and GW190814 dark sirens, GW170817 bright siren, the combination of them, the distance ladder, and the Planck CMB observation. Right panel: posterior distributions for $H_0$ using GW170817 bright siren, 46 dark sirens with an empty catalog, combination of 46 dark sirens with \texttt{GLADE+} $K$-band galaxy catalog and bright siren GW170817, 46 dark sirens ($K$-band), Planck CMB, and the SH0ES distance ladder. The figures are from Refs.~\cite{DES:2020nay} and \cite{LIGOScientific:2025jau} and reproduced by permission of the AAS.}
\label{gw190814&o4h0}
\end{figure*}

\subsection{Constraints from dark sirens}\label{sec3.2}
GW170814~\cite{LIGOScientific:2017ycc} was the first BBH event for which the dark siren method was applied to infer cosmological parameters, yielding a constraint of \( H_0 = 75^{+40}_{-32}~\mathrm{km\,s^{-1}\,Mpc^{-1}} \)~\cite{DES:2019ccw}. 
GW170814 was localized within a volume of \(2\times 10^6~\rm Mpc^{3}\)~\cite{Palmese:2021mjm} and was associated with \(\sim 77{,}000\) potential galaxies in the Dark Energy Survey data~\cite{DES:2019ccw}. In the upper left panel of Fig.~\ref{fig:gw170814local}, we show the GW170814 localization region at 50\% and 90\% confidence levels and the used stellar mass distribution of the DES galaxies. The bottom panel shows the galaxies' stellar mass distribution in RA and redshift, projected over DEC. The right panel of Fig.~\ref{fig:gw170814local} shows the distribution of the DES galaxy redshifts within the region of interest and the luminosity distance in HEALPIX pixels from the distance likelihood. The posterior distributions for $H_0$ from the GW170814 dark siren, GW170817 bright siren, Planck CMB, and SH0ES distance ladder are shown in Fig.~\ref{fig:gw170814h0}. As seen, in the case that only one data point is employed in $H_0$ measurement, the measurement of the GW170814 dark siren is significantly less precise than that of the GW170817 bright siren. The error comes from the inherent limitations of the dark siren analysis, which relies on the statistical associations between poor GW localization and a large number of potential host galaxies, rather than a uniquely identified host galaxy.

For comparison, the best localized BBH merger GW190814~\cite{LIGOScientific:2020zkf} was localized to a volume roughly 100 times smaller than that of GW170814 and was associated with $\sim 2000$ potential galaxies~\cite{DES:2020nay}. GW190814 provides a slightly better constraint on $H_0$ with $H_0=78^{+57}_{-13}~\rm km~s^{-1}~Mpc^{-1}$ (a $45\%$ measurement)~\cite{DES:2020nay} than that by GW170814 (a $48\%$ measurement). The joint analysis combining GW170814 and GW190814 further improve the constraint on $H_0$ to 44\%~\cite{DES:2020nay}. 
In the left panel of Fig.~\ref{gw190814&o4h0}, we show the posterior distributions for $H_0$ using GW190814 dark siren, GW170814+GW190814 dark sirens, GW170817 bright siren, the combination of them, the distance ladder measurement, and the Planck CMB measurement. We can see that GW190814 alone provides a rather poor constraint on $H_0$ and the addition of GW170814 provides a slight improvement. The constraint result using the combination of both dark sirens and bright sirens is mainly driven by the bright siren GW170817. 

% \FloatBarrier
\begin{figure*}[!t]
\centering
\includegraphics[width=1.5\columnwidth]{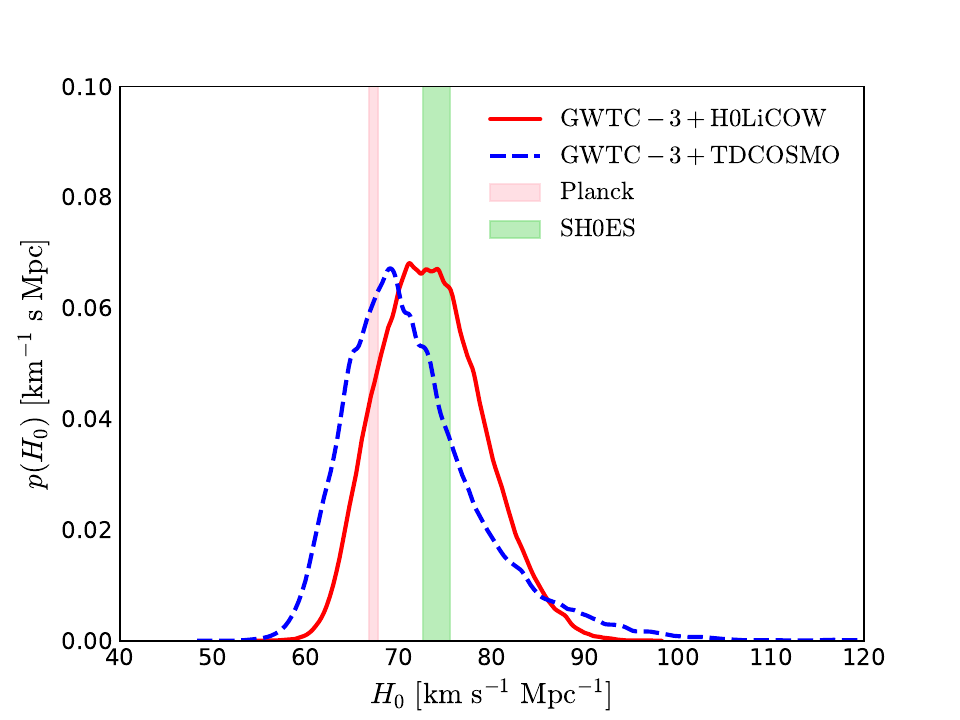}
\caption{Posterior distributions for $H_0$ using 47 GW events from GWTC-3, the SGL system RXJ1131-1231 from H0LiCOW (red) and TDCOSMO (blue) in the polynomial model, Planck CMB, and SH0ES distance ladder. The figure is from Ref.~\cite{Song:2025ddm} and reproduced by permission of the AAS.}
\label{fig: GWTC3RXJ1131-1231}
\end{figure*}

Currently, the most precise measurement of $H_0$ using dark sirens is $H_0=67^{+13}_{-12}~\rm km~s^{-1}~Mpc^{-1}$ given by 46 dark sirens~\cite{LIGOScientific:2021aug} with $K$-band galaxy from \texttt{GLADE+}.
% \blue{However, this result is sensitive to the BH population assumptions that contribute to the out-of-catalog likelihood terms (when the galaxy catalog is not complete) and the in-catalog terms (when a large number of galaxies are in the GW sky localization).}
The combination of 46 dark sirens and a bright siren GW170817 offers $H_0=68^{+8}_{-6}~\rm km~s^{-1}~Mpc^{-1}$ with a 10\% precision~\cite{LIGOScientific:2021aug}. \blue{Recently, the LIGO--Virgo--KAGRA collaboration reported constraints on the Hubble constant from O4a using multiple standard-siren approaches \cite{LIGOScientific:2025jau}: dark sirens alone give 
$H_{0} = 81.6^{+21.5}_{-15.9}\,\mathrm{km\,s^{-1}\,Mpc^{-1}}$, 
a re-analysis result of the bright siren GW170817 (with a wider $H_{0}$ prior and consistent selection effects) yields 
$H_{0} = 78.4^{+25.7}_{-12.0}\,\mathrm{km\,s^{-1}\,Mpc^{-1}}$, 
spectral sirens give 
$H_{0} = 76.4^{+23.0}_{-18.1}\,\mathrm{km\,s^{-1}\,Mpc^{-1}}$, 
and combining the dark-siren posterior with the re-analysis bright siren results in 
$H_{0} = 76.6^{+13.0}_{-9.5}\,\mathrm{km\,s^{-1}\,Mpc^{-1}}$ 
(all 68\% confidence levels, see Table~\ref{tab:H0_measurements} for more detailed results of the current $H_0$ measurements from standard sirens).}
% The result highlight the importance of the accumulation of dark sirens, which can statistically reduce the measurement accuracy of $H_0$. 
In the right panel of Fig.~\ref{gw190814&o4h0}, we show the posterior distributions for $H_0$ using re-analysis GW170817 bright siren, spectral sirens from the FULLPOP-4.0 mass model, 141 dark sirens with \texttt{GLADE+} $K$-band and the FULLPOP-4.0 mass model, combination of dark and bright siren GW170817, Planck CMB measurement, and distance ladder measurement. With the accumulation of dark sirens, the constraint on $H_0$ using dark sirens is comparable with that given by the bright siren GW170817. Their combination of dark and bright sirens gives a more concentrated constraint which has a 40\% improvement compared to GW170817 bright siren result. Although the precision is still much worse than those given by either CMB or distance ladder measurements, it shows the potential of standard siren in help resolving the Hubble tension.

In addition to using the GW dark siren alone, recently, Song et al.~\cite{Song:2025ddm} used 47 GW standard sirens from the third Gravitational-Wave Transient Catalog (GWTC-3) to calibrate the distances in the strong gravitational lensing (SGL) system RXJ1131-1231 and constrain $H_0$ without assuming a specific cosmological model. For the case of $\Omega_{\rm k}=0$, they obtained $H_0=73.22^{+5.95}_{-5.43}~\rm km~s^{-1}~Mpc^{-1}$ (a 7.8\% measurement) and $H_0=70.40^{+8.03}_{-5.60}~\rm km~s^{-1}~Mpc^{-1}$ (a 9.7\% measurement) by breaking the mass-sheet transform using lens galaxy’s mass models (H0LiCOW) and stellar kinematics (TDCOSMO), respectively, with $H_0$ posteriors shown in Fig.~\ref{fig: GWTC3RXJ1131-1231}. In the future, as more high-redshift GW dark sirens are observed, this approach is expected to enable high-precision, model-independent constraints on $H_0$.
% In the future, with next-generation sky surveys (e.g., DESI, LSST, CSST) providing more complete catalogs and next-generation GW detectors delivering improved spatial localization, this approach will have significant potential for cosmological applications \cite{Song:2022siz}. 

The spatial cross-correlation between the GW sources and the galaxy can also be applied to infer cosmological parameters~\cite{Oguri:2016dgk,Bera:2020jhx,Mukherjee:2020hyn,Ghosh:2023ksl}. Since GW sources and galaxies trace the matter density, they are spatially correlated through the underlying matter field. By analyzing the three-dimensional spatial correlation between GW sources and redshift-known galaxies, we can infer the host redshift shells for GW sources. A constraint of $H_0=68.2^{+26.0}_{-6.2}~\rm km~s^{-1}~Mpc^{-1}$ was obtained~\cite{Mukherjee:2022afz} using the cross-correlation method from the 8 best-localized BBH events from GWTC-3~\cite{KAGRA:2021vkt}. Recent forecast~\cite{Diaz:2021pem} showed that $H_0$ could be measured to a precision of $\mathcal{O}(1)\%$ with 5-year observation of the designed LVK, dark energy spectroscopic instrument~\cite{DESI:2016fyo}, and SPHEREx~\cite{SPHEREx:2014bgr}.

\begin{table*}[!htbp]
\centering
\small
\setlength{\tabcolsep}{10pt}
\renewcommand{\arraystretch}{1.6}
\caption{\label{tab:H0_measurements} \blue{Recent $H_0$ constraints from GW standard sirens using bright sirens, dark sirens, spectral sirens, and their combination. Here CBC represents compact binary coalescence and Precision is defined as the average of the upper and lower 68\% credible interval bounds (relative error), divided by the reported central value. To reflect the original sources accurately, the interval types are preserved as reported: most entries use 68\% highest density intervals, while the O4a results use symmetric 68\% credible intervals.}
}
\begin{adjustbox}{width=\textwidth}
\begin{tabular}{>{\raggedright\arraybackslash}p{7.2cm} c c c c}
\hline
\blue{Description} & \blue{Dataset} & \blue{Reference} & \blue{$H_0\ [\mathrm{km\,s^{-1}\,Mpc^{-1}}]$} & \blue{Precision} \\
\hline
\blue{Bright siren}                                      & \blue{GW170817}                         & \blue{\cite{LIGOScientific:2017adf}} & \blue{$70^{+12}_{-8}$}     & \blue{14.29\%} \\
\blue{Bright siren (updated with O4a priors and selection consistent with the O4a dark and spectral siren analyses)}                                      & \blue{GW170817}                         & \blue{\cite{LIGOScientific:2025jau}} & \blue{$78.4^{+25.7}_{-12.0}$}     & \blue{24.04\%} \\
\blue{Dark sirens (fixed population model)}              & \blue{O1+O2+O3 [46 CBCs]}               & {\cite{LIGOScientific:2021aug}} & \blue{$67^{+13}_{-12}$}         & \blue{18.66\%} \\
\blue{Dark sirens}                                       & \blue{O1+O2+O3+O4a [141 CBCs]}          & {\cite{LIGOScientific:2025jau}} & \blue{$81.6^{+21.5}_{-15.9}$}   & \blue{22.92\%} \\

\blue{Spectral sirens}                                   & \blue{O1+O2+O3 [42 CBCs]}               & {\cite{LIGOScientific:2021aug}} & \blue{$50^{+37}_{-30}$}         & \blue{67.00\%} \\
\blue{Spectral sirens}                                   & \blue{O1+O2+O3+O4a [141 CBCs]}          & {\cite{LIGOScientific:2025jau}} & \blue{$76.4^{+23.0}_{-18.1}$}   & \blue{26.90\%} \\
\blue{Bright + dark sirens (fixed population model)}     & \blue{GW170817+O1+O2+O3 [47 CBCs]}      & {\cite{LIGOScientific:2021aug}} & \blue{$68^{+8}_{-6}$}           & \blue{10.29\%} \\
% \blue{Bright + dark sirens}                              & GW170817+O1+O2+O3 [43 CBCs]      & \blue{\cite{Mastrogiovanni:2023emh}} & \blue{$68^{+13}_{-7}$}     & \blue{14.71\%} \\
% \blue{Bright + dark sirens (fixed BNS population model)} & GW170817+O1+O2+O3 [47 CBCs]      & \blue{\cite{Gray:2023wgj}} & \blue{$69^{+12}_{-7}$}               & \blue{13.77\%} \\
\blue{Bright (updated) + dark sirens}                              & \blue{GW170817+O1+O2+O3+O4a [142 CBCs]} & {\cite{LIGOScientific:2025jau}} & \blue{$76.6^{+13.0}_{-9.5}$}    & \blue{14.69\%} \\

\blue{Bright + spectral sirens}                          & \blue{GW170817+O1+O2+O3 [43 CBCs]}      & {\cite{LIGOScientific:2021aug}} & \blue{$68^{+12}_{-8}$}          & \blue{14.71\%} \\
\blue{Bright (updated) + spectral sirens}                          & \blue{GW170817+O1+O2+O3+O4a [142 CBCs]} & {\cite{LIGOScientific:2025jau}} & \blue{$74.6^{+13.4}_{-9.1}$}    & \blue{15.08\%} \\
\hline
\end{tabular}
\end{adjustbox}
\end{table*}

\begin{figure*}[!htbp]
\includegraphics[width=0.8\textwidth]{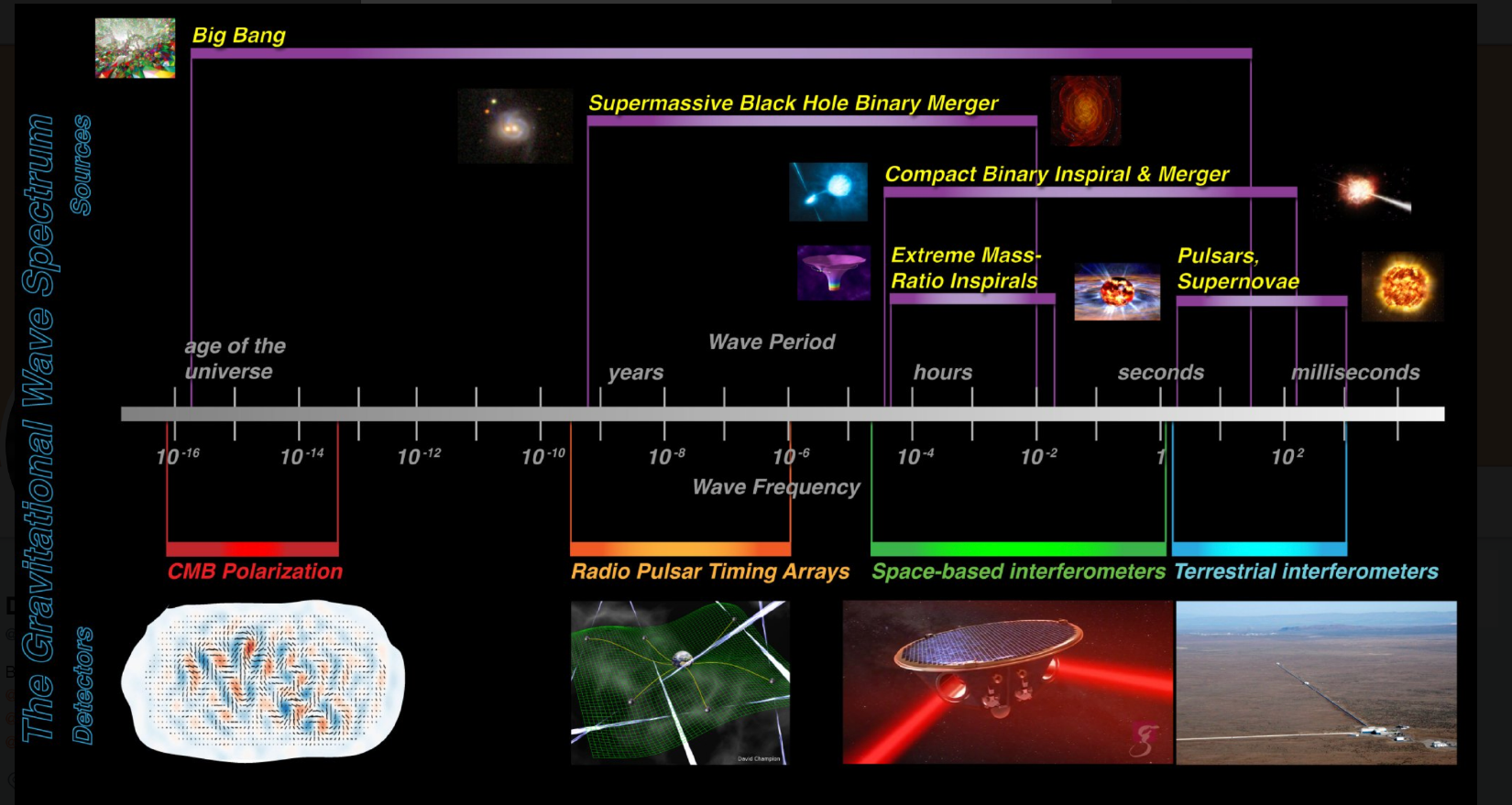}
\centering
\caption{GW sources and corresponding GW detectors ranging from $10^{-16}$ Hz to $10^4$ Hz. Reproduced from Ref.~\cite{Holley-Bockelmann:2019qte} (Credits: K. Jani, Bicep-Keck: \url{https://bicepkeck.org/}, NANOGrav: \url{https://nanograv.org/}, LISA: \url{https://lisa.nasa.gov/}, and LIGO: \url{https://www.ligo.caltech.edu/}).}
\label{fig:gwband}
\end{figure*}

\section{Prospects for measuring $H_0$ and other cosmological parameters with future standard siren observations}\label{sec4}
Although significant progress has been made in the field of GW standard siren cosmology in recent years, current measurements of $H_0$ using bright siren, dark sirens, or even their combination are still far from precisely measuring $H_0$ (achieving sub-percent level) which is required for standard siren to play a third-party and independent role to address the Hubble tension. In particular, the best current constraint although increasingly competitive, still has a 5\%--10\% measurement precision, which limits their role as an independent arbitrator between early-universe and late-universe measurement of $H_0$.
However, the methodology itself is highly robust which is independent of the distance ladder and could provide self-calibrated distance information. Therefore, it is worthwhile to discuss the prospects of future standard siren observations for measuring $H_0$. Moreover, the future GW standard siren observations will be in operation across from nanohertz to hundreds of hertz, which will greatly expand the number of detectable events and improve the localization accuracies of GWs. Furthermore, a deeper galaxy catalog could be obtained with the help of the next-generation galaxy survey projects. All the above facts would significantly improve the ability of standard sirens in constraining cosmological parameters, making them that can not only provide measurement on $H_0$, but also provide help constrain other cosmological parameters.
In the following section, we shall briefly introduce GW in various frequency bands and their forecasted constraints on $H_0$.

\subsection{Future GW observatories}\label{sec4.1}
Similar to traditional EM waves, GW frequency bands span nearly 20 orders of magnitude, ranging from approximately $10^{-16}$ Hz to $10^4$ Hz. Each frequency band correspond to different GW sources, as shown in Fig.~\ref{fig:gwband}. In this subsection, we shall provide an overview of the future GW observatories that will operate across the GW spectrum.

\subsubsection{Ground-based GW observatories}\label{sec4.1.1}
Ground-based GW observatories are primarily sensitive to GW in the $\sim 10$--$10^3$ Hz range, corresponding to the compact binary coalescence such as BNS, stellar-mass BBH, and NSBH systems. 
Currently, the operational ground-based GW observatories include the advanced LIGO detectors in the US~\cite{LIGOScientific:2014pky}, Virgo in Europe~\cite{VIRGO:2014yos}, KAGRA in Japan~\cite{Somiya:2011np}, and GEO 600 in Germany~\cite{Willke:2002bs}. They have reported 90 merger events during the first, second, and third observing runs~\cite{LIGOScientific:2018mvr,LIGOScientific:2020ibl,KAGRA:2021vkt}. 
% Among these detections, two are BNS mergers, three are NSBH mergers, and the remaining events are BBH mergers or uncertain classification. 

Building upon the technological advances and operational experience from these ground-based GW detectors, the next-generation ground-based GW observatories have been proposed. These include the 2.5-generation (2.5G) detectors such as LIGO Voyager in the US~\cite{LIGO:2020xsf} and Neutron Star Extreme Matter Observatory (NEMO) in Australia~\cite{Ackley:2020atn}. Moreover, the third-generation (3G) GW detectors such as Cosmic Explorer (CE) in the US~\cite{LIGOScientific:2016wof} (with a proposed southern hemisphere site in Australia~\cite{Evans:2021gyd}), and Einstein Telescope (ET)~\cite{Punturo:2010zz} in Europe are also proposed.
These future ground-based GW detectors are designed to improve the strain sensitivity by a factor of approximately 5–10 over current ground-based GW detectors, significantly enhancing the detection range, the number of observable GW events, and the GW parameter estimation. 

\subsubsection{Space-based observatories}\label{sec4.1.2}

Massive black hole binary (MBHB) mergers emit GWs mainly in the millihertz frequency band. 
The detection of such low-frequency GWs requires interferometers with arm lengths on the order of millions of kilometers, a scale which cannot be achieved on Earth.
Several space-based GW observatories have been proposed to access this unexplored regime, such as LISA (The Laser Interferometer Space Antenna)~\cite{LISA:2017pwj,Robson:2018ifk,LISACosmologyWorkingGroup:2022jok}, Taiji~\cite{Ruan:2018tsw,Wu:2018clg,Hu:2017mde}, TianQin~\cite{Liu:2020eko,Wang:2019ryf,TianQin:2015yph,Luo:2020bls,Milyukov:2020kyg,TianQin:2020hid,Luo:2025ewp}, and DECIGO\footnote{The Japanese proposed the space-based GW observatory DECIGO (The Deci-hertz Interferometer Gravitational Wave Observatory) is designed to detect GWs in the deci-hertz frequency band. It will consist of four clusters of spacecraft in heliocentric orbits, with each interferometer arm length being 1000 km.}~\cite{Kawamura:2011zz}. 
LISA, a joint ESA-NASA mission, will consist of three identical drag-free satellites configured in an equilateral triangle with an arm length of 2.5 million kilometers, following a heliocentric orbit trailing the Earth by about 20 degrees. It is expected to launch between 2030 and 2035, with an initial mission duration of 4 years, extendable up to 10 years.
In a design similar to LISA, the Taiji mission, led by the Chinese Academy of Sciences, will also employ a triangular formation of three satellites. However, it will have an arm length of 3 million kilometers and will be positioned in a heliocentric orbit leading the Earth by approximately 20 degrees. Meanwhile, TianQin is designed for geocentric orbit operation with an arm length of about $\sqrt{3}\times 10^5$ kilometers. Recently, successful flights of pathfinder missions, such as LISA pathfinder, Taiji-1, and TianQin-1, have demonstrated the feasibility of drag-free technology and interferometric distance measurement in space~\cite{Liu:2021ocy,Luo:2020bls,LISAPathfinder:2017khw}, validating the key technologies required for the full-scale observatories.

As there is an overlapping launch window and mission durations, the concept of a space-based detector network has been proposed. The space-based GW network aims not only to be able to localize GW sources more precisely~\cite{Ruan_2020,Ruan:2019tje}, but also beneficial for the cosmological analysis~\cite{Wang:2020dkc,Wang:2021srv} (see Refs.~\cite{Jin:2023sfc,Cai:2023ywp,Zhu_2022_2,Lyu_2023,Torres-Orjuela:2023hfd,Wanggang_2021} and references therein for related discussions).

However, the application of standard sirens from space-based GW detectors into cosmology also highly relies on the redshift inference of MBHB.
Several works suggest that EM radiations in the radio/optical could be emitted during the mergers of MBHBs~\cite{Palenzuela:2010nf,OShaughnessy:2011nwl,Moesta:2011bn,Kaplan:2011mz,Shi:2011us,Blandford:1977ds,Meier:2000wk,Dotti:2011um}. Such emissions are expected to be detectable by EM telescopes. However, due to the limited field of view of these telescopes, only GW events with a localization angle of $\Delta\Omega\leq 10~\mathrm{deg^2}$ are likely to be viable candidates for concurrent EM wave detection. Additionally, since the SNR of the space-based GW detection is high, the dark siren approach remains a promising and realistic method for redshift inference, even in the absence of EM counterparts or when no EM emission occurs. Deep and complete photometric and spectroscopic surveys from e.g., CSST~\cite{2011SSPMA..41.1441Z,Gong:2019yxt}, will be important for the dark siren analysis.

\FloatBarrier
\begin{figure*}[!htbp]
\centering
\includegraphics[width=1.8\columnwidth]{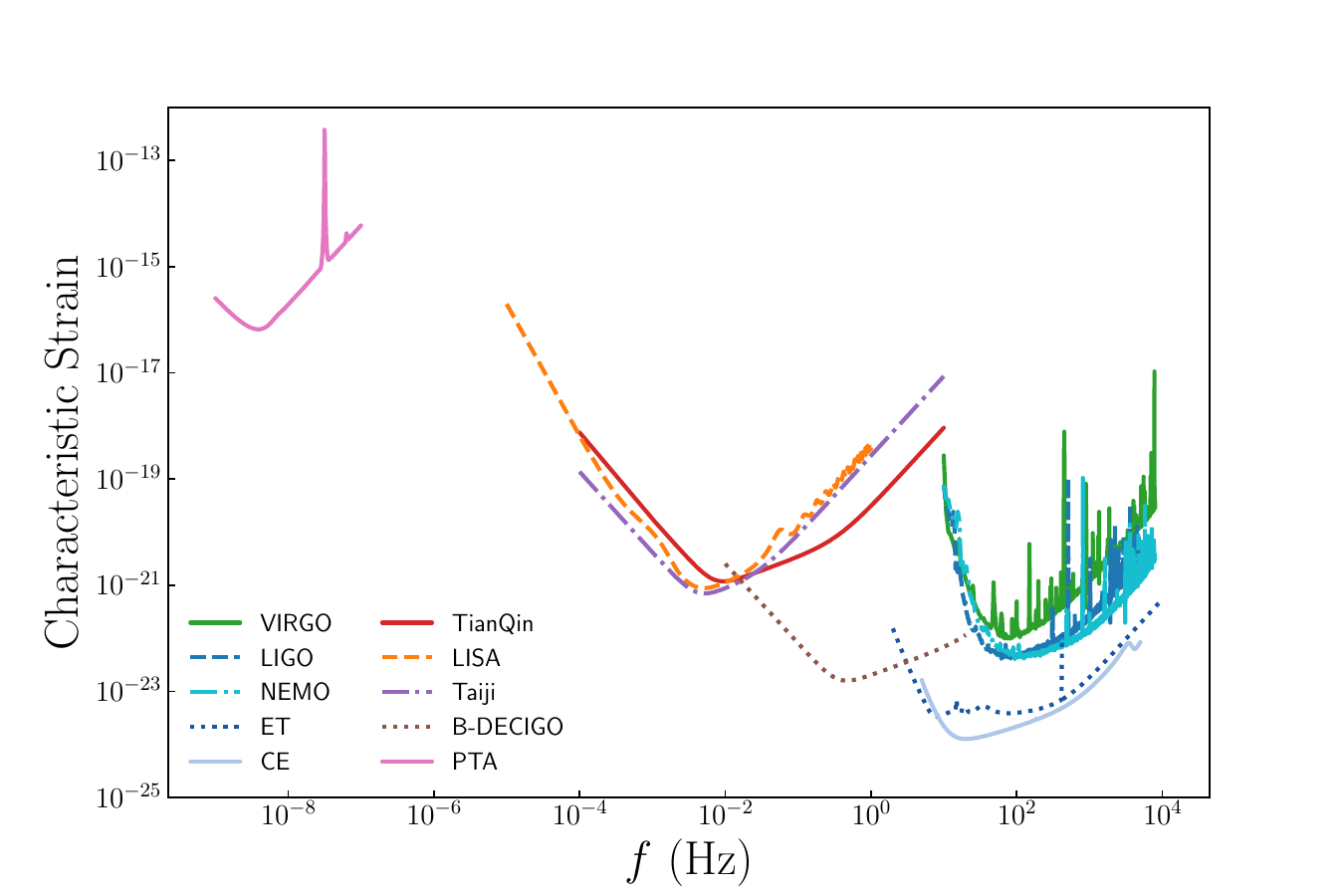}
\caption{Characteristic strains of the current ground-based GW detectors and future proposed GW detectors, LIGO~\cite{KAGRA:2021vkt}, Virgo~\cite{KAGRA:2021vkt}, NEMO \cite{Ackley:2020atn}, ET~\cite{Hild:2010id}, CE (https://cosmicexplorer.org/sensitivity.html), TianQin~\cite{Wang:2019ryf}, LISA~\cite{Klein:2015hvg}, Taiji~\cite{Ruan:2018tsw}, B-DECIGO~\cite{Isoyama:2018rjb}, and the 10-year observation of SKA-era PTAs (100 pulsars with 100-ns white noise).}
\label{fig:strains}
\end{figure*}

\subsubsection{Pulsar timing array observatories}\label{sec4.1.3}
GWs in the nanohertz band ($\sim10^{-9}$–$10^{-7}$ Hz) originate mainly from inspiraling supermassive black hole binaries (SMBHBs) in the early stages of coalescence\footnote{Note that other potential sources include curvature perturbations~\cite{Kawasaki:2013xsa,Kohri:2018awv,Wang:2019kaf}, first-order phase transition~\cite{Witten:1984rs,Hogan:1986qda,Kosowsky:1991ua,Kosowsky:1992vn,Kosowsky:1992rz,Kamionkowski:1993fg,Caprini:2007xq,Huber:2008hg,Caprini:2009yp,Binetruy:2012ze,Hindmarsh:2013xza,Caprini:2015zlo,Hindmarsh:2015qta,Athron:2023xlk}, domain walls~\cite{Vilenkin:1981zs,Gleiser:1998na,Hiramatsu:2013qaa,Kawasaki:2011vv,Saikawa:2017hiv,Ferreira:2022zzo}, cosmic strings~\cite{Vachaspati:1984gt,Damour:2001bk,Damour:2004kw,Siemens:2006yp,DePies:2007bm,Kawasaki:2010yi,Olmez:2010bi,Sanidas:2012ee,Sanidas:2012tf,Kuroyanagi:2012wm,Matsui:2016xnp,Chen:2022azo,Bian:2022tju}, and so on. However, this review specifically focuses on standard sirens, and thus we will primarily discuss SMBHBs.}. Due to their extremely long wavelengths, they are undetectable by both ground-based and space-based GW observatories, but can be probed via pulsar timing arrays (PTAs), which is composed of a network of millisecond pulsars that serve as highly stable cosmic clocks. 

PTAs monitor deviations in the pulse arrival times (known as timing residuals) caused by passing GWs.
As GWs travel between pulsars and the Earth, they can modify the distance separating the pulsars from the Earth. This alteration leads to changes in the times of arrival (ToAs) of the pulsars' radio pulses, either causing a delay or an advance. These variations in the ToAs carry information about the nanohertz GWs.
By analyzing the signals from a single pulsar, it is possible to model both the timing parameters and the noise parameters specific to each pulsar. 
A stochastic GW background (SGWB), potentially formed by the superposition of many unresolved SMBHBs or other early-universe processes, imprints a spatial correlation pattern known as the Hellings-Downs curve~\cite{Hellings:1983fr} on the cross-correlation of pulsar timing residuals.
Therefore, detecting SGWB is one of the most important scientific goals of the current International PTA collaboration which includes the North American Observatory for Gravitational Waves (NANOGrav)~\cite{McLaughlin:2013ira}, the European Pulsar Timing Array (EPTA)~\cite{Kramer:2013kea}, and the Parkes Pulsar Timing Array (PPTA)~\cite{Manchester:2012za}. The Indian Pulsar Timing Array (InPTA)~\cite{Tarafdar:2022toa} has recently become a full member, while the Chinese Pulsar Timing Array (CPTA)~\cite{2016ASPC..502...19L} and MeerKAT PTA~\cite{Miles:2022lkg} are observer members.

In 2023, the CPTA, NANOGrav, EPTA, and PPTA collaborations reported the evidence of a SGWB~\cite{NANOGrav:2023gor,Reardon:2023gzh,EPTA:2023sfo,Xu:2023wog}. These observations include spatial correlation signals consistent with such a background. Notably, the CPTA collaboration has detected a quadrupolar correlation signal attributable to the GWB, achieving a confidence level of $4.6\sigma$, which marks a significant step in nanohertz GW detection. Related issues have been extensively discussed in the literature~\cite{NANOGrav:2023hde,NANOGrav:2023tcn,NANOGrav:2023hvm,NANOGrav:2023pdq,NANOGrav:2023ctt,NANOGrav:2023hfp,NANOGrav:2023icp,NANOGrav:2023ygs,Zic:2023gta,Reardon:2023zen,InternationalPulsarTimingArray:2023mzf,EPTA:2023akd,EPTA:2023fyk,EPTA:2023gyr,EPTA:2023xxk,Zhu:2023gmx,Franciolini:2023pbf,Liu:2023ymk} and references therein. Although resolving GW signals from individual SMBHBs remains challenging, continued observations are expected to improve the sensitivity and spectral characterization of the SGWB.

In the upcoming era of the Square Kilometer Array (SKA), the potential to detect nanohertz GW signals from individual SMBHB events becomes increasingly feasible with its unprecedented sensitivity and timing precision. By analyzing the GW waveform, it is expected that the luminosity distance of these SMBHB events could be determined. If redshifts can also be determined either through direct EM identification or statistical association with galaxy catalogs, the distance-redshift relationship could be used to perform cosmological analysis. 

\subsection{Simulations of standard sirens}\label{sec4.2}
In Fig.~\ref{fig:strains}, we show the characteristic strains for the current and future GW observatories across the GW spectrum. The standard siren data should be simulated based on these strains. For the cosmological analysis, the redshift, luminosity distance, and the errors of luminosity distances and location angle are required for performing the Bayesian inference of cosmological parameters, as introduced in Sec.~\ref{sec2.3}.

\FloatBarrier
\begin{figure*}[!htbp]
\centering
\includegraphics[width=\columnwidth]{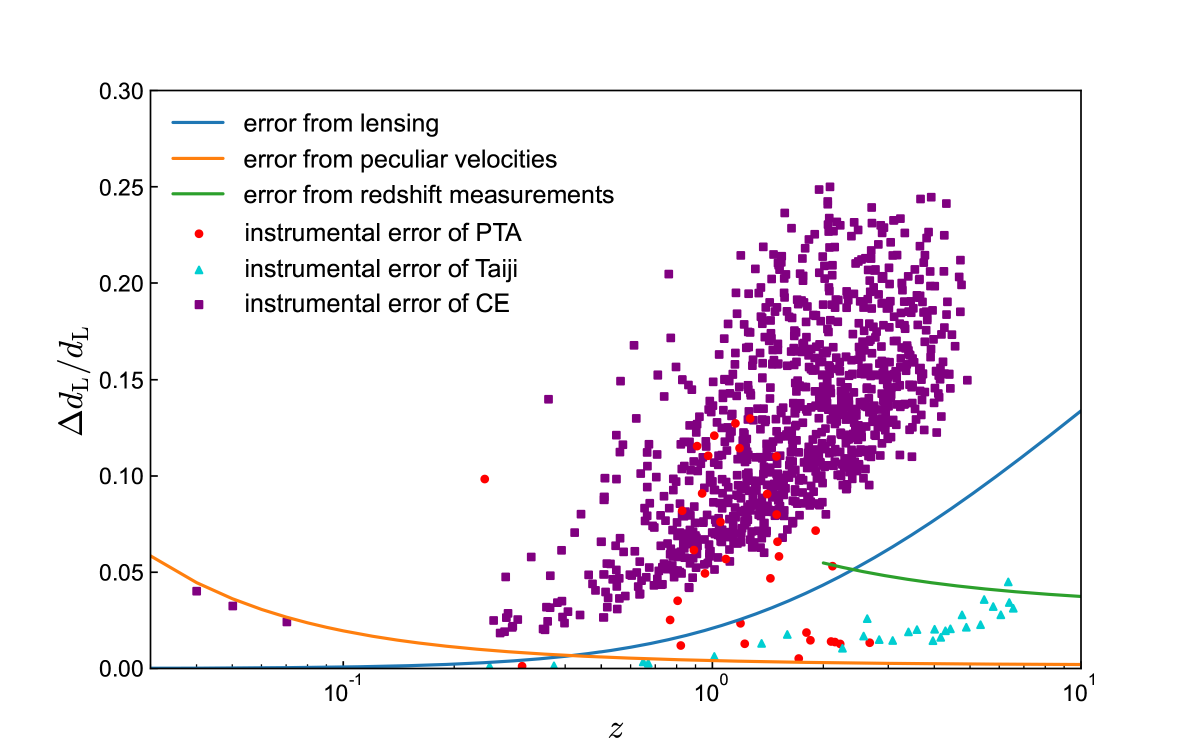}
\includegraphics[width=\columnwidth]{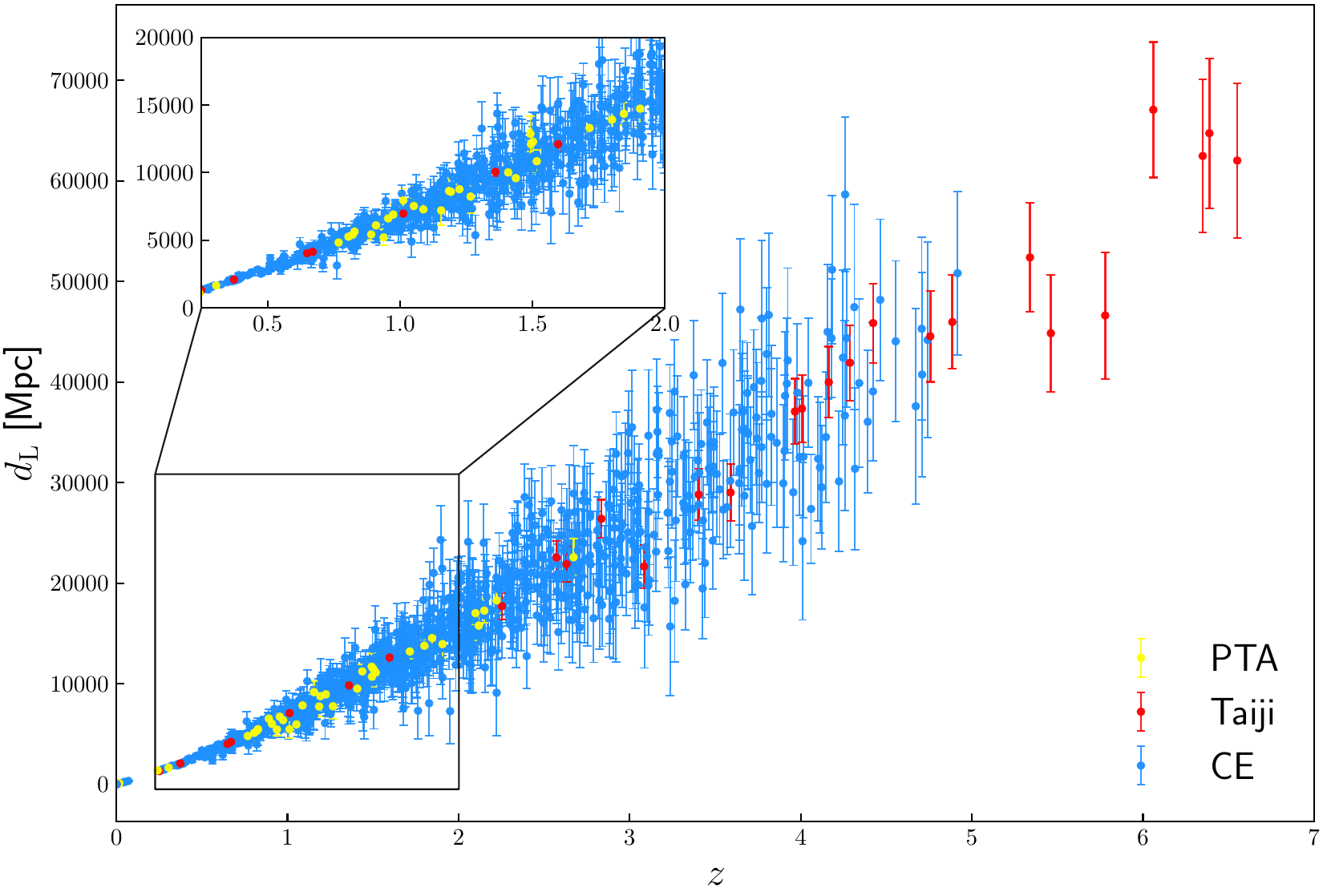}
\caption{The simulated measurements of luminosity distances from 10-year observation of SKA, 5-year observation of Taiji, and 10-year observation of CE. Left panel: relative errors of luminosity distances from lensing, peculiar velocities, redshift measurements (only for GW
standard sirens at $z > 2$ from Taiji by considering error propagation), and instrumental errors
of PTA, Taiji, and CE. Right panel: simulated GW bright siren data points observed by PTA, Taiji, and CE. Reproduced from Ref.~\cite{Jin:2023zhi} with permission.}
\label{fig:dlerror}
\end{figure*}

Since $h \propto 1/d_{\rm L}$, the precision of $d_{\rm L}$ estimated from GW detection is inversely proportional to the SNR of the GW signal, $\Delta d_{\rm L}/d_{\rm L} \propto 2/\mathrm{SNR}$ adopted in e.g., Refs.~\cite{Zhao:2010sz,Li:2013lza,Cai:2016sby,Zhang:2018byx,Cai:2017plb,Zhang:2019ple,Jin:2020hmc,Zhao:2020ole,Qiu:2021cww,Zhang:2023gye,Zhang:2019loq}, which is a rough estimate to account for the correlation between the GW parameters. A more robust analysis should be applied to obtain a reliable estimate~\cite{Cutler:1994ys,Finn:1992wt,Vallisneri:2007ev}. In general, the Fisher information matrix is widely adopted to estimate the GW parameter errors by considering the correlations between the GW parameters. Since this simulated $d_{\rm L}$ error is based on the instrument, it is generally referred to as instrumental error. Moreover, the errors of other GW parameters, e.g., sky location angles can also be inferred using Fisher information matrix. 

In addition, other factors, e.g., peculiar velocity~\cite{Kocsis:2005vv,Gordon:2007zw} and weak gravitational lensing~\cite{Holz:2005df,Takahashi:2003ix,Markovic:1993cr,Wang:1996as}, can also affect the measurements of GW amplitude during their propagation. Since all objects in the universe possess peculiar motion, the peculiar velocity of a GW source introduces a Doppler shift, modifying the observed GW frequency. This shift, along with its time derivative, plays a crucial role in determining the intrinsic parameters of the binary system, particularly its mass. These mass estimates in turn influence the inferred amplitude of the GW signal, thereby introducing uncertainties in the derived luminosity distance. Since the peculiar velocities of the GW sources are difficult to measure, it can usually be accounted for statistically~\cite{Kocsis:2005vv,Gordon:2007zw}. While for the weak gravitational lensing, it also should be accounted for statistically. Since the spatial number density of galaxies in the universe is approximately 0.01 Milky-way equivalent galaxies per $\mathrm{Mpc}^3$~\cite{Kopparapu:2007ib}. The probability that the emitted GW is affected by the gravitational potentials of galaxies during propagation is proportional to the distance of the GW source. Gravitational lensing will inevitably cause amplification or de-amplification of the GW amplitude, affecting the measurements of luminosity distances. There is a widely adopted fitting formula of the luminosity distance error caused by the weak gravitational lensing in Ref.~\cite{Hirata:2010ba}. Furthermore, several other environmental factors, dark matter~\cite{Karydas:2024fcn,Kavanagh:2024lgq}, the resolution of the GW waveform~\cite{Jan:2023raq}, and the non-stationary~\cite{Edy:2021par,Kumar:2022tto} and non-Gaussian~\cite{Steltner:2021qjy} characteristic of GW detector noise, may also affect measurements of $d_{\rm L}$. For the following analysis, the luminosity distance errors caused by instrument, peculiar velocity, and weak gravitational lensing are considered.

In the left panel of Fig.~\ref{fig:dlerror}, for demonstration, we show the relative errors of the luminosity distance from weak gravitational lensing, peculiar velocity, redshift measurements (only for $z>2$ events for Taiji), and the instrumental errors of 10-year observation of PTA from SKA, 5-year observation of Taiji, and 10-year observation of CE. We can clearly see that for the ground-based detector CE, the instrumental errors dominate the error of $d_{\rm L}$. Here note that a fixed redshift distribution of GW source is assumed and the EM counterpart selections are not taken into account. This assumption enables the inclusion of many bright sirens at high redshifts, which may not be feasible in realistic observational scenarios.
While for Taiji, we can see that the errors of $d_{\rm L}$ are mainly dominated by weak gravitational lensing and the instrumental errors are much smaller than those of CE due to the much higher SNR of Taiji. Similar to CE, for the 10-year observation of PTA from SKA, the errors of $d_{\rm L}$ are mainly dominated by instrumental error and a part of them are dominated by weak gravitational lensing. In the right panel of Fig.~\ref{fig:dlerror}, the simulated standard siren data points are shown. CE has the largest error of $d_{\rm L}$ at the same redshift due to the lowest SNR of GW events. Taiji and PTA show comparable errors of $d_{\rm L}$. 

\subsection{Forecasts for $H_0$}\label{sec4.3}
The simulated GW standard siren data can then be used to perform cosmological analysis.
In this subsection, we shall introduce the prospects for $H_0$ estimations using future GW standard sirens from ground-based, space-based, and PTA observatories.

\subsubsection{Forecasts from ground-based observatories}\label{sec4.3.1}

Chen et al.~\cite{Chen:2020zoq} and Jin et al.~\cite{Jin:2023tou} forecasted that bright sirens observed by A+, Voyager, and NEMO could constrain $H_0$ to the percent level, assuming several years of observation. However, the identification of EM counterparts remains challenging, with only about 0.1\% of BNS mergers in the 3G GW detector era are expected to yield detectable EM counterparts. 
{Given that the BNS merger rate is estimated to be $\mathcal{O}(10^5)$, 
{pioneering works by Zhao et al.~\cite{Zhao:2010sz} and Cai et al.~\cite{Cai:2016sby}} demonstrated that using 1000 simulated bright sirens from ET allows $H_0$ to be constrained to sub-percent precision under a 10-year observation. After that, Refs.~\cite{Cai:2017aea,Cai:2017cbj,Zhang:2018byx,Cai:2017plb,Zhang:2019ple,Zhang:2019loq,Zhao:2020ole,Qiu:2021cww,Zhang:2021kqn,Su:2024avk,Zhang:2023gye} conducted relevant studies based on this hypothesis. Jin et al.~\cite{Jin:2020hmc} used 1000 simulated bright sirens from a 10-year observation of CE, constraining $H_0$ to sub-percent precision.}
% Given that the BNS merger rate is estimated to be $\mathcal{O}(10^5)$, several preliminary forecasts~\cite{Zhao:2010sz,Cai:2016sby,Zhang:2018byx,Cai:2017plb,Zhang:2019ple,Jin:2020hmc,Zhao:2020ole,Qiu:2021cww,Zhang:2023gye,Zhang:2019loq} used 1000 simulated bright sirens from the individual GW observatories such as CE and ET to estimate $H_0$ without EM counterpart selections, achieving sub-percent level precision under a 10-year observation. 
Several other studies incorporating EM counterpart modeling have reported similar constraints~\cite{Belgacem:2019tbw,Chen:2020zoq,Hou:2022rvk,Han:2023exn,Han:2025fii}.
In contrast, the use of dark sirens for $H_0$ estimation is strongly limited by the completeness of galaxy catalogs. Although high-redshift GW events can be observed by 3G detectors, only low-redshift events associated with complete galaxy catalogs are suitable for cosmological inference. Using these dark sirens, Yu et al.~\cite{Yu:2020vyy}, Song et al.~\cite{Song:2022siz}, and Zhu et al.~\cite{Zhu:2023jti} found that the constraints on $H_0$ could achieve sub-percent level precision.
Furthermore, considering detector networks can substantially improve the constraints due to enhanced SNRs and source localization capabilities. 
Borhanian et al.~\cite{Borhanian:2020vyr}, Yu et al.~\cite{Yu:2020vyy}, Yu et al.~\cite{Yu:2021nvx}, Song et al.~\cite{Song:2022siz}, Yu et al.~\cite{Yu:2023ico}, and Han et al.~\cite{Han:2023exn,Han:2025fii} found that $H_0$ could potentially reach a sub-percent level precision using the 3G detector network. Recently, Han et al.~\cite{Han:2025fii} conducted a comprehensive cosmological parameter estimation by considering the multi-messenger observation of BNS mergers from 3G ground-based GW detectors and their gamma-ray burst and kilonovae from future surveys. They found that $H_0$ could be measured to a precision of 0.1\%, which shows the strong potential of GW standard sirens in resolving the Hubble tension.

\subsubsection{Forecasts from space-based observatories}\label{sec4.3.2}
For the simulations of standard sirens from MBHBs in the millihertz band, Tamanini et al.~\cite{Tamanini:2016zlh}, Zhao et al.~\cite{Zhao:2019gyk}, Wang et al.~\cite{Wang:2019tto}, and Zhu et al.~\cite{Zhu:2021aat} forecasted the constraint results of $H_0$ using single space-based GW observatories including Taiji, TianQin, and LISA. Given the uncertainties in the formation mechanisms of MBHs, these studies considered three representative population models:
\begin{itemize}
  \item \textbf{Q3d}: a heavy-seed scenario assuming MBHs originate from the direct collapse of protogalactic disks, with seed masses around $10^5~M_{\odot}$incorporating a time delay between MBH mergers and their host galaxy mergers~\cite{Klein:2015hvg}.
  \item \textbf{pop III}: a light-seed scenario where MBHs form from the remnants of Population III stars, with initial seed masses around $100~M_{\odot}$~\cite{Madau:2001sc,Volonteri:2002vz}.
  \item \textbf{Q3nod}: identical to Q3d but without accounting for time delays between black hole and galaxy mergers~\cite{Ferrarese:2000se}.
\end{itemize}
These studies found that $H_0$ can achieve percent-level precision and might reach sub-percent precision under optimistic assumptions in a several-year observation by considering bright sirens. Moreover, Ruan et al.~\cite{Ruan:2020smc} proposed that Taiji and LISA observatories could form a space-based GW detector network owing to their overlapping observational windows, thereby improving the identification and localization of GW sources. They found that the Taiji-LISA network with a configuration angle of $40^\circ$ could rapidly and accurately localize GW sources. Cai et al. \cite{Cai:2023ywp} provided a comprehensive summary of the capabilities of space-based gravitational-wave detector networks, emphasizing their potential for precise sky localization, cosmological studies, tests of parity violation in the stochastic background, efficient subtraction of galactic foregrounds, and enhanced observations of stellar-mass binary black holes. Based on this configuration, Wang et al.~\cite{Wang:2020dkc}, Wang et al.~\cite{Wang:2021srv}, Zhu et al.~\cite{Zhu:2021aat}, and Jin et al.~\cite{Jin:2023sfc} used the mock standard siren data from the Taiji-LISA network, the TianQin-LISA network, and the Taiji-TianQin-LISA network to constrain $H_0$. These studies demonstrated that $H_0$ could be constrained to the 1\% level using bright or dark sirens, highlighting the potential of space-based GW detector networks to resolve the Hubble tension. In general, the pop III and Q3nod models yield similar constraints, while the Q3d model results in significantly weaker constraints. This discrepancy arises from differences in the merger rates, mass distributions, and redshift distributions among the models, with the pop III and Q3nod scenarios predicting a greater number of detectable MBHB mergers compared to Q3d.

\subsubsection{Forecasts from PTA observatories}\label{sec4.3.3}
Wang et al.~\cite{Wang:2022oou} found that constraints on $H_0$ could be constrained to percent-level precision by considering various pulsar configurations and root mean square of timing residual using bright or dark sirens assuming a 10-year observation. In an optimistic scenario, the constraint precision of $H_0$ could be reduced to approximately 1\%, which meets the criteria of precision cosmology. In the analysis considering dark sirens, the matter density $\Omega_{\rm m}$ is treated as a fixed parameter, whereas it is allowed to vary as a free parameter in the bright siren analysis. 
%The constraint results using dark sirens are shown in Fig.~\ref{fig_pta_dark} as a representative of nanohertz standard siren constraints. We see that considering more pulsars or reducing the root mean square (rms) of timing residual can provide smaller measurement errors for $H_0$. 
Here, the 2MASS catalog ($z<0.05$) is considered complete and the red noise is not considered.
The constraint precisions of $H_0$ are in the ranges of $1.31\%$--$5.00\%$. 
In addition, Xiao et al.~\cite{Xiao:2024nmi} considered a quasar-based SMBHB population model to constrain $H_0$ using dark sirens and compared the results with those based on the galactic major merger model. They found that modeling the SMBHB population has a great impact on the analysis of SMBHB dark sirens. Based on the quasar-based model, the precision of $H_0$ could be close to 1\% with a 10-year observation of 100 pulsars and a white noise level of 20 ns, or a 5-year observation of 200 pulsars.

\subsection{Constraints on other cosmological parameters} \label{sec4.4}
\subsubsection{$\Lambda$CDM model}\label{sec4.4.1}
In fact, the high-redshift GW standard sirens could not only measure $H_0$, but also provide the measurements of other cosmological parameters. Usually, due to the limitations in the completeness of the galaxy catalogs, the redshifts of dark sirens are low. Therefore, in some works, cosmological parameters other than $H_0$ are fixed in e.g., Refs.~\cite{Borhanian:2020vyr,Yu:2020vyy,Song:2022siz} since the constraints on other cosmological parameters are limited. 

As shown in Refs.~\cite{Jin:2020hmc,Yu:2021nvx,Belgacem:2019tbw,Safarzadeh:2019pis,Zhang:2019ple,Zhang:2019loq,Han:2023exn,Jin:2023zhi,Wu:2022dgy,Jin:2021pcv,Ye:2021klk,You:2020wju,Zhang:2018byx,Liao:2017ioi,Cai:2016sby,Zhao:2010sz,Wei:2018cov,Liao:2019hfl,Zheng:2020tau,Wang:2020dbt,Cao:2021jpx,He:2021rzc,Wang:2022rvf,Pan:2023omz}, BNS mergers with EM counterparts, as detected by 3G GW detectors, can be used to estimate the values of the cosmological density parameters $\Omega_{\rm m}$ and $\Omega_{k}$, but with large errors. Considering the $\Lambda$CDM model, constraints on $\Omega_{\rm m}$ are expected to reach a few percent level in the optimistic scenario. While these results may not be as competitive as those obtained from EM observations, they nonetheless offer an independent avenue for measurement, contributing valuable data to the field.

The constraint precision of $\Omega_{\rm m}$ using the single space-based GW observatory can achieve more than 10\% level by considering several-year observation of bright or dark sirens~\cite{Tamanini:2016uin,Wang:2019tto,LISACosmologyWorkingGroup:2019mwx,Zhao:2019gyk} and further improve to a few percent level~\cite{Jin:2023sfc,Jin:2023zhi,Jin:2021pcv,Wang:2021srv} when considering the GW detector network.

The bright sirens of SMBHBs with potential EM counterparts can be used to constrain the matter density $\Omega_{\rm m}$ as well. Wang et al.~\cite{Wang:2022oou} found that the constraint precision of $\Omega_{\rm m}$ can achieve more than a 10\% level.

\subsubsection{$w$CDM model}\label{sec4.4.2}

The $w$CDM model is the simplest dynamical dark energy model and the EoS parameter of dark energy $w$ is a constant. 

The study of dark energy using standard sirens from the ground-based GW detectors has been detailedly investigated in Refs.~\cite{Han:2023exn,Jin:2023tou,Jin:2023zhi,Jin:2022qnj,Wu:2022dgy,Jin:2021pcv,Ye:2021klk,Zhao:2020ole,Jin:2020hmc,Zhang:2019loq,Zhang:2018byx,Ding:2018zrk}. The bright sirens from the 2.5G GW detectors could give the constraint on $w$ being above 50\%~\cite{Jin:2023tou}. A better constraint could indeed be obtained if considering bright sirens from the 3G GW detectors, but the constraints on $w$ are also not enough to probe the nature of dark energy, with the precision level from a few percent to dozens of percent. However, the dark siren method, which relies on the tidal deformation of NS to obtain the redshift information from solely GW observations is widely discussed in its cosmological application~\cite{Jin:2022qnj,2022arXiv221213183D,DelPozzo:2015bna,Wang:2020xwn,Chatterjee:2021xrm,Ghosh:2022muc,Li:2023gtu}. Since such method relies on the precise measurement of the tidal phase which can only be achieved in the 3G detector era. Moreover, the detected number of BNS mergers could reach the level of hundreds of thousands to millions per year in the 3G detector era. 
The constraint precision of $w$ using such dark sirens can potentially reach $1\%$ level assuming a 3-year observation, which is expected to usher in the era of dark-energy precision cosmology.
Nevertheless, the effectiveness of such dark sirens highly depends on assumptions about the EoS of NS, which is modeled using the SLy model. Additionally, extracting redshift information from tidal effects requires precise Bayesian parameter estimation for each individual BNS event, particularly of the phase shift induced by tidal deformation of NS. This poses both observational and computational challenges that must be addressed to fully realize the potential of this method in the 3G era.

\begin{figure*}[!htbp]
\centering
\includegraphics[width=\columnwidth]{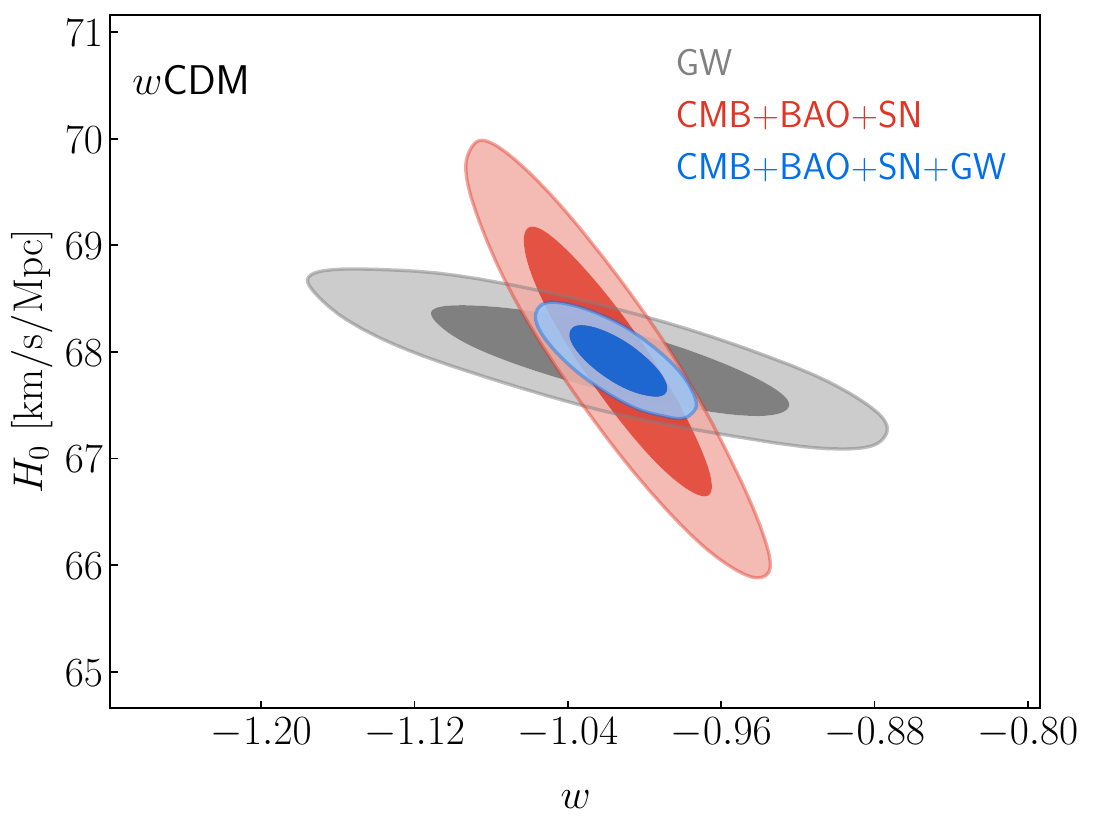}
\includegraphics[width=\columnwidth]{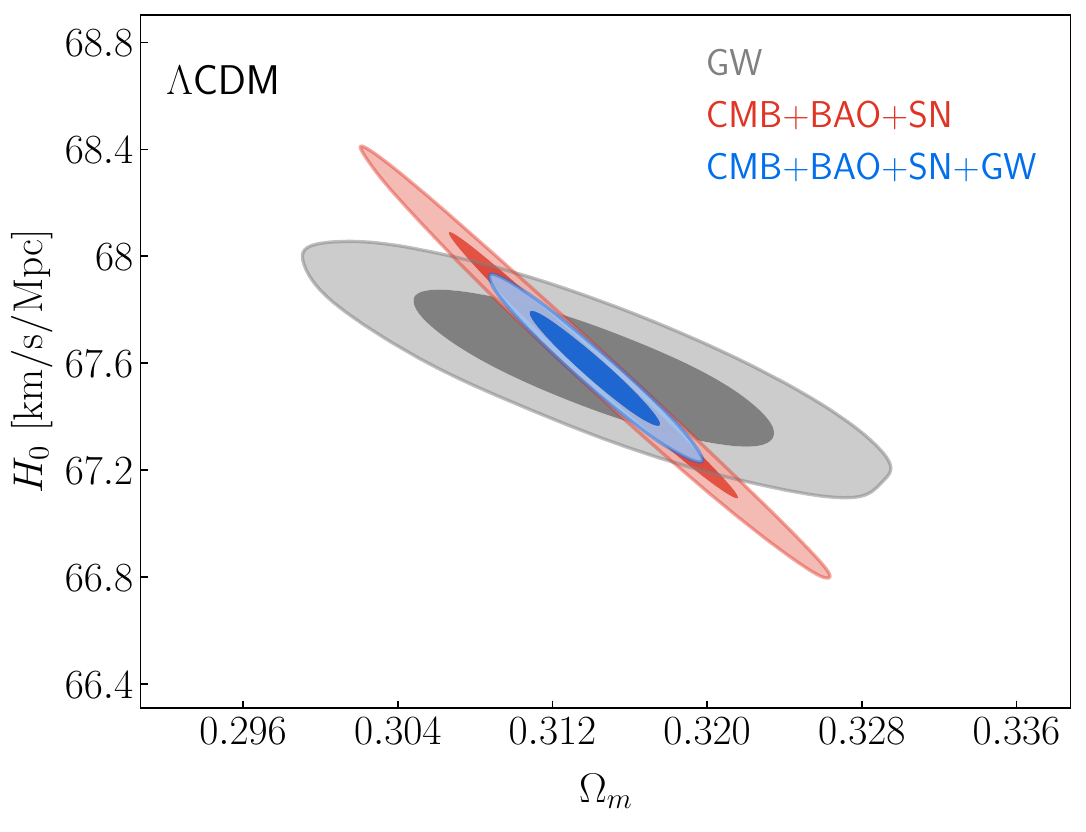}
\caption{Two-dimensional marginalized posterior contours (68.3\% and 95.4\% confidence level) in the $\Omega_{\rm m}$--$H_0$ (the $\Lambda$CDM model) and $w$--$H_0$ (the $w$CDM model) planes using 10-year GW bright sirens from CE, CMB+BAO+SN (BAO is baryon acoustic oscillation and SN is supernova), and their combination. Left panel: constraints on $w$ and $H_0$ in the $w$CDM model. Right panel: constraints on $\Omega_{\rm m}$ and $H_0$ in the $\Lambda$CDM model. Reproduced from Ref.~\cite{Jin:2020hmc} with permission.}
\label{fig:break_degeneracy}
\end{figure*}

Several works used the mock bright siren data from the space-based GW detectors to constrain dark energy~\cite{Jin:2023sfc,Jin:2023zhi,Chen:2022bqi,Wang:2021srv,Zhao:2019gyk,Wang:2019tto}. Also, the single 5-year GW mock bright siren data could not give tight constraint on $w$ with the precisions being a few dozens percent level using the single GW observatory or even detector network. The GW bright sirens alone are basically not able to constrain dark energy with $w(z)$ being constant. 

{Using the mock bright siren data of the 10-year observation of SKA-era PTAs, Yan et al.~\cite{Yan:2019sbx} first proposed this approach by considering different pulsar configurations and root mean square of timing residual. In their work, they demonstrated that the constraint on dark-energy EoS parameter $w$ could reach a few percent level by fixing other cosmological parameters. Following this methodology, Wang et al.~\cite{Wang:2022oou} further explored the constraints and found that the precision decreases by more than 50\% when considering different configurations and allowing other cosmological parameters to vary freely, based on the same 10-year observation period. Therefore, while Yan et al.~\cite{Yan:2019sbx} introduced the method of using nanohertz standard sirens in cosmological constraints, current evidence suggests these sirens alone are insufficient to precisely constrain the EoS parameter of dark energy or other cosmological parameters, and they primarily offer tight constraints on $H_0$.}

% Using the mock bright siren data of the 10-year observation of SKA-era PTAs by considering different pulsar configurations and root mean square of timing residual, 
% the constraint on dark-energy EoS parameter $w$ could reach a few percent level~\cite{Yan:2019sbx} by fixing other cosmological parameters.
% Wang et al.~\cite{Wang:2022oou} found that the constraint precisions are more than 50\% considering different configurations, setting other cosmological parameters free based on the 10-year observation. Therefore, nanohertz standard sirens alone are insufficient to precisely constrain the EoS parameter of dark energy or other cosmological parameters, they primarily offer tight constraints on $H_0$.

\subsubsection{$w_0w_a$CDM model}\label{sec4.4.3}
It is still challenging to constrain EoS parameter of dark energy by solely using the bright sirens from future ground-based GW detector, see e.g., Refs.~\cite{Jin:2020hmc,Zhang:2019loq,Han:2023exn,Jin:2023tou,Zhang:2023gye,Wu:2022dgy,Zheng:2022gfi,Jin:2021pcv}. However, the dark sirens with NS tidal effect also perform well in constraining the $w_0w_a$CDM model. 
The constraints on $w_0$ and $w_a$ could potentially reach $\sigma(w_0)=0.02$ and $\sigma(w_a)=0.13$~\cite{Jin:2022qnj} by setting other cosmological parameters free and can potentially improve to $\sigma(w_0)=0.0006$ and $\sigma(w_a)=0.004$~\cite{Wang:2020xwn} by fixing other cosmological parameters.

\begin{figure*}[!htbp]
\centering
\includegraphics[width=\columnwidth]{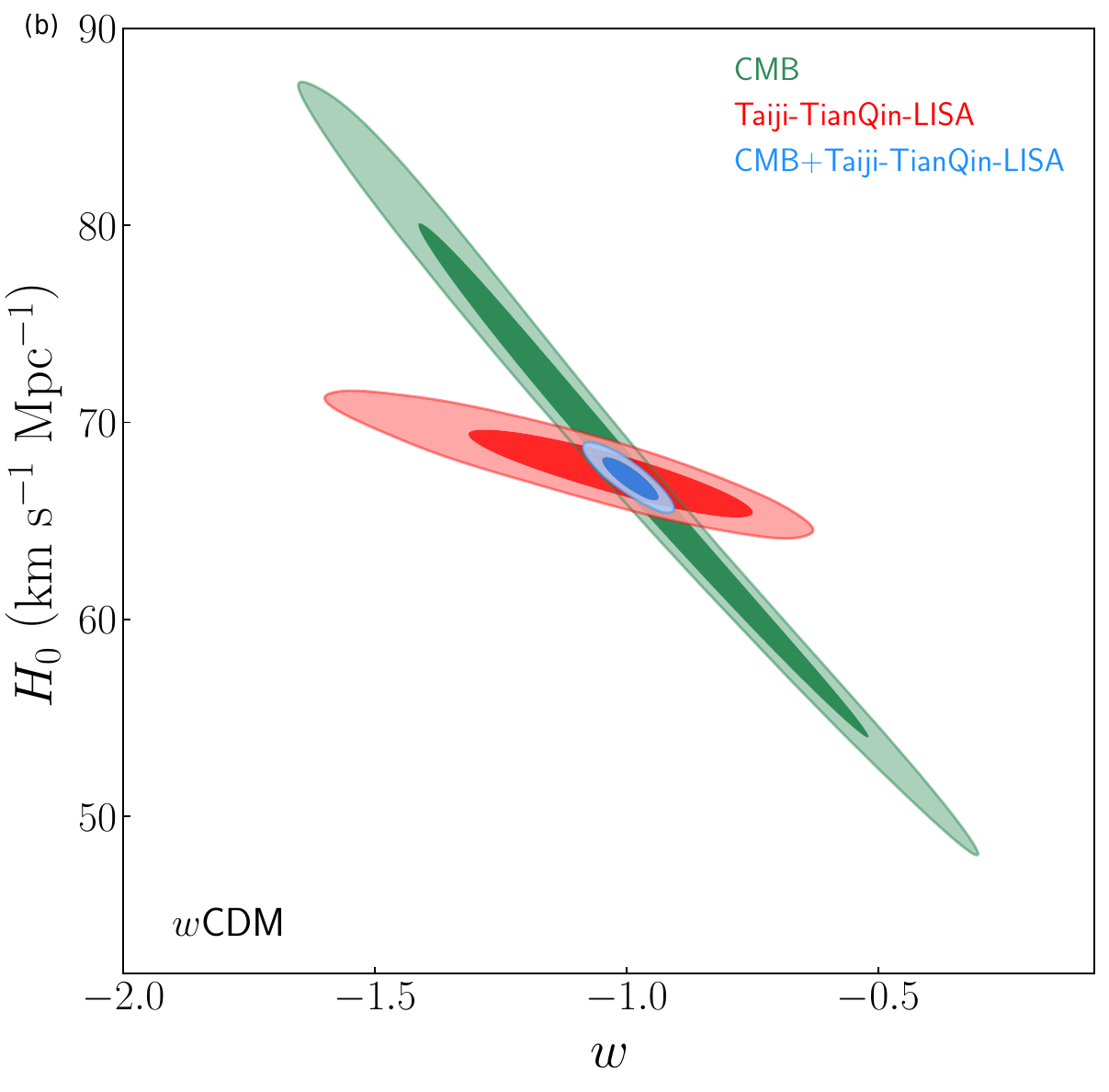}
\includegraphics[width=\columnwidth]{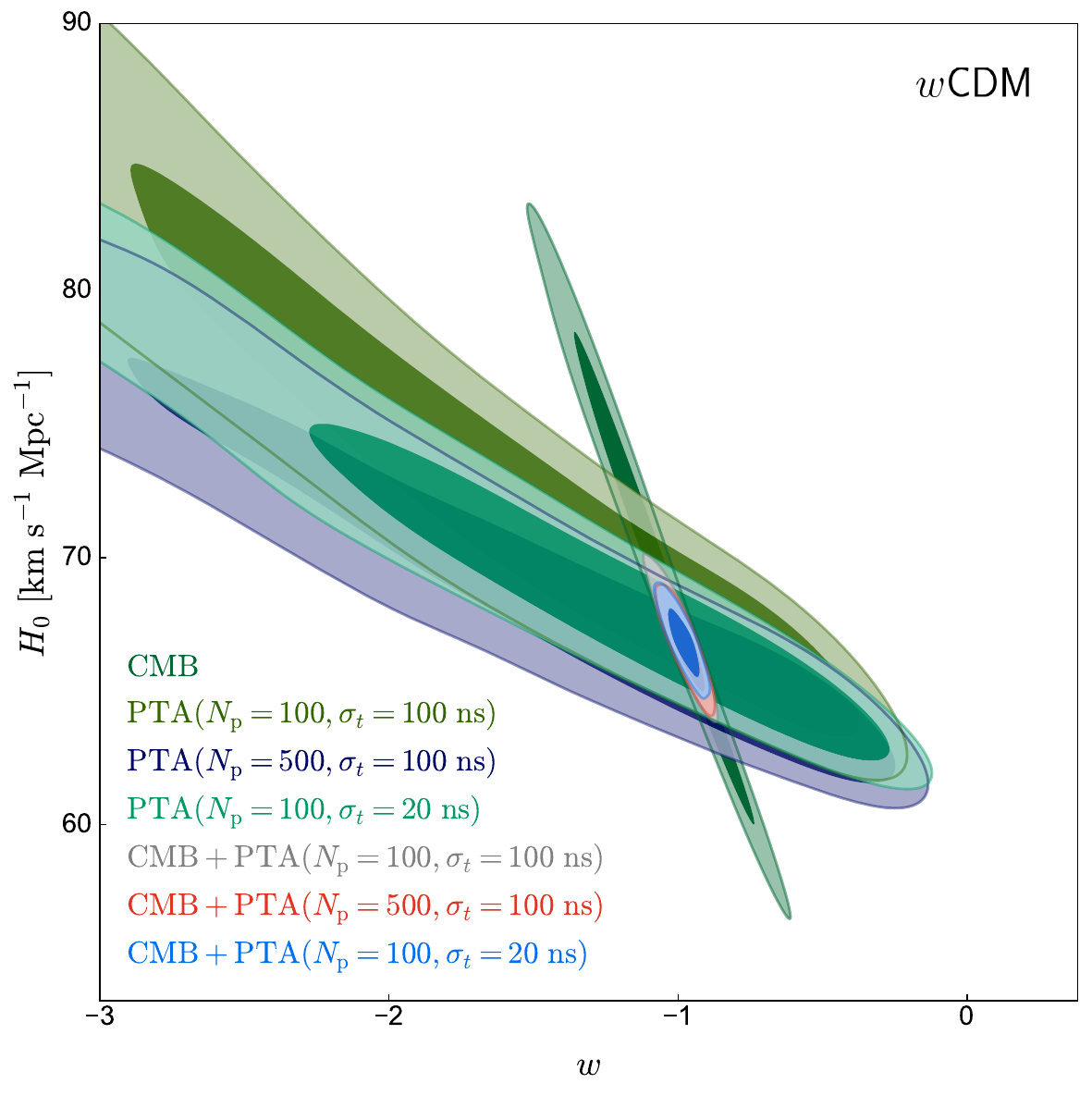}
\caption{Two-dimensional marginalized posterior contours in the $w$--$H_0$ plane for the $w$CDM model. Left panel: constraints from CMB, the simulated 5-year millihertz bright siren data from the Taiji-TianQin-LISA network based on the pop III model. Right panel: constraints from CMB, the simulated nanohertz GW standard siren data from various-configuration PTAs, and their combination. Reproduced from Refs.~\cite{Jin:2023sfc} and \cite{Wang:2022oou} with permission.}
\label{fig:WCDM_CMB}
\end{figure*}

\subsubsection{Other dark energy models}\label{sec4.4.4}
%Although the $\Lambda$CDM model is now commonly viewed as the standard model of cosmology, it faces the fine-tuning \cite{Weinberg:1988cp} and coincidence problems \cite{Zlatev:1998tr,Huey:2004qv,Velten:2014nra} on the theoretical level. Some extended models are proposed to theoretically solve these problems, for example, the holographic dark energy (HDE) model and the interaction dark energy (IDE) models. The HDE model is constructed by combining the holographic principle of quantum gravity with the effective quantum field theory. While the IDE models assume that there have direct and non-gravitational interaction between dark energy and dark matter. 
The study of the holographic dark energy (HDE) and interaction dark energy (IDE) models using standard sirens from the 3G GW detectors has been investigated in Refs.~\cite{Cai:2017plb,Zhang:2019ple,Bachega:2019fki,Zhang:2019loq,Burns:2019byj,Li:2019ajo,Bonilla:2021dql,Jin:2022tdf,Hou:2022rvk,Han:2023exn,Li:2023gtu}. For example, Zhang et al.~\cite{Zhang:2019ple} used 1000 mock bright sirens from ET to constrain the HDE model. GW bright sirens can also provide tight constraints on $H_0$, but are poor at measuring other cosmological parameters, consistent with other cosmological models. 

Li et al.~\cite{Li:2023gtu} used the mock dark siren data from the 3G GW detectors to investigate the IDE models. Since the interaction between dark energy and dark matter is unknown, only phenomenological forms of the interaction are assumed which are related to the dimensionless coupling parameter $\beta$, $H(z)$ or $H_0$, and energy densities of dark energy or cold dark matter. Bright sirens give poor constraints on the IDE models. In addition, dark sirens with NS tidal effect can provide tight constraints on $\Omega_{\rm m}$ and $H_0$, but it is challenging to measure $\beta$ by solely using such dark sirens with constraint errors being 0.013--0.056.

{Cai et al.~\cite{Cai:2017yww} predicted the ability of the high-redshift MBHBs observed by LISA as standard sirens to reconstruct the interaction of dark energy and dark matter through Gaussian processes, assuming that the EM counterpart can be observed. Wang et al.~\cite{Wang:2022llq} forecasted the ability of constraining cosmological parameters in the IDE model using bright sirens observed by PTAs in the SKA era.}

In conclusion, it is quite difficult for future bright sirens to probe the nature of dark energy, with precisions being on the order of dozens of percent. However, redshift measurements based on the tidal deformation of NS can potentially precisely measure the EoS parameter of dark energy $w$ with a precision of better than 1\%. For the other dark energy models, GW bright sirens can also tightly measure $H_0$, but are poor at measuring other cosmological parameters. Dark sirens with tidal effect are anticipated to provide precise measurement on the dark energy EoS parameter $w$, but provides uncompetitive constraint on the coupling parameter $\beta$ in the IDE models. 

\section{Breaking cosmological parameter degeneracies in EM observations}\label{sec5}
As discussed above, future GW standard sirens can serve as a powerful cosmological probe, due to their ability to provide absolute and self-calibrating luminosity distance measurements directly from the GW waveform without relying on the traditional distance ladder. 
% As discussed above, future GW standard sirens can be served as a powerful cosmological probe that can independently measure $H_0$ with high precision, due to their ability to provide absolute and self-calibrating luminosity distance measurements directly from the GW waveform without relying on the traditional distance ladder. 
Since the luminosity distance is inversely proportional to $H_0$, GW standard sirens are particularly sensitive to $H_0$ which may address the Hubble tension.
% Since the GW waveform amplitude scale is inversely proportional to luminosity distance, GW is particularly sensitive to $H_0$ which may address the Hubble tension. 

However, it remains challenging to tightly constrain other cosmological parameters by solely using GW standard sirens. This is because the cosmological parameter degeneracies between these parameters cannot be broken using GW data alone.

Zhang~\cite{Zhang:2019ylr} made a significant advancement by systematically investigating the ability of GW standard sirens to break the cosmological parameter degeneracies generated by the EM observations and how the combination of them can improve the cosmological parameter estimations. In particular, Zhang~\cite{Zhang:2021} proposed the synergy of GW and other cosmological probes to forge precise cosmological probes for exploring the late universe. In this section, we shall introduce the role of standard sirens in this context and highlight how GW observations can complement EM observations in joint cosmological analyses. In addition, the synergy between GW and other promising late-universe cosmological probes is also discussed.

\begin{figure*}[!htbp]
\centering
\includegraphics[width=\columnwidth]{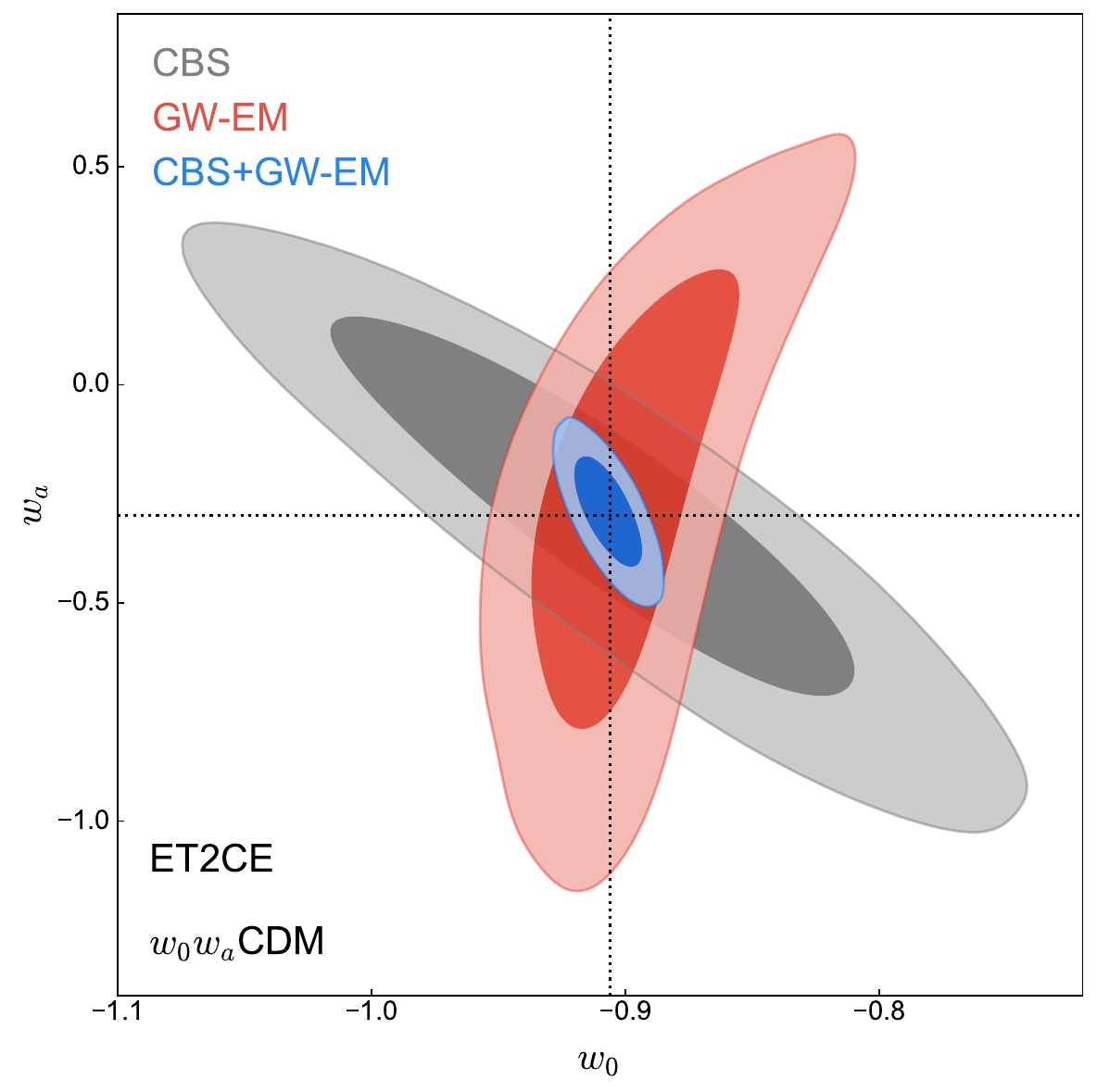}
\includegraphics[width=\columnwidth]{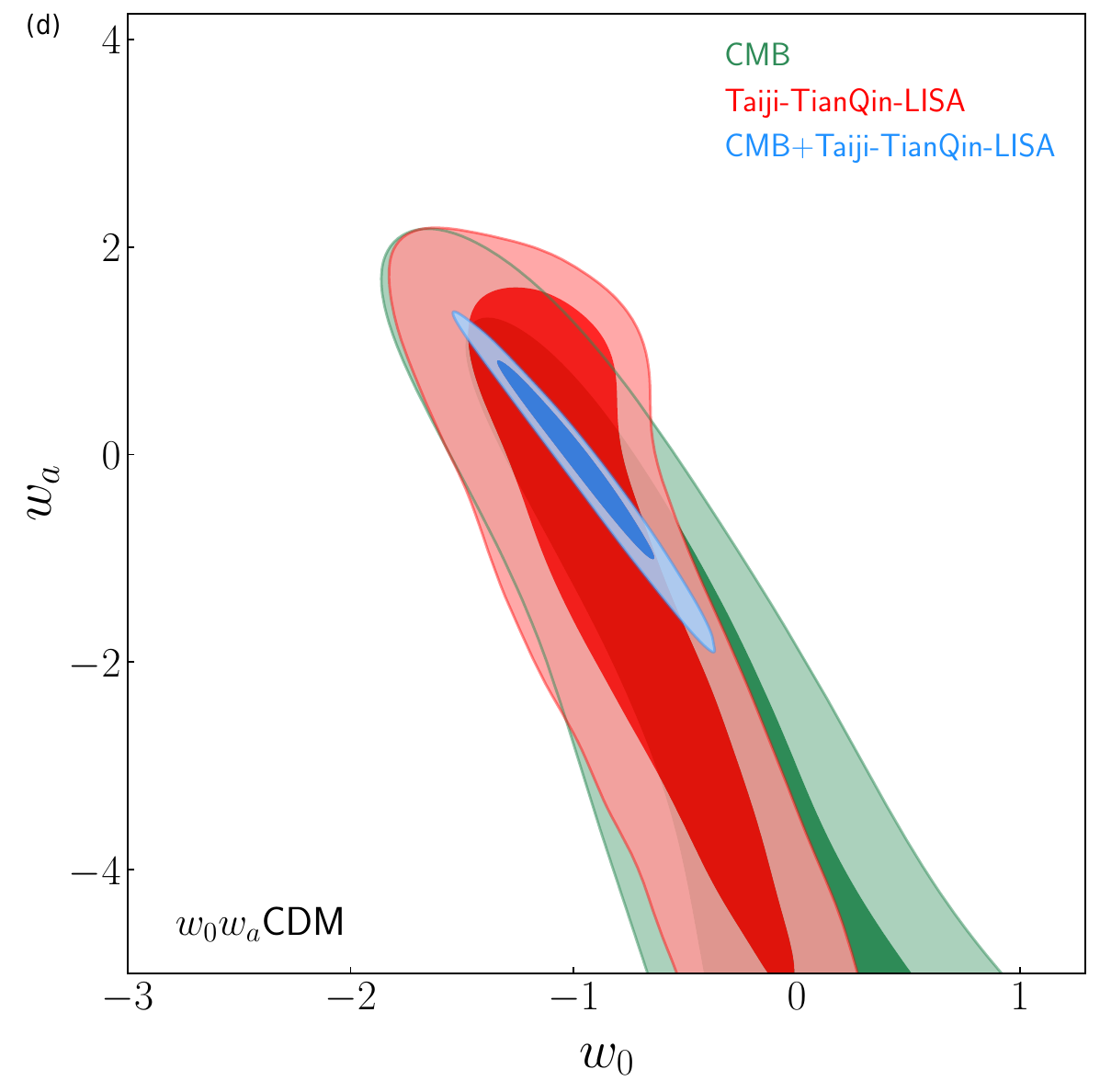}
\caption{Two-dimensional marginalized contours in the $w_0$--$w_a$ plane for the $w_0w_a$CDM model. Left panel: constraints from the CMB+BAO+SN (denoted as CBS in this figure) data, the simulated 10-year bright siren data from the ET2CE network (multi-messenger observation of GW-GRB and GW-kilonova from the GW detection network consisting of one ET, one CE in the US, and one CE in Australia), and their combination. Right panel: constraints from CMB, the simulated 5-year bright siren data from the Taiji-TianQin-LISA network based on the pop III model, and their combination. Reproduced from Refs.~\cite{Han:2025fii} and \cite{Jin:2023sfc} with permission.}
\label{fig:cpl_break}
\end{figure*}

\subsection{Synergy with current mainstream cosmological probes}\label{sec5.1}

Precisely measuring the dark energy EoS $w$ is a key step toward understanding the fundamental nature of dark energy. 
Currently, the most precise measurement of dark energy EoS parameter $w$ is given by the CMB+BAO+SN data, with a precision of about 3\%~\cite{DES:2025bxy}. Such precision is insufficient for uncovering the fundamental physics behind dark energy. 
As mentioned above, only GW dark sirens (tidal effect) can potentially precisely measure the EoS parameter of dark energy, and it is quite challenging to measure $w$ with bright sirens. However, bright sirens can still help constrain $w$ through breaking degeneracies in cosmological parameters when combined with other cosmological probes. To show the ability of bright sirens to break the cosmological parameter degeneracies, in Fig.~\ref{fig:break_degeneracy}, we show the cases in the $w$CDM and $\Lambda$CDM model by considering the 10-year observation of CE and CMB+BAO+SN. From the left panel of Fig.~\ref{fig:break_degeneracy}, we can clearly see that the cosmological parameter degeneracy orientations of GW and CMB+BAO+SN are quite different and thus the combination of them could significantly improve the constraints on both $w$ and $H_0$. In this case, with the addition of GW, the constraints on $w$ and $H_0$ could be improved by 46.9\% and 73.5\%, respectively. Even in the standard $\Lambda$CDM model, the cosmological parameter degeneracy orientations of GW and CMB+BAO+SN are still quite different, as shown in the right panel of Fig.~\ref{fig:break_degeneracy}. Constraints on $\Omega_{\rm m}$ and $H_0$ could be improved by 56\% and 57.6\%, respectively, after combining the GW data with the CMB+BAO+SN data. 

\begin{figure*}[!htbp]
\centering
\includegraphics[width=\columnwidth]{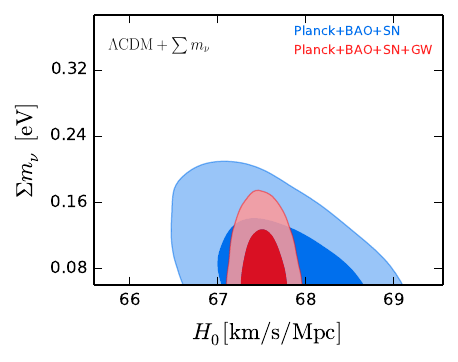}
\includegraphics[width=\columnwidth]{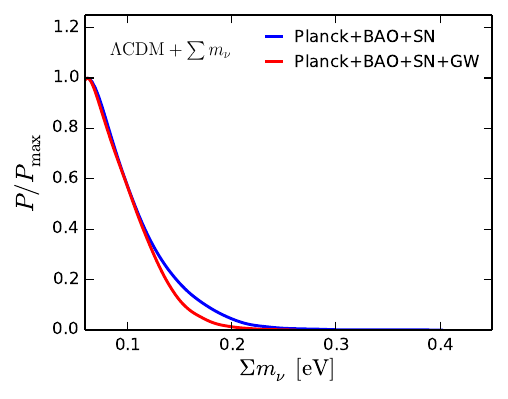}
\caption{Constraints in the $\Lambda$CDM+$\sum m_{\nu}$ model. Left panel: two-dimensional marginalized posterior contours (68\% and 95\% confidence level) in the $H_0$--$\sum m_{\nu}$ plane using CMB+BAO+SN and CMB+BAO+SN+GW. Here the simulated GW data are based on the 10-year observation of ET. Right panel: One-dimensional marginalized posterior distributions of $\sum m_{\nu}$ using CMB+BAO+SN and CMB+BAO+SN+GW. Reproduced from Ref.~\cite{Wang:2018lun} with permission.}
\label{fig:break_mnu}
\end{figure*}

In Fig.~\ref{fig:WCDM_CMB}, we show the cases considering CMB, the space-based, and SKA-era PTA observatories in the $w$CDM model. 
As is known, CMB data can only precisely measure the standard $\Lambda$CDM model. If other new physics is considered by extending the $\Lambda$CDM model, the newly introduced parameters will severely degenerate with other cosmological parameters when the CMB data alone are used to constrain the extended model. 
As clearly shown, either GW or CMB can provide rather poor constraints on $w$. Nevertheless, the cosmological parameter degeneracy orientations of GW and CMB are roughly orthogonal, which means that GW can thoroughly break the cosmological parameter degeneracies generated by CMB. In the left panel, the bright sirens from the 5-year observation of the Taiji-TianQin-LISA detection network are used to perform the cosmological analysis. Concretely, the constraints on $\Omega_{\rm m}$, $H_0$, and $w$ can be improved by 89.1\%, 90.3\%, and 85.6\%, respectively, after adding the GW data to the CMB data. Moreover, the constraints precision of $w$ using the data combination is expected to reach 3.6\%, which is close to the latest constraint result by the CMB+SN data~\cite{DES:2025bxy}. In the right panel, the 10-year observations of SKA-era PTAs are considered in different configurations to perform the joint cosmological analysis with CMB. No matter which cases are considered, the ability of GW to break the cosmological parameter degeneracies is strong. Moreover, the combination results of CMB and PTAs with different configurations show a slight difference, which means that the conservative case also shows great potential to provide precise measurements in the $w$CDM model. Concretely, by considering the conservative (100 pulsars and root mean square of the timing residual $\sigma_t=100$ ns) and optimistic cases (100 pulsars and $\sigma_t=20$ ns) into CMB, the constraints on $w$ and $H_0$ could be improved by 75.5\% (81.5\%) and 78.3\% (85\%), respectively. These results highlight the robustness of nanohertz GW standard sirens in breaking cosmological degeneracies, even under conservative assumptions.
It can also be seen that the improvements from the Taiji-TianQin-LISA network or SKA-era PTA observations are more than those from CE. This is because the EM dataset used in the latter case (CMB+BAO+SN) already partially breaks parameter degeneracies by the BAO+SN data, thus reducing the marginal benefit of adding GW data.

If the dynamical dark energy with parameterization is considered, GW can still break the cosmological parameter degeneracies. As clearly seen from the left panel of Fig.~\ref{fig:cpl_break}, the cosmological parameter degeneracy orientation of CMB+BAO+SN (CBS) and GW from the one ET and 2CE (one CE in the US and another one in Australia) network is orthogonal. 
Their combination gives significant improvements on $w_0$ and $w_a$. Specifically, the constraints on $w_0$ and $w_a$ could be improved by 86.8\% and 69.0\%, respectively, when adding the GW data to the CBS data. Note that as mentioned above, the parameter degeneracy has been broken by the BAO and SN data to some extent since there are strong cosmological parameter degeneracies when the CMB alone data are used to constrain the $w_0w_a$CDM model. The results also show the strong ability of GW bright sirens to break the cosmological parameter degeneracies in the $w_0w_a$CDM model. For comparison, in the right panel of Fig.~\ref{fig:cpl_break}, we show the constraints on $w_0$ and $w_a$ by using CMB, GW bright sirens from the Taiji-TianQin-LISA network, and their combination. We could see that either CMB or GW provides rather loose constraints, but their combination could give much better constraints on $w_0$ and $w_a$. Concretely, in this case, after adding the GW data, $w_0$ could be improved by 62.3\%. Since CMB cannot give a constraint on $w_a$, the constraint on $w_a$ could be improved by 62.9\% after adding the CMB data to GW.

\begin{figure*}[!htbp]
\centering
\includegraphics[width=\columnwidth]{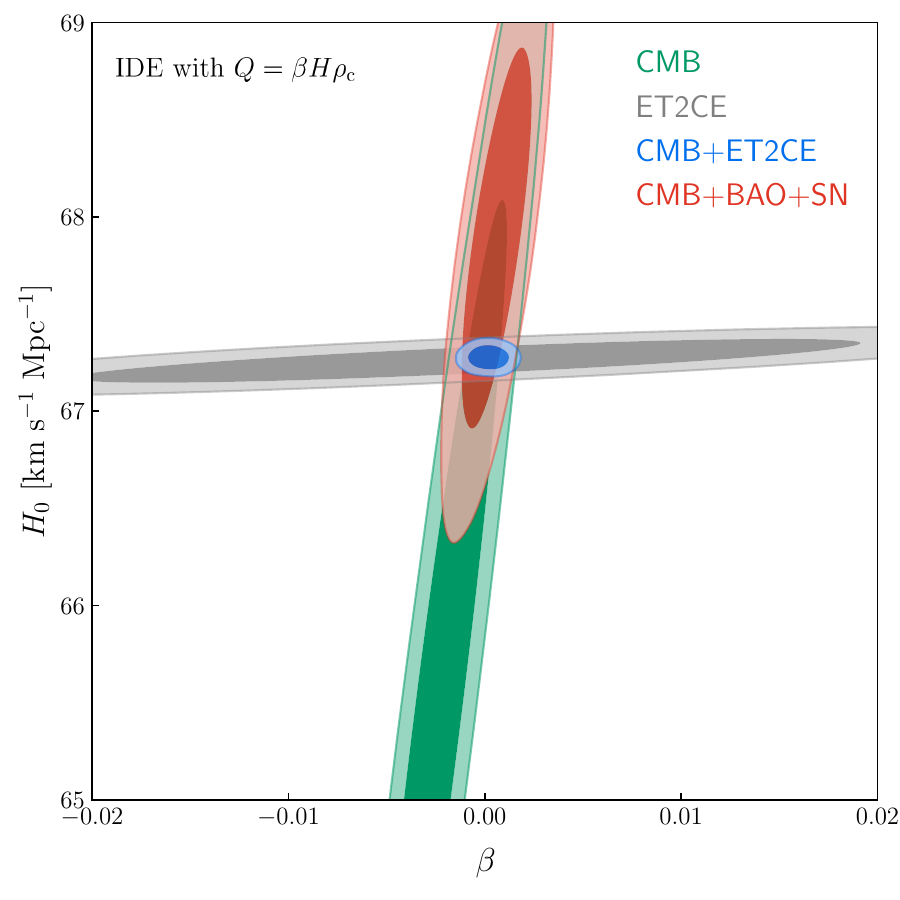}
\includegraphics[width=\columnwidth]{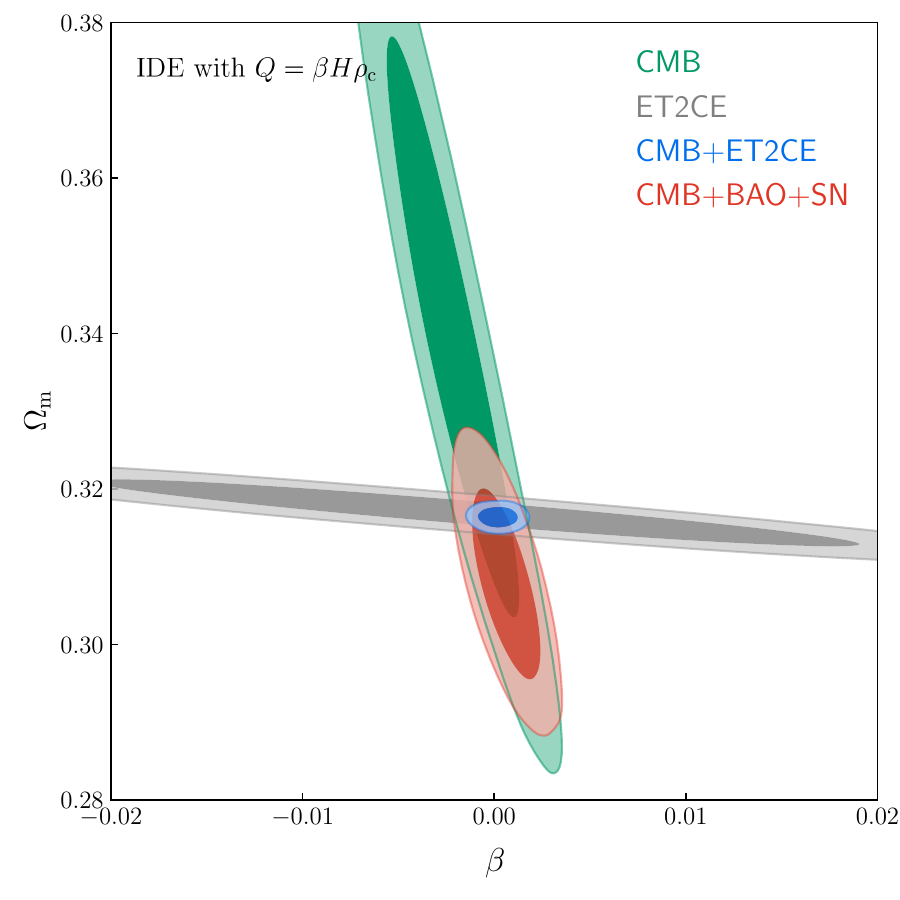}
\caption{Constraints in the IDE model using CMB, the mock ET2CE data, CMB+ET2CE, and CMB+BAO+SN. Left panel: two-dimensional marginalized posterior contours (68.3\% and 95.4\% confidence level) in the $\beta$--$H_0$ plane from CMB, the simulated dark siren data (considering tidal effect of BNS mergers) from the ET2CE network, CMB+ET2CE, and CMB+BAO+SN in the IDE model. Right panel: two-dimensional marginalized posterior contours (68.3\% and 95.4\% confidence level) in the $\beta$--$\Omega_{\rm m}$ plane from CMB, ET2CE, CMB+ET2CE, and CMB+BAO+SN. Reproduced from Ref.~\cite{Li:2023gtu} with permission.}
\label{fig:break_beta}
\end{figure*}

For the study of cosmology, neutrinos play an important role in helping shape the large-scale structure and the expansion history of the universe. The phenomenon of neutrino oscillation indicates that neutrinos have masses~\cite{Lesgourgues:2006nd} and that there are mass splittings between different neutrino species. While oscillation experiments can determine the squared mass differences and mixing angles, they are insensitive to the absolute mass scale of neutrinos and the mass hierarchy. The solar and reactor experiments give $\Delta m_{21}^2 \simeq 7.5\times 10^{-5}~\mathrm{eV}^2$ and the atmospheric and accelerator beam experiments give $\Delta m_{31}^2 \simeq 2.5\times 10^{-3}~\mathrm{eV}^2$~\cite{ParticleDataGroup:2014cgo,Xing:2020ijf}, which indicates that there are two possible mass hierarchies of the neutrino mass spectrum: the normal hierarchy with $m_1<m_2\ll m_3$ and the inverted hierarchy with $m_3\ll m_1 < m_2$.
Although the particle physics experiments can hardly give a constraint on the total neutrino mass $\sum m_{\nu}$, the cosmological observations can provide a constraint on $\sum m_{\nu}$. Using the current cosmological observations, an upper limit of $\sum m_{\nu}$ could be obtained. 

For weighing total neutrino mass $\sum m_{\nu}$, GW bright sirens can also play an important role. Although GW cannot constrain $\sum m_{\nu}$, it can break the cosmological parameter degeneracy generated by the EM observation. Since there are three possible mass hierarchies of the neutrino mass spectrum, only the normal hierarchy is shown in the following analysis.
In Fig.~\ref{fig:break_mnu}, we show the constraints on $H_0$ and $\sum m_{\nu}$ using the CMB+BAO+SN and the combination of CMB+BAO+SN and simulated 10-year bright sirens from ET. From the left panel of Fig.~\ref{fig:break_mnu}, we can see that with the addition of GW, the upper limit of $\sum m_{\nu}$ is reduced and a tighter constraint on $H_0$ is obtained. The primary cause is that GW can provide a tight constraint on the $H_0$ and thus break cosmological parameter degeneracies. From the right panel of Fig.~\ref{fig:break_mnu}, the upper limit of $\sum m_{\nu}$ is improved to some extent with the addition of GW. Concretely, after adding the GW data, the upper limit of $\sum m_{\nu}$ could be reduced by 13.7\%. For other cosmological parameters $\Omega_{\rm m}$ and $H_0$ could be improved by 62.8\% and 67.7\%, respectively.

\begin{figure*}[!htbp]
\centering
\includegraphics[width=\columnwidth]{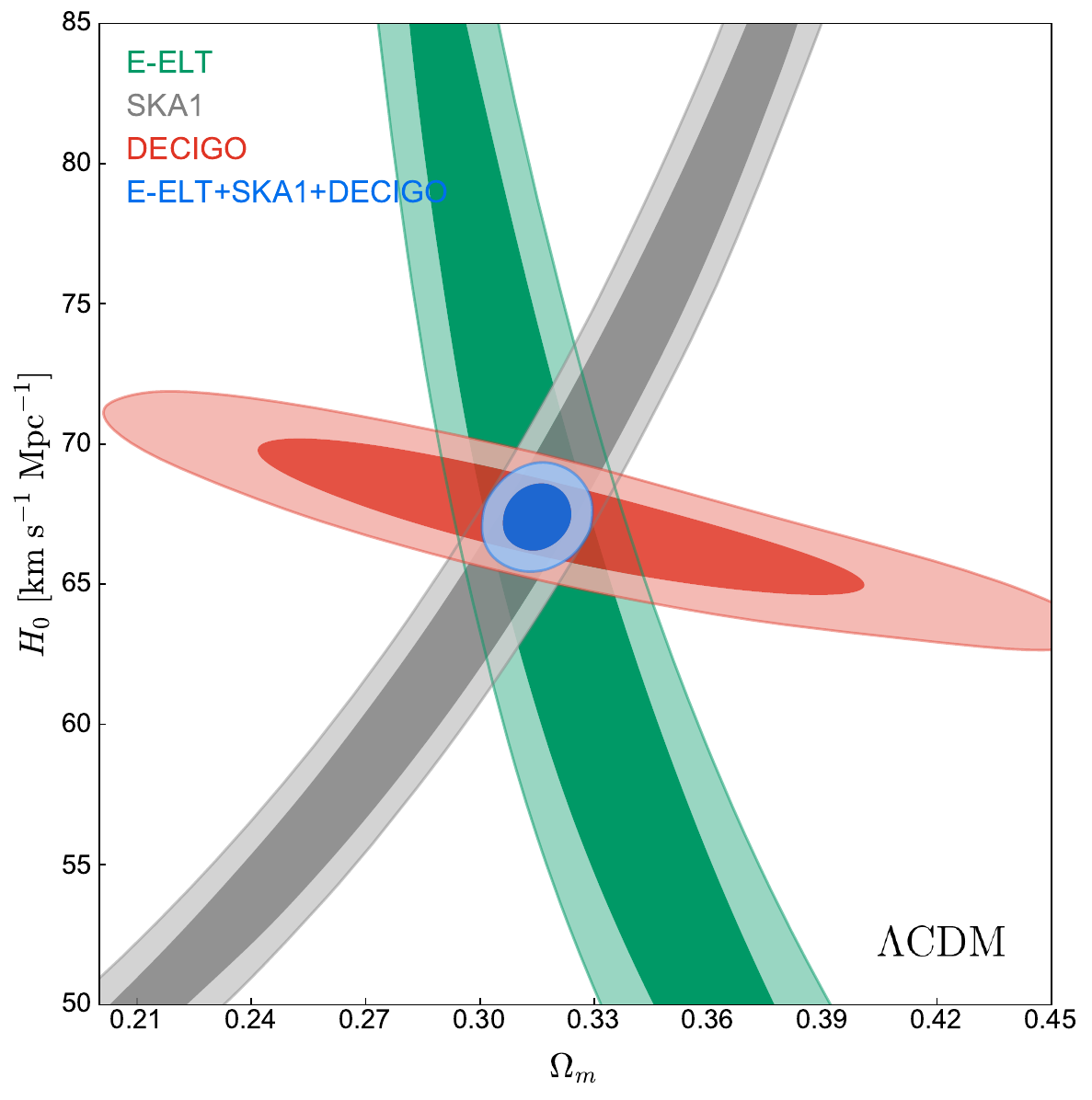}
\includegraphics[width=\columnwidth]{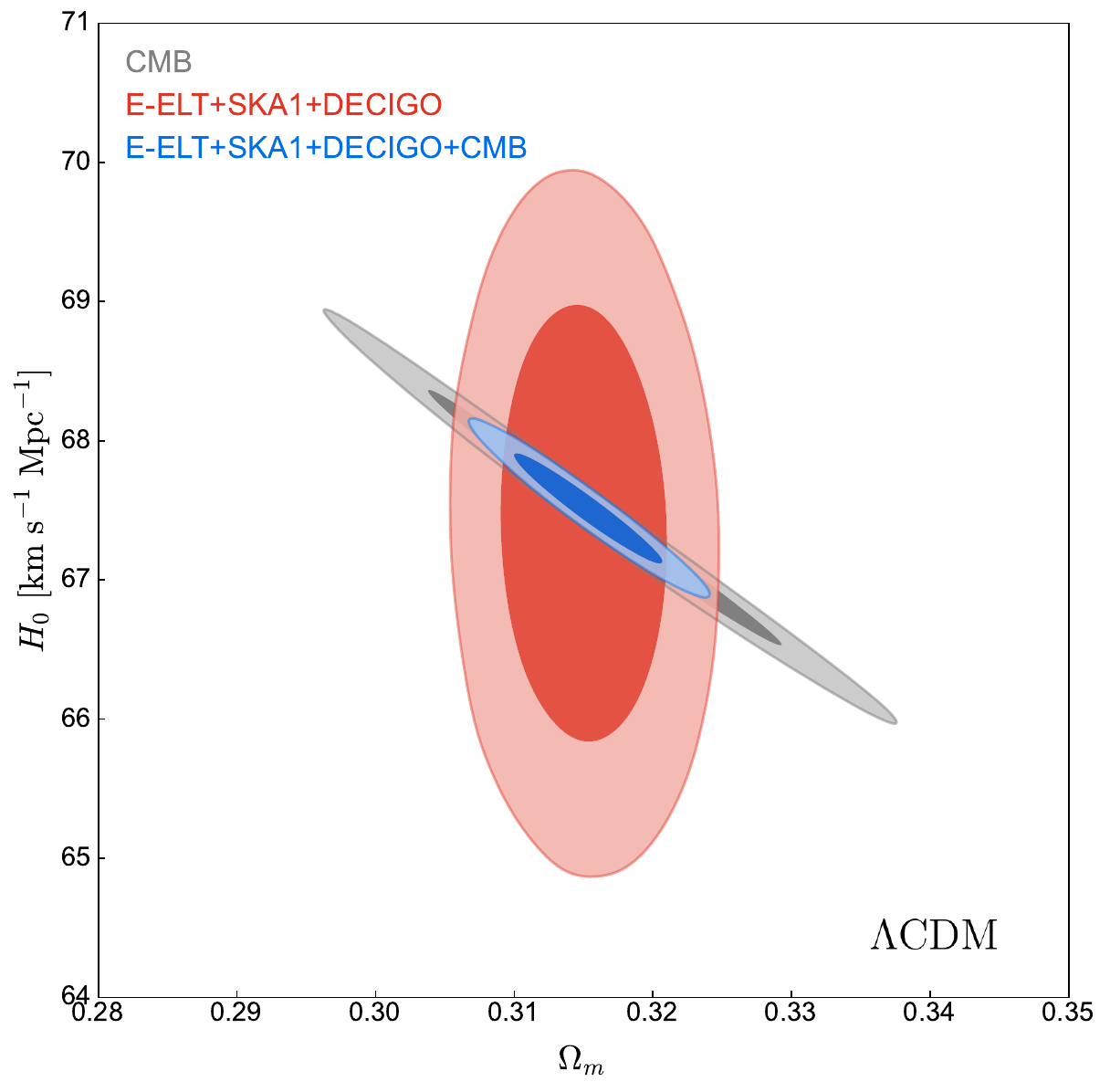}
\caption{Two-dimensional marginalized posterior contours (68.3\% and 95.4\% confidence level) in the $\Omega_{\rm m}$--$H_0$ plane for the $\Lambda$CDM model. Left panel: constraints from E-ELT, SKA1, DECIGO, and their combination. Right panel: constraints from CMB, E-ELT+SKA1+DECIGO, and their combination. Reproduced from Ref.~\cite{Qi:2021iic} with permission.}
\label{fig:break_degeneracy_synergy}
\end{figure*}

In the extended $\Lambda$CDM models, there is one important theoretical possibility that dark energy and dark matter may directly interact with each other via some unknown scalar field degrees of freedom, also referred to as ``the fifth force''. Importantly, IDE models can not only help alleviate the Hubble tension~\cite{Guo:2018ans}, but also help alleviate the cosmic coincidence problem by exhibiting attractor solutions~\cite{Cai:2004dk,Zhang:2005rg}. For the phenomenological IDE models, the form of $Q$, which denotes the phenomenological interaction term describing the energy transfer rate between dark energy and dark matter due to the interaction, is usually assumed to be proportional to the energy density of dark energy or cold dark matter~\cite{Amendola:1999qq,Billyard:2000bh}. Moreover, to balance the dimensions, it must be multiplied by a quantity with units of the inverse of time. The Hubble parameter could provide an analytical solution for the conservation equations.
In Fig.~\ref{fig:break_beta}, we show the constraints in the IDE model. The phenomenological IDE model with $Q=\beta H\rho_{\rm c}$ is considered. Here $\beta$ is a dimensionless coupling parameter that describes the interaction strength between dark energy and dark matter. $\beta>0$ indicates cold dark matter decaying into dark energy, $\beta<0$ indicates dark energy decaying into cold dark matter, and $\beta=0$ indicates no interaction between dark energy and cold dark matter. Note that the dark sirens, considering the tidal effect of BNS, are simulated based on the three-year observation of the ET2CE network.
As clearly seen, the parameter degeneracy orientations of CMB and ET2CE are orthogonal, and thus the combination of them can effectively break the cosmological parameter degeneracies. Moreover, in the IDE model, the role of CMB and GW is complementary since GW can provide a tight measurement on $H_0$ and $\Omega_{\rm m}$ and CMB can tightly constrain $\beta$, making the combination of them tightly constrain these cosmological parameters simultaneously. Concretely, the constraints on $\beta$, $H_0$, and $\Omega_{\rm m}$ could be improved by 69.6\%, 97.6\%, and 96.4\%, respectively. Additionally, the contour of CMB+ET2CE is much smaller than that of CMB+BAO+SN, showing the potential of such dark sirens in helping probe the IDE models.

We conclude that GW standard sirens are not only crucial for addressing the Hubble tension, but also hold significant potential for improving constraints on other cosmological parameters by breaking the degeneracies inherent in current EM observations. While this subsection presents representative constraint results, we refer the reader to the corresponding literature for details regarding the underlying assumptions and analysis methodologies. The capability of GW standard sirens to break cosmological parameter degeneracies inherent in the current EM observations has also been explored in Refs.~\cite{Zhang:2018byx,Jin:2020hmc,Qi:2021iic,Jin:2021pcv,Wu:2022dgy,Jin:2023zhi,Jin:2023sfc,Li:2023gtu,Jin:2023tou,Han:2023exn,Zheng:2024mbo,Han:2025fii}.

\subsection{Synergy with other promising cosmological probes}\label{sec5.2}

As shown in the above subsection, GW performs rather well in breaking cosmological parameter degeneracies inherent in the current EM observations within the mainstream cosmological models. Looking ahead, the next few decades will witness the rise of several other promising cosmological probes, such as fast radio bursts (FRBs), 21 cm intensity mapping (IM), and the SGL\footnote{The cosmological applications of these cosmological probes are detailedly investigated in e.g., Refs.~\cite{Wu:2022dgy,Zhao:2022bpd,Liu:2022bmn,James:2022dcx,Hagstotz:2021jzu,Wu:2021jyk,Deng:2013aga,Gao:2014iva,Zhou:2014yta,Yang:2016zbm,Li:2017mek,Walters:2017afr,Jaroszynski:2018vgh,Liu:2019jka,Liu:2019ddm,Zhang:2020btn,Qiang:2021bwb,Dai:2021czy,Zhao:2021jeb,Zhu:2022mzv,Bhandari:2021thi,Xiao:2021omr,Petroff:2021wug,Caleb:2021xqe,Shao:2023agv,Sun:2024ywb,Xu:2020uws,Zhang:2019ipd,Wu:2022jkf,Jin:2021pcv,Wu:2021vfz,Zhang:2021yof,Zhang:2019dyq,Bull:2015esa,Qi:2022kfg,Wang:2022rvf,Cao:2021zpf,Wang:2019yob,H0LiCOW:2019pvv,Kelly:2014mwa,H0LiCOW:2016tzl,Wong:2016dpo,Birrer:2018vtm,Birrer:2015fsm,Chen:2016fwu,Kumar:2014vvy,Sereno:2013ona} and references therein.}. As a promising and powerful cosmological probe, the question of how GWs can synergize with other potential late-universe cosmological probes to forge a precise cosmological probe to explore the late universe remains an important topic for further investigation. In this subsection, we shall address this question in detail.

\begin{figure*}[!htbp]
\centering
\includegraphics[width=\columnwidth]{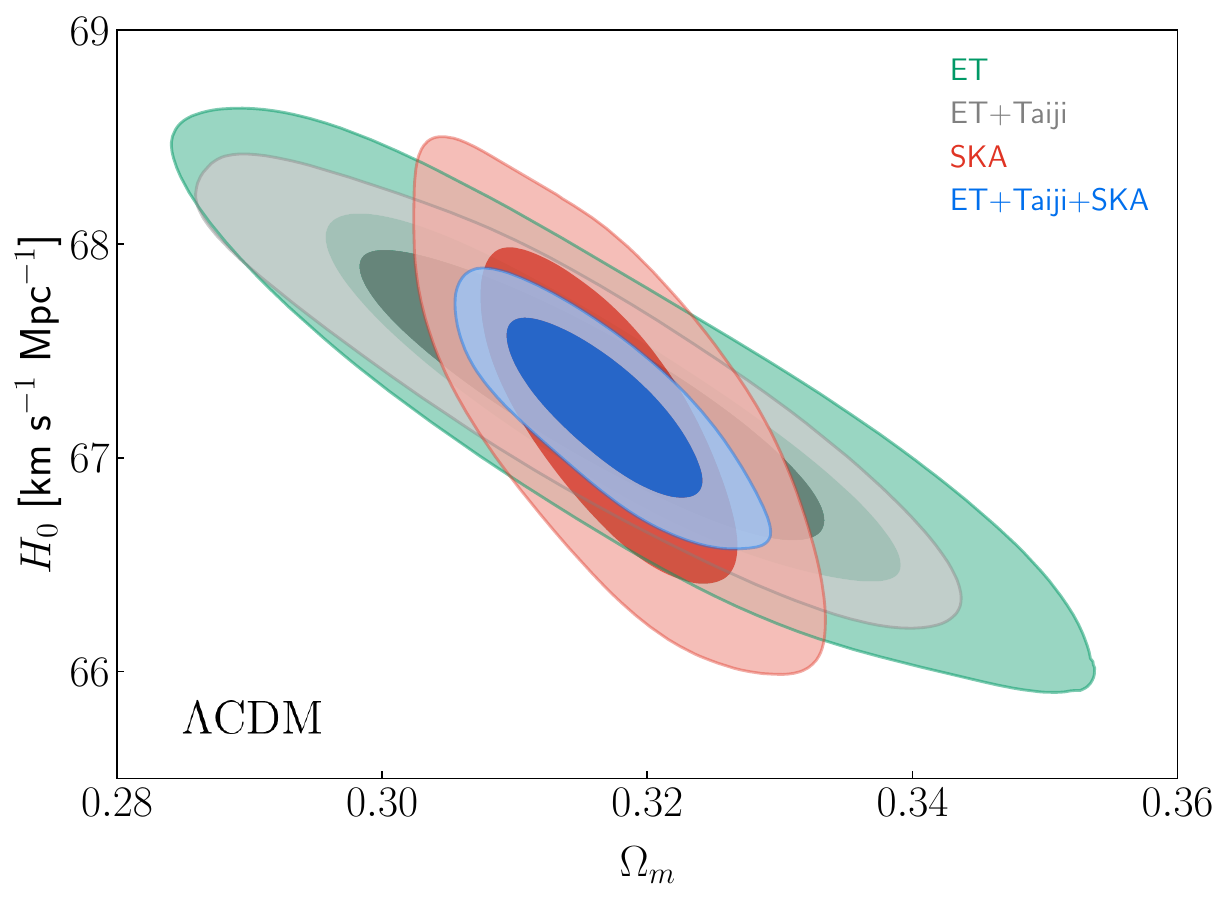}
\includegraphics[width=\columnwidth]{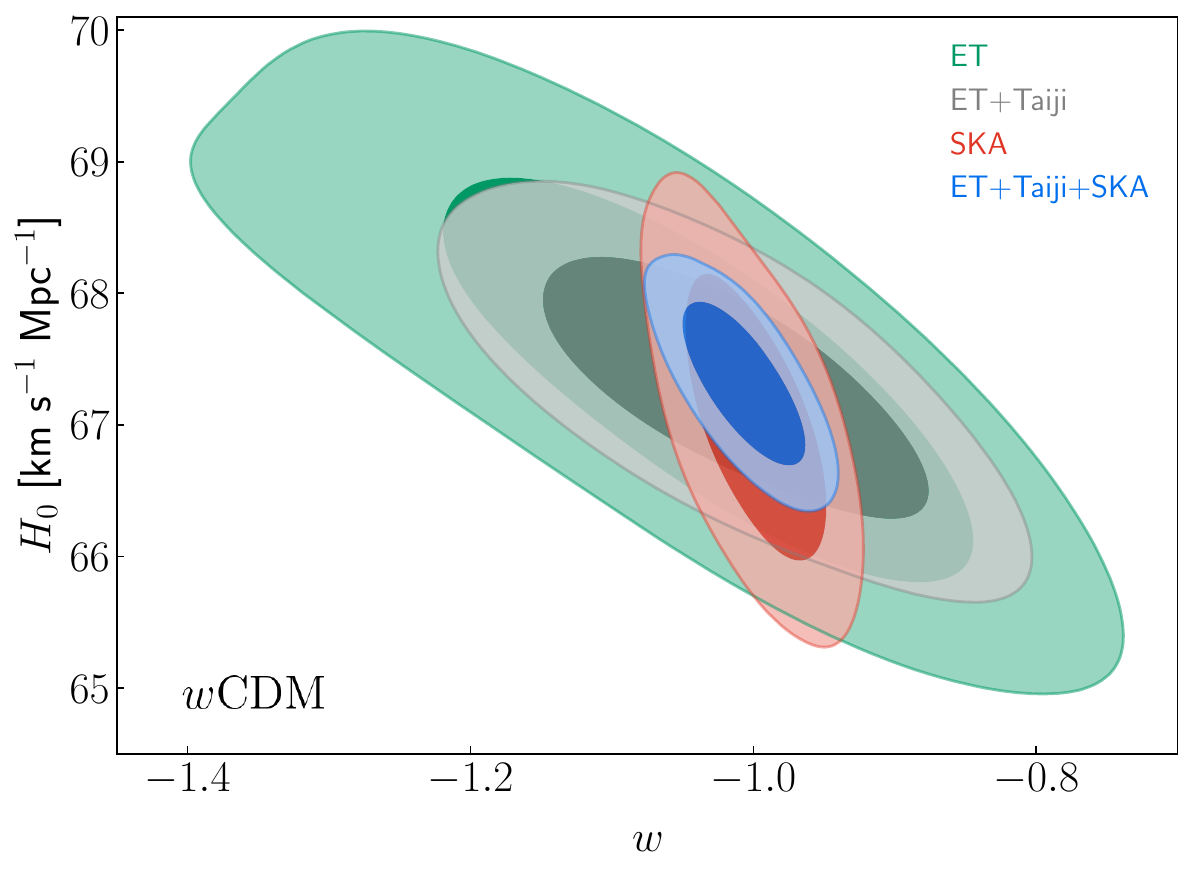}
\caption{Two-dimensional marginalized contours (68.3\% and 95.4\% confidence level) using bright sirens from ET and ET+Taiji, 21 cm IM from SKA, and their combination.
Left panel: constraints in the $\Omega_{\rm m}$--$H_0$ plane for the $\Lambda$CDM model. Right panel: constraints in the $w$--$H_0$ plane for the $w$CDM model. Reproduced from Ref.~\cite{Jin:2020hmc} with permission.}
\label{fig:break_degeneracy_21}
\end{figure*}

Actually, GW standard siren is not the most optimal way to measure the EoS parameter of dark energy $w(z)$. Since there are two integrals in the expression of luminosity distance relating to $w(z)$, the information is lost when one infers $w(z)$ using the distance-redshift relation, shown in Eq.~(\ref{eq:hz_wz}). If $H(z)$ could be directly measured, it would be one of the best ways to constrain $H(z)$ directly. There are several ways to directly measure $H(z)$ from current EM observations, such as the differential age method~\cite{Stern:2009ep} and radial BAO method~\cite{Gaztanaga:2008xz}. While for the future $H(z)$ measurement, redshift drift is one of the most promising ways, also known as the Sandage-Loeb test \cite{Sandage:1962ApJ}. The precise measurement of the redshift drift is expected to be detected by the European Extremely Large Telescope
(E-ELT) and SKA. In addition, Seto et al.~\cite{Seto:2001qf} proposed that $H(z)$ can be measured by the 10-year observation of BNS merger from DECIGO since DECIGO can provide a precise measurement of the phase and frequency evolutions of the GW signal, which is proportional to $H(z)$. Although this method is highly relies on the waveform knowledge. Another method to measure $H(z)$ using GW is to use dipole components of luminosity distance coming from the matter inhomogeneities of large-scale structure and the local motion of the observer~\cite{Sasaki:1987ad}. Nishizawa et al.~\cite{Nishizawa:2010xx} further developed this method using GW standard sirens. By simulating future redshift drift observations from E-ELT and SKA and such dipole measurement of $H(z)$, Qi et al.~\cite{Qi:2021iic} investigated the ability of the multi-messenger and multi-wavelength observations to break the cosmological parameter degeneracies. In the left panel of Fig.~\ref{fig:break_degeneracy_synergy}, the constraints in the $\Lambda$CDM model are shown. As clearly seen, the constraints from E-ELT, SKA1, and DECIGO are rather weak, but they are highly complementary with each other. Therefore, the combination of them could effectively break the cosmological parameter degeneracies and significantly improve the constraints on both $\Omega_{\rm m}$ and $H_0$. Concretely, the constraint accuracy of $H_0$ of E-ELT or SKA1 is improved from $\sigma(H_0)=20~\rm km~s^{-1}~Mpc^{-1}$ to $\sigma(H_0)=0.78~\rm km~s^{-1}~Mpc^{-1}$. While for the EoS parameter of dark energy, after adding GW to SKA1, the constraint on $w$ could be improved by 88\%. From the right panel of Fig.~\ref{fig:break_degeneracy_synergy}, the combination of E-ELT+SKA1+DECIGO also shows a rather different cosmological parameter degeneracy direction from that of CMB, and thus the combination can break the cosmological parameter degeneracies again. Concretely, the constraints on $H_0$ in the $\Lambda$CDM model and $w$ in the $w$CDM model could be improved by 48.0\% and 96.2\%, respectively. Moreover, the constraint precision of $w$ is 1.3\%, close to the standard of precision cosmology.

As another promising and powerful cosmological probe, neutral hydrogen (HI) 21 cm radio observations may also play a crucial role in cosmological parameter estimation, which is widely discussed in Refs.~\cite{Mao:2008ug,Bharadwaj:2008yn,Wuensche:2018alk,Zhang:2019dyq,Jin:2021pcv,Zhang:2021yof,Xiao:2021nmk,Billings:2021yqk,Scelfo:2021fqe,Wu:2021vfz,Wu:2022dgy,Wu:2022jkf,Moresco:2022phi,Berti:2022ilk,Berti:2023viz,Ibitoye:2024qip,Pan:2024xoj}. In the post-reionization epoch of the universe, HI is thought to reside primarily in dense gas clouds embedded within galaxies, making it an effective tracer of the galaxy distribution. However, it is quite challenging to detect a sufficient number of individual HI-emitting galaxies for a precise cosmological analysis. Thus, the 21 cm IM method is proposed by measuring the collective HI emission across relatively large angular scales without resolving individual galaxies. Using such a method, the scale of the standard ruler BAO can be measured, and then the late-universe expansion history of the universe can be probed. We shall first focus on the synergy between GW standard sirens and 21 cm IM. In Fig.~\ref{fig:break_degeneracy_21}, the constraints in the $\Lambda$CDM and $w$CDM models using GW standard sirens from ET, ET+Taiji, 21 cm IM from SKA, and their combination. As clearly seen, 21 cm IM from SKA (denoted as SKA in the following discussion) can provide rather tight constraints on $\Omega_{\rm m}$ and $w$, while GW can tightly constrain $H_0$. Therefore, the cosmological parameter degeneracy orientations of them are different, and the combination of them could significantly break the cosmological parameter degeneracies. In the $\Lambda$CDM model, the constraints on $\Omega_{\rm m}$ and $H_0$ could be improved by 16.7\% and 45.1\%, respectively, with the addition of the GW data. While for the $w$CDM model, the improvements of $\Omega_{\rm m}$, $H_0$, and $w$ are 28.6\%, 40.3\%, and 15.2\%, respectively. Moreover, the joint constraints of GW and 21 cm IM could give $\sigma(w)=0.028$, which is comparable with the result from the latest CMB+BAO+SN constraint~\cite{DES:2025bxy}. Note that for the 21 cm IM, there is a strong degeneracy between $H_0$ and $r_d$ (the sound horizon at the drag epoch where baryons decouple from photons) and thus 21 cm IM can measure $H_0r_d$. In this analysis, the Planck best-fit $\Lambda$CDM model is chosen as the fiducial model to generate the 21 cm IM data, in which the Planck best-fit value of $r_d$ is adopted in order to break the strong degeneracy between $H_0$ and $r_d$. Thus, the constraints from 21 cm IM is contributed by CMB to some extent. If the actual data would be used to resolve the Hubble tension, any parameter priors from CMB should be avoided.

\begin{figure*}[!htbp]
\centering
\includegraphics[width=\columnwidth]{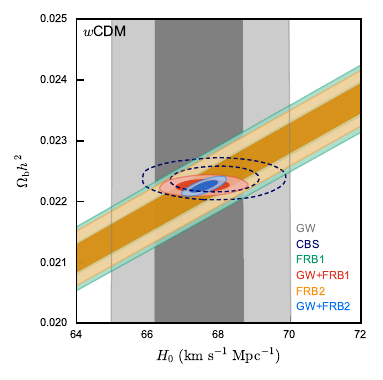}
\caption{Two-dimensional marginalized posterior contours (68.3\% and 95.4\% confidence level) in the $H_0$--$\Omega_{\rm b}h^2$ plane for the $w$CDM model using 10-year GW bright sirens from ET, CBS, the simulated FRB1 ($10^5$) data, the combination of FRB1 and GW, the simulated FRB2 ($10^6$) data, and the combination of FRB2 and GW. Reproduced from Ref.~\cite{Zhang:2023gye} with permission.}
\label{fig:break_degeneracy_frb}
\end{figure*}

FRBs are a class of millisecond-duration, high-energy, and transient astronomical radio pulses. A propagating FRB could have interactions with free electrons in the plasma and generate dispersion, which results in its lower-frequency signal being delayed. The amount of dispersion is characterized by the dispersion measure (DM), which quantifies the integrated column density of free electrons along the line of sight. Since the value of DM is related to the baryonic and distance information along the FRB signal's path, the relation between dispersion measure and redshift has been established, known as the Macquart relation~\cite{Macquart:2020lln}. By using such relation, the cosmological parameter estimations could be performed. 
In Fig.~\ref{fig:break_degeneracy_frb}, we show the constraint results of the $w$CDM model in the $H_0$ and $\Omega_{\rm b}h^2$ plane by using 10-year GW bright sirens from ET, CBS, the simulated FRB1 ($10^5$) data, the combination of FRB1 and GW, the simulated FRB2 ($10^6$) data, and the combination of FRB2 and GW. As seen, GW cannot provide a constraint on $\Omega_{\rm b}h^2$, but can provide a tight constraint on $H_0$. While FRB can constrain $\Omega_{\rm b}h^2$ to some extent, but cannot provide well constraint on $H_0$ because the dispersion measure is proportional to $H_0\Omega_{\rm b}$. Due to this, FRB provides a rather different cosmological parameter degeneracy direction from that of GW. Therefore, they are complementary in constraining cosmological parameters, and the combination could effectively break the cosmological parameter degeneracies. The contours of both GW+FRB1 and GW+FRB2 are smaller than those of CBS, showing the strong potential of GW and FRB in precisely measuring cosmological parameters. Moreover, the constraint precision of $w$ using GW+FRB2 is 1.6\%, which is much better than the current measurements.

\begin{figure*}[!htbp]
\centering
\includegraphics[width=\columnwidth]{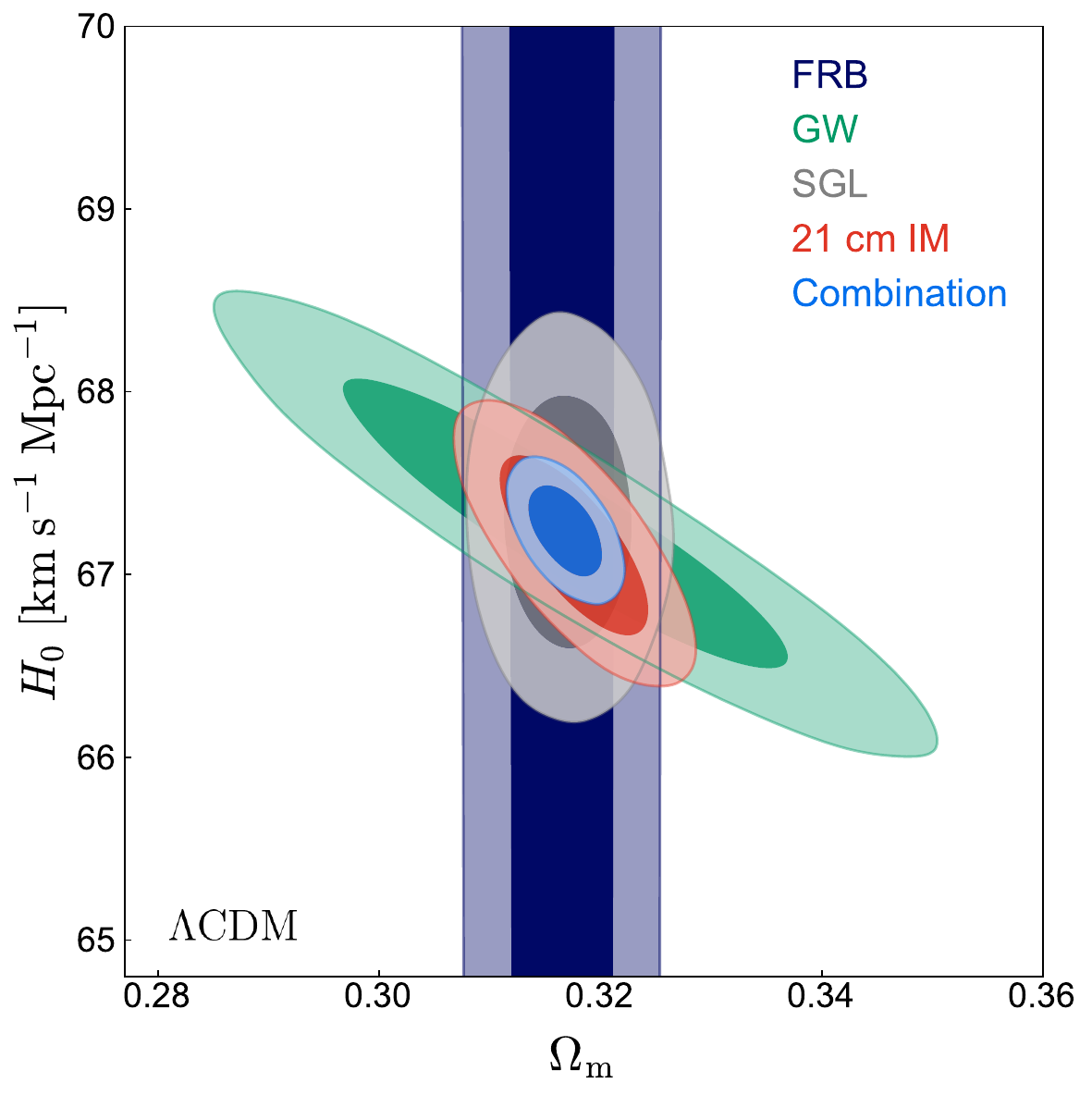}
\caption{Two-dimensional marginalized posterior contours (68.3\% and 95.4\% confidence level) in the $\Omega_{\rm m}$--$H_0$ plane for the $\Lambda$CDM model using the $10^5$ simulated FRB, 1000 simulated bright sirens from ET, simulated SGL (8000 VD events and 55 TD events), 21 cm IM data from HIRAX, and their combination.
Reproduced from Ref.~\cite{Wu:2022dgy} with permission.}
\label{fig:break_degeneracy_all}
\end{figure*}

As photons from a distant source travel toward detectors on Earth, their trajectories are deflected by the gravitational fields of intervening mass overdensities, such as galaxies, galaxy groups, and clusters. Such SGL systems have been discovered in the past few decades, which greatly enhance two cosmological applications. One is the velocity dispersion (VD) method~\cite{Futamase:2000hnr,Grillo:2007iv}, which combines SGL observations with stellar dynamics in elliptical galaxies. The method is to estimate the mass enclosed within the Einstein radius by measuring the Einstein angle from lensing geometry or by determining the central stellar velocity dispersion through spectroscopic observations. Once the lens mass model is determined, a relation between the Einstein angle and the stellar velocity dispersion can be established to constrain cosmological parameters. Another is the time delay (TD) method~\cite{Refsdal:1964nw,Treu:2016ljm}, which considers the differences in arrival times of light between multiple images of a strongly lensed variable source. 
The measured time delays highly depend on the gravitational potential of the lensing galaxy and a combination of angular diameter distances between the observer, lens, and source, commonly referred to as the time-delay distance. one can infer this distance ratio, which is directly related to cosmological parameters, including $H_0$, by accurately modeling the lens mass distribution and measuring these time delays. The H0LiCOW collaboration used six lensed quasars to measure $H_0$ with a precision of 2.4\%~\cite{H0LiCOW:2019pvv}. In the future LSST era, it is expected that more than 8000 SGL systems with accurately measured velocity dispersion and several dozen systems with accurately measured time delays will be observed~\cite{Oguri:2010ns,Collett:2015roa}. 

In Fig.~\ref{fig:break_degeneracy_all}, 8000 SGL systems with velocity dispersion measurements, 55 SGL systems with TD measurements, $10^5$ FRBs, 1000 bright sirens based on ET, and 21 cm IM data based on the HIRAX experiment~\cite{Newburgh:2016mwi} are simulated for cosmological analysis. As seen, although FRBs cannot provide good constraints on $H_0$, GW, SGL, and 21 cm IM can provide tight constraints on $H_0$. Nevertheless, the cosmological parameter degeneracy orientations of them are different, which means the combination of them can still break the cosmological parameter degeneracies. Moreover, the combination gives $\sigma(w)=0.020$ with a 2.0\% precision. We can also see that such precision is a little bit higher than that of GW+FRB2 in the previous paragraph due to the inclusion of $10^6$ FRBs in FRB2. But a 2\% constraint precision of $w$ in this simulation is a conservative scenario. In addition, several works have explored the use of these cosmological probes for constraining cosmological parameters; see, e.g., Refs.~\cite{Sereno:2011ty,Cao:2011bg,Suyu:2012aa,Yuan:2015npa,Cardone:2015nra,Linder:2016bso,H0LiCOW:2016tzl,Magana:2017gfs,Taubenberger:2019qna,DES:2019fny,Wucknitz:2020spz,Caminha:2021iwo,Wu:2022dgy,Qi:2022kfg,Gao:2022ifq,Jana:2022shb,Grillo:2024rhi,Hu:2024niv,Sahu:2025kts,Zhang:2025yhi} and references therein.

\section{Challenges and the role of machine learning in future standard siren analysis}\label{sec6}

Although GW standard sirens show great potential in helping precisely constraining cosmological parameters, several challenges must be addressed before they can fully realize their potential.
In this section, we shall briefly introduce the systematic uncertainties on standard sirens, the detection challenges, and the emerging role of machine learning techniques in addressing some of these challenges.

\subsection{Systematic uncertainties in standard siren measurement}\label{sec6.1}
% In order to resolve the $H_0$ tension, the systematic uncertainty in the standard siren method has to be well understood. 
In the real GW observations, systematic uncertainties on standard sirens are pivotal to the application of cosmological research. 
First, the measurements of luminosity distances are affected by the uncertainty of the detector calibration~\cite{Karki:2016pht,Sun:2020wke,Huang:2022rdg} and the accuracy of the GW waveform template~\cite{LIGOScientific:2018hze,Cutler:2007mi,Ohme:2013nsa,Samajdar:2018dcx,Berry:2014jja}. 
Second, high-redshift GW sources will be affected by the weak lensing uncertainties~\cite{Cusin:2020ezb,Hirata:2010ba,Holz:2004xx}, while low-redshift GW sources will be affected by the peculiar velocities of the potential host galaxy. 
The methods of reducing the peculiar velocity uncertainties of galaxies have been investigated recently, the updated $H_0$ posterior distributions of GW170817 are reported in Refs.~\cite{Mukherjee:2019qmm,Nicolaou:2019cip,Howlett:2019mdh} (see also Refs.~\cite{He:2019dhl,Nimonkar:2023pyt} for the impacts for future standard siren observations).
Third, as mentioned, luminosity distance has strong degeneracy with the inclination angle, although the degeneracy can be broken by long duration signals, observing radio jet of the EM counterparts, or measuring higher harmonics~\cite{Baibhav:2020tma,Cotesta:2018fcv} (see e.g., Ref.~\cite{Chen:2020dyt} and references therein for related discussions), it remains a limiting factor. 
\blue{Moreover, host galaxy identification and catalog completeness would introduce additional systematic errors: the wrong association of the host galaxy or incomplete galaxy catalogs can bias the inferred redshift distribution, which could then propagate into cosmological parameter estimations (see also Sec.~\ref{sec2.3.2} for the discussion).}
Finally, observational selection effects, especially those related to the detectability of EM counterparts, pose another critical challenge to GW standard siren cosmology. In fact, selection biases associated with EM follow-up are expected to be a dominant source of systematic error~\cite{Chen:2020dyt}. 

\blue{Here are the potential mitigation strategies for these effects.}
\begin{itemize}
    \item \blue{Calibration and waveform systematics: (i) treat calibration and waveform-model errors as nuisance parameters in the inference, (ii) cross-check with multiple waveform templates, (iii) reduce amplitude or phase bias by considering a larger and better-oriented detector network, and including higher-order modes.}
    
    \item \blue{Weak lensing: (i) reduce lensing bias by reweighting with shear maps and large-scale-structure reconstructions from external galaxy surveys (delensing) and (ii) marginalize the redshift-dependent lensing-induced scatter term.}
    
    \item \blue{Peculiar velocities: (i) correct host redshifts with peculiar velocity field models and dense peculiar velocity measurements from galaxy surveys, (ii) consider larger redshift error when velocity flows are uncertain, and (iii) focus on higher-redshift events where peculiar velocity corrections are subdominant, which has shown in the left panel of Fig.~\ref{fig:dlerror}.}

    \item \blue{Host identification and catalog completeness: (i) incorporate a completeness function to account for the missing fraction of galaxies in the catalog, (ii) employ deeper galaxy catalogs with fainter apparent magnitude limits, and (iii) introduce large-scale-structure information to replace the commonly used catalog-completion assumption that galaxies are uniformly distributed in comoving volume, to complete the galaxy catalog better.}
    
    % \item \blue{Host identification and catalog completeness: (i) use strict multi-band (and when possible kinematic) criteria for host galaxy association, (ii) measure the depth and redshift completeness of the used galaxy catalog, and (iii) include the completeness function and photo-$z$ errors in the likelihood.}
    
\end{itemize}

\subsection{From systematics to solutions: the need for machine learning}\label{sec6.2}

As mentioned above, the realization of GW standard siren cosmology faces a number of systematic uncertainties, which ranges from instrumental calibration and waveform modeling to lensing distortions, host galaxy peculiar motions, GW parameter degeneracies, and EM selection effects. Although these challenges are significant, they are not beyond resolution. With the development of next-generation GW detectors and the anticipated increase in event rates, traditional data analysis methods such as matched filtering and Markov Chain Monte Carlo (MCMC) are likely to face serious challenges such as scalability and accuracy limitations, especially in regimes where noise is non-stationary, signals overlap, or other model assumptions break down.

To overcome these barriers and fully exploit the potential of future GW observations, new tools are needed that are both computationally efficient and robust to the complexities of real data. In this context, deep learning and normalizing flow (NF) techniques have emerged as a powerful approach. These methods have demonstrated promising results not only in signal detection and classification but also in parameter inference under non-ideal conditions.

In the following subsections, we shall briefly explore how deep learning is employed to expand the detection volume of standard sirens, enhance the reliability of parameter estimation, and address the computational bottlenecks of standard siren cosmology in the high-event-rate era.

\begin{figure*}[!htbp]
\centering
\includegraphics[width=\columnwidth]{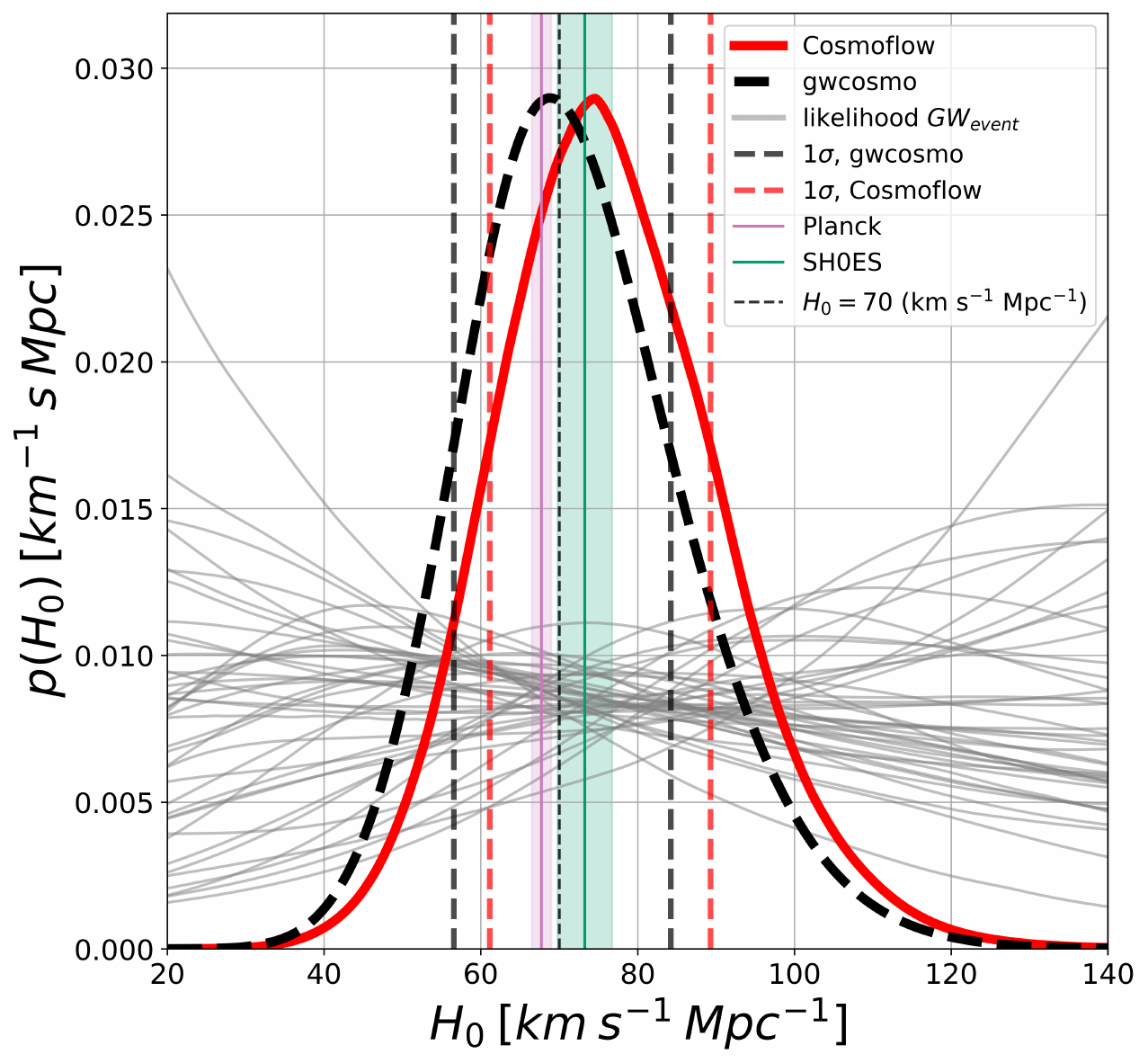}
\caption{Posterior distributions of $H_0$ inferred using the deep learning-based dark siren method. The single event likelihoods for deep learning are plotted in the background in gray. Current Planck and SH0ES estimates of $H_0$ are also plotted in pink and green respectively, with 3$\sigma$ uncertainties. The 1$\sigma$ boundaries are also plotted for both deep learning and Bayesian inference. Reproduced from Ref.~\cite{Stachurski:2023ntw} with permission.}
\label{fig:dl}
\end{figure*}
%%%%%%%%%%%%%%%%%%%%%%%%%%%%%%%%%%%%%%%%%%%%%%%%%%%%%%
\subsection{The role of deep learning}\label{sec6.3}
To enhance the precision and reliability of GW standard sirens, it is crucial to increase the number of detected GW events and improve the accuracy and reliability of GW source parameter inference. Deep learning plays a pivotal role in addressing these challenges, providing advanced techniques for signal detection and parameter inference. \blue{In this subsection we focus on machine-learning contributions that directly serve standard sirens: enlarging the usable event yield, securing well-calibrated posteriors for distance and sky localization, and accelerating hierarchical cosmological inference.}

\subsubsection{Searching for more signals}\label{sec6.3.1}

\blue{For standard sirens, the most direct way to improve cosmological parameter precision is to increase the number of detected signals. For bright sirens, deep learning can rapidly classify signal types~\cite{Cabero:2020eik,Abbott:2021cuf,Qiu:2022wub,Raza:2023gyv,Chan:2024kzu,Raza:2025ouw,Tsukamoto:2025vuu} and provide low-latency sky localization~\cite{Chatterjee:2019gqr,Kolmus:2021buf,Chatterjee:2022ggk,Chatterjee:2022dik,Dax:2024mcn}, which facilitates the prompt identification of BNS candidates and enables rapid EM follow-up. Both of these capabilities rely on the integration of deep learning with high-performance computing, which makes the efficient handling of large data streams possible~\cite{Huerta:2020xyq,Gunny:2021gne}. For 3G GW detectors and space-based detectors, the number and complexity of signals will surpass those of current GW observatories; deep-learning frameworks, which have the potential to accelerate inference by several orders of magnitude, will be essential for data analysis in these facilities~\cite{Alhassan:2022sis,Papalini:2025exy,Ma:2025ulw,hu2025hierarchical}.} 

\blue{For the more numerous dark sirens, expanding the usable sample relies on the generalization ability of deep learning. Neural networks can achieve near-optimal sensitivity within the parameter ranges covered by training, matching or even exceeding template-based searches at fixed false-alarm rates, and have demonstrated some ability to detect signals outside the training distribution in controlled tests. This enables them to extend beyond existing templates and enlarge the discovery space for standard-siren analyses~\cite{Mukund:2016thr,Gabbard:2017lja,Fan:2018vgw,Krastev:2020skk,Wang2024waveformer,George:2016hay,Xia:2020vem,Zhao:2022qob,Beveridge:2023bxa,Wang:2024oei,Wang:2019zaj,Marx:2024wjt,Koloniari:2024kww,Marx:2025jol}.}

\blue{In contrast to conventional template-based methods, deep learning approaches can identify signals that do not conform to pre-established models, including those not present in the training set, thereby further expanding the discovery space for new GW events. Notably, deep learning has been shown to detect signals beyond general relativity, revealing potential new physics~\cite{Wang:2024oei}. However, these capabilities require careful validation through injection studies and cross-checks with standard pipelines to ensure reliability.}

\subsubsection{Ensuring reliable parameter inference}\label{sec6.3.2}
A core feature of deep learning is its capacity for complex modeling, which enhances robustness—especially when applied to parameter inference. As gravitational-wave signal detection and parameter estimation advance, they face numerous challenges such as waveform distortion~\cite{Zhang:2021tdt}, event overlap~\cite{Wang:2023ldq}, and non-stationary noise~\cite{Edy:2021par}, where conventional Bayesian methods often struggle, particularly in non-Gaussian regimes. In these demanding scenarios, deep learning pipelines exhibit the following two major advantages over traditional approaches.

\textbf{\textbf{End-to-end modeling.}} \blue{In addition to serving as effective denoisers that mitigate the impact of non-stationary noise~\cite{Wei:2019zlc,Chatterjee:2021lit,Murali:2022sba,Wang2024waveformer,Xu:2024jlv}, deep-learning architectures learn directly from raw strain data and, crucially, establish a learnable mapping between the data and the physical parameters of interest. This enables them to automatically identify, extract, model, and fit the underlying patterns and features present in the data distribution, thereby achieve more accurate and robust results.} When non-stationary noise degrades data quality or multiple events overlap, Bayesian inference often falters because it assumes Gaussian and stationary noise and focuses on a single signal. Neural networks, by contrast, can disentangle signal and noise with far greater efficiency, making them a powerful candidate for improved gravitational-wave parameter estimation~\cite{Langendorff:2022fzq,Sun:2023vlq,hu2025hierarchical}.

\textbf{Extrapolation and robustness.} Even beyond the parameter ranges covered by the training set, neural networks retain impressive extrapolative power: they can identify binaries whose spins or other parameters lie outside the sampled domain, and they remain stable when non-Gaussian glitches contaminate the strain data~\cite{Xiong:2024gpx,Qin:2025mvj}. \blue{To quantitatively assess the reliability of these extrapolative predictions, Kolmogorov–Smirnov tests have been employed, showing that the resulting posteriors are statistically consistent with expectations. Nevertheless, it is important to note that the posteriors obtained by neural networks in extrapolation regimes are not strictly equivalent to those from Bayesian inference with an explicitly extended parameter space. This subtle but systematic difference underscores the need for careful validation and cross-comparison to ensure the robustness of deep learning-based inference in gravitational-wave cosmology.}

Across these studies, NF networks stand out, and a recent demonstration shows that they can deliver real-time sky-localization alerts and parameter posteriors for binary-neutron-star mergers~\cite{Dax:2024mcn}, which provides a solution for the quick search for EM counterparts. When transferred to GW standard-siren analyses, these strengths allow the network to learn the true posterior distribution from the detectors, tighten the luminosity-distance posterior, and thereby safeguard the reliability of the standard siren method itself.

\subsubsection{Challenges of deep learning in GW standard siren inference}\label{sec6.3.3}

As GW observations enter the high-event-rate era, deep learning—especially NF generative models—has emerged as a promising tool to address the computational bottleneck in standard siren cosmological inference. Recent studies have demonstrated that NF-based frameworks can dramatically accelerate the inference process: for example, NF networks have been applied to the dark siren and spectral siren methods, achieving second-level posterior inference for $H_0$ on real O3 events~\cite{Stachurski:2023ntw} and analyzing hundreds of GW events within minutes~\cite{Leyde:2023iof}. These results highlight the remarkable efficiency of NF networks and suggest the possibility of near real-time standard siren cosmology in the future. However, a critical challenge has also emerged: the $H_0$ posterior distributions produced by NF-based deep learning methods do not always align with those obtained from traditional Bayesian (MCMC) pipelines. As shown in Fig.~\ref{fig:dl}, while the overall $H_0$ constraints are broadly consistent, the detailed structure of individual event posteriors—particularly in the low-$H_0$ region—can differ significantly~\cite{Stachurski:2023ntw}. This discrepancy underscores a fundamental limitation of current deep learning approaches in fully reproducing the results of classical methods.

A key reason for this limitation lies in the ``black box'' nature of deep learning models~\cite{Liang2025SBI}. The inference process of NF networks lacks interpretability, making it difficult to trace how input uncertainties propagate to the final $H_0$ posterior. In practice, this means that it is challenging to disentangle the contributions of statistical and systematic uncertainties in the resulting $H_0$ distribution. Moreover, different NF models trained on the same data can yield noticeably different $H_0$ posteriors, reflecting the sensitivity of deep learning outcomes to training convergence and model architecture. This lack of transparency and robustness complicates the assessment of uncertainty budgets and hinders the identification of the sources of error in cosmological inference. Therefore, before deep learning methods can be convincingly used to search for new physics or claim discoveries beyond the reach of traditional pipelines, it is essential that they first demonstrate the ability to reliably reproduce Bayesian posteriors and provide clear, interpretable uncertainty estimates. Only then can their potential for uncovering novel phenomena in gravitational-wave cosmology be fully realized.

\section{Conclusion}\label{sec7}
Precisely measuring $H_0$ and EoS parameter of dark energy is one of the most crucial missions in the cosmological research. The current cosmological measurements of these two parameters either exhibit significant tension (Hubble tension) or lack the precision required to unveil the fundamental nature of dark energy. GW standard sirens offer a promising and independent method to address these issues.

GW standard siren observations hold high promise that they can become a powerful cosmological probe. The directly obtained luminosity distance from the GW waveform analysis and redshift measured from the EM counterparts, inferred statistically, or others, enable the establishment of the distance-redshift relation, and thereby the measurements of cosmological parameters. Due to the limitations of the current GW detector sensitivity, GW standard sirens can now only measure $H_0$ with large uncertainties.
The best measurement of $H_0$ using the current GW standard siren observations comes from the combination of the bright siren event GW170817, afterglow LC and centroid motion through VLBI, $H_0=68.3^{+4.4}_{-4.3}~\rm km~s^{-1}~Mpc^{-1}$ with a 6.7\% precision, which is much larger than both CMB and distance ladder measurements.

In the next decades, ground-based GW observatories, such as ET and CE, are expected to provide precise measurements of $H_0$, reaching sub-percent level precision through 10-year observations of $\mathcal{O}(10^3)$ BNS mergers. As also an important supplement to GW detection, space-based GW observatories such as Taiji, TianQin, and LISA, operating in the millihertz band, are also expected to measure $H_0$ to sub-percent level precision in the 5-year observation. Furthermore, PTAs aiming nanohertz GW detections from SMBHBs will also potentially offer a rather tight constraint on $H_0$ with the precision close to 1\%. Nevertheless, the ability of the standard sirens to precisely measure other cosmological parameters is limited. In contrast, dark sirens considering the tidal effect in BNS mergers, have the potential to measure the EoS parameter of dark energy $w$ to a sub-percent level precision. However, it requires tracking and making precise tidal-phase measurements of millions of BNS mergers, as well as assuming a reliable EoS model of NS. 

Fortunately, it is found that GW can break the cosmological parameter degeneracies inherent in traditional EM probes, offering a new avenue for joint analyses. In this review, we highlight the importance of combining GW standard sirens with other emerging late-time cosmological probes such as FRBs, 21 cm IM, and SGL to forge a precise cosmological probe for exploring the late universe. In particular, we comprehensively summarize how they can jointly break the cosmological parameter degeneracies and thus improving the measurements on the cosmological parameters. For example, the constraint precision of EoS parameter of dark energy $w$ is expected to be 1.6\% by using the combination of the simulated $10^6$ FRBs from SKA and 1000 bright sirens from ET, representing a promising step toward precision cosmology. We conclude that it is worth expecting that GW can provide a great help for precisely measuring $H_0$ and can forge a precise cosmological probe for exploring the late universe by combining with FRB, SGL, and 21 cm IM.

%%%%%%%%%%%%%%%%%%%%%%%%%%%%%%%%%%%%%%%%%%%%%%%%%%%%%%%
%%% Acknowledgements. 
%%%%%%%%%%%%%%%%%%%%%%%%%%%%%%%%%%%%%%%%%%%%%%%%%%%%%%%
\Acknowledgements{We thank Yue-Yan Dong, Yue Shao, Geng-Chen Wang, Tian-Nuo Li, and Tao Han for helpful discussions. This work was supported by the National SKA Program of China (Grants Nos. 2022SKA0110200 and 2022SKA0110203), the National Natural Science Foundation of China (Grants Nos. 12533001, 12575049, 12473001, and 12305058), the China Manned Space Program (Grant No. CMS-CSST-2025-A02), the 111 Project (Grant No. B16009), the China Scholarship Council, and JST ASPIRE Program of Japan (Grant No. JPMJAP2320).}

%%%%%%%%%%%%%%%%%%%%%%%%%%%%%%%%%%%%%%%%%%%%%%%%%%%%%%%
%%% Conflict of interest. ????????????
\InterestConflict{The authors declare that they have no conflict of interest.}
%%%%%%%%%%%%%%%%%%%%%%%%%%%%%%%%%%%%%%%%%%%%%%%%%%%%%%%

%%%%%%%%%%%%%%%%%%%%%%%%%%%%%%%%%%%%%%%%%%%%%%%%%%%%%%%
%%% Supplements. 
%%%%%%%%%%%%%%%%%%%%%%%%%%%%%%%%%%%%%%%%%%%%%%%%%%%%%%%
%\Supplements{}

%%%%%%%%%%%%%%%%%%%%%%%%%%%%%%%%%%%%%%%%%%%%%%%%%%%%%%%
%%% Reference section. 
%%% citation in the content using "some words~\cite{1,2}".
%%% ~ is needed to make the reference number is on the same line with the word before it.
%%%%%%%%%%%%%%%%%%%%%%%%%%%%%%%%%%%%%%%%%%%%%%%%%%%%%%%

%%%%%%%%%%%%%%%%%%%%%%%%%%%%%%%%%%%%%%%%%%%%%%%%%%%%%%%
%%% Appendix sections. 
%%%%%%%%%%%%%%%%%%%%%%%%%%%%%%%%%%%%%%%%%%%%%%%%%%%%%%%

\bibliographystyle{scpma}
\bibliography{gw_review}
\end{multicols}
\end{document}